**Biological plausibility and stochasticity in scalable VO$_2$ active memristor neurons**

Wei Yi*, Kenneth K. Tsang, Stephen K. Lam, Xiwei Bai, Jack A. Crowell, Elias A. Flores

HRL Laboratories, 3011 Malibu Canyon Rd, Malibu, CA 90265, USA

Neuromorphic networks of artificial neurons and synapses can solve computational hard problems with energy efficiencies unattainable for von Neumann architectures. For image processing, silicon neuromorphic processors outperform graphic processing units (GPUs) in energy efficiency by a large margin, but they deliver much lower chip-scale throughput. The performance-efficiency dilemma for silicon processors may not be overcome by Moore's law scaling of complementary metal-oxide-semiconductor (CMOS) field-effect transistors. Scalable and biomimetic active memristor neurons and passive memristor synapses form a self-sufficient basis for a transistorless neural network. However, previous demonstrations of memristor neurons only showed simple integrate-and-fire (I&F) behaviors and did not reveal the rich dynamics and computational complexity of biological neurons. Here we show that neurons built with nanoscale vanadium dioxide active memristors possess all three classes of excitability and most of the known biological neuronal dynamics, and are intrinsically stochastic. With the favorable size and power scaling, there is a path toward an all-memristor neuromorphic cortical computer.

*Corresponding authors: wyi@hrl.com

The gap between artificial intelligence (AI) and mammal-level intelligence lies in both the architecture and the building blocks. It is unlikely that AI implemented on conventional computing platforms will eventually fill both gaps. Even if the brain's connectivity were reproduced, artificial neurons and synapses built with non-biomimetic CMOS circuits are not capable of emulating the rich dynamics of biological counterparts without sacrificing the energy consumption and size. CMOS-based neuromorphic computing (NMC) hardware suffer from the cost-fidelity dilemma i.e., scalability and biological fidelity are not simultaneously achievable. Although spike domain algorithms are energy savvy, their performance is handicapped by the poor scalability of neuron and synapse building blocks. A survey of chip-scale deep-learning image inference (See Supplementary Fig. 1) reveals that GPUs are the state-of-the-art (SOA) in throughput. However, the higher throughput comes at the cost of lower energy efficiency (EE). By contrast, NMC processors are the SOA in EE, but their throughput is much lower than GPUs'. Regardless of architecture, a universal boundary looks to exist for the throughput·EE product of all CMOS processors, which is likely limited by the CMOS device physics.

Memristors provide an alternative approach to advance NMC. The nonvolatile, stochastic and adaptive passive memristor offers an electronic analogue to biological synapses. The superb scalability of memristor crossbars projects towards the synapse density of the brain ($10^{10}/cm^2$)[1,2]. Recently, biologically plausible self-learning and spike-timing dependent plasticity (STDP) were demonstrated[3,4]. A complementary device, the active memristor, can be used to construct an electronic equivalent of biological neurons. Active memristors show volatile resistive switching and are locally active within a hysteretic negative differential resistance (NDR) regime in current-voltage characteristics. The NDR provides signal gain needed for signal processing. Recently, active memristor based spiking neurons were demonstrated[5] with biomimetic properties such as all-or-nothing spiking, refractory period, and tonic spiking and bursting. However, these demonstrations were interpreted by leaky I&F models[6]. I&F neurons possess much fewer neuro-computational properties[7] than biologically-accurate models, e.g. the Hodgkin Huxley (HH) model[8]. Network-wise, most of the prior art pursued hybrid approaches that combine passive memristors with software neurons or CMOS neurons[9-12]. Such hybrid approaches promise bio-competitive synaptic scalability, but still suffer the poor size and power scalability of Si neurons (See Supplementary Fig. 2). The lack of built-in stochasticity for CMOS neurons is a handicap for achieving complex computational tasks, e.g. Bayesian inference, that require stochastic neuronal populations[13].

In this article, using scalable $VO_2$ active memristors, we show that memristor neurons possess most of the known biological neuronal dynamics. 23 types of biological neuronal behaviors are experimentally demonstrated, including tonic spiking and bursting, phasic spiking (class 3 excitability) and bursting, mixed-mode spiking, spike frequency adaptation, class 1 and class 2 excitabilities, spike latency, subthreshold oscillations, integrator, resonator, rebound spike and burst, threshold variability, bistability, depolarizing after-potential, accommodation, inhibition-induced spiking and bursting, all-or-nothing firing, refractory period, and excitation block. The built-in stochasticity is demonstrated by stochastic phase-locked firing, aka skipping. Finally, our simulations show that the dynamic and static power scaling of memristor neurons project toward biologically competitive neuron density and EE.

**Locally-active memristors**

Chua's memristive theorem[14] proves that a pinched hysteresis in the I-V loci is the only required fingerprint of a memristor. Although a canonical memristor is a passive one-port (two-terminal) circuit element, the same theorem can be applied to a class of one-port devices that exhibit a hysteretic negative differential resistance (NDR, i.e. $\frac{dv}{di} < 0$) in certain region of the I-V loci. If the circuit operating point lies within the NDR regime, e.g. when a resistor load line intersects with the I-V of the nonlinear device in the NDR regime, the device becomes locally active (see Fig. 1e). A locally-active (active hereinafter) memristor can produce an a.c. signal gain greater than 1 and serve as an amplifier, or excite oscillations in appropriate circuits having reactive elements (see Supplementary Fig. 3 and Note 1). Therefore, active memristors can be used as scalable gain elements in information processing. Local activity, together with edge of chaos, are two basic properties for neurons. Chua showed that the locally active domain in the activity diagram of a HH cell is the origin of spikes[15], and it is derived from the locally active regime in the I-V loci of voltage-gated ion channels.

We limit the discussions to active memristors that show current-controlled NDR ("S"-shaped I-V loci in current sweeps), since they are at a high-resistance state when powered off, thus offering low standby power dissipation. Not every hysteretic NDR device is an active memristor, though. A counter example is silicon thyristor, which exhibits a hysteretic NDR but the I-V loci does not pass through the origin (not "pinched"), and therefore it is not a genuine memristor[16]. In contrast to a passive memristor, in an active memristor the hysteresis collapses before the external voltage is removed, therefore the memory effect is transient. Several mechanisms can produce "S"-NDR. It may show up in the electroforming of a passive oxide memristor[17] due to self-heating induced conductivity instability. However, such an NDR is irreversible and vanishes after the device is electroformed. Ovonic threshold switches (OTS) made of amorphous chalcogenides have reversible "S"-NDR[18,19], which can be explained by trap-limited conduction that increases exponentially under high field[20]. OTS are being exploited as an access selector device in memristor crossbars to mitigate the sneak-path issue[19], but their endurance is limited by material degradation due to field stress. A more promising class of "S"-NDR devices is Mott memristors based on thermodynamically-driven Mott insulator-to-metal transition (IMT) in certain transition metal oxides. Mott memristors are more robust since there is no high field or chemical redox reaction involved in the quantum phase transition. Nanoscale $NbO_2$ Mott memristor electroformed from amorphous $Nb_2O_5$ was reported[21]. However, we found that such an electroforming process, in our case $VO_2$ from amorphous $V_2O_5$, produced void in the oxide film and electrode damage, likely due to $O_2$ gas released in the reduction of $V_2O_5$. Electroformed $VO_2$ devices showed poor yield and large variations in switching characteristics, and hence are impractical for circuit applications[22]. In this work, we have developed electroform-free $VO_2$ active memristors on CMOS-compatible $SiN_x$-coated silicon substrates with typical yield >98 % (See Methods and Supplementary Figs. 4-8). These electroform-free $VO_2$ nano-crossbar devices show low device-to-device variability with <13 % coefficient of variation in switching threshold voltage for devices with critical dimension from 50-600 nm, and high switching endurance of >26.6 million cycles without discernible change in device I-V characteristics. The electroform-free $VO_2$ device technology expedited the development of active memristor neuron circuitries that can emulate most of the known neuronal dynamics and cleared the path toward large-scale integrated circuit (IC) implementations. Moreover, $VO_2$ is a superior Mott memristor than $NbO_2$ in both switching speed and switching energy. Simulated Mott transition in $VO_2$ is 100 times faster than in $NbO_2$, and only consumes about one-sixth (16 %) of the energy (See Supplementary Fig. 9, Table 1, Note 1 and Note 3).

**Circuit topology and spiking behaviors of $VO_2$ neurons**

Schematic structure and action potential generation mechanism in a biological neuron is shown in Fig. 1a and 1b. The basic circuit topology of a single-compartment $VO_2$ active memristor neuron is shown in Fig. 1c. The prototype circuit consists of two resistively coupled relaxation oscillators, each having a d.c. biased active memristor ($X_1$ or $X_2$), a parallel membrane capacitor ($C_1$ or $C_2$), and a load resistor ($R_{L1}$ or $R_{L2}$). The oppositely-energized (polarized) memristors $X_1$ and $X_2$ emulate the voltage gated $Na^+$ and $K^+$ membrane protein ion channels, respectively. The basic operation steps of action potential generation is described in Supplementary Fig. 10 and Note 2. Similar circuit concepts emerged in early 1960s, e.g. the "Neuristor" axon first proposed by H. Crane[23,24], but the scalability of these early proposals were poor due to the needs of either inductors[25] or bipolar thyristors[26]. Scalable Neuristor circuitries can be realized by Mott memristors due to their superior $4F^2/N$ ($F$: half pitch, $N$: number of stacked layers) scalability[5]. In our design, the two membrane capacitors are grounded instead of d.c. biased[5], so that the voltages across them are the actual local membrane potential across the nerve cell membrane. This is consistent with the original HH neuron model, except that the single membrane capacitor is divided into two, each closely coupled with a voltage-gated membrane ion channel. Unbiased capacitors also offer more flexibility in IC design. The two-stage circuit has the same dimensionality as the HH model. Its dynamics is described by four coupled first-order differential equations that solve four state variables ($u_1$, $u_2$, $q_1$, $q_2$), wherein $u_1$ and $u_2$ are the normalized metallic channel radii of the memristors, $q_1$ and $q_2$ are the charges stored on the capacitors[5]. Since $q_1$ and $q_2$ are connected to the local membrane potentials $V_{Na}$ and $V_K$ by the linear relationships of $q_1 = C_1 V_{Na}$ and $q_2 = C_2 V_K$, the four state variables can be rewritten as ($u_1$, $V_{Na}$, $u_2$, $V_K$) (See Supplementary Note 4). A benefit of this transformation is that $V_{Na}$ and $V_K$ are straightforward to measure experimentally. Some characters of spiking dynamics, e.g. limit cycle oscillation and bifurcation, can be revealed in the two-dimensional $V_{Na}$–$V_K$ phase plane[6]. Several groups have used a single-stage Pearson-Anson relaxation oscillator as a leaky I&F neuron[27-30]. Having only two state variables, such neurons may provide some simple spiking functionalities, but lack the heterogeneity and ergodicity of neuronal dynamics needed for more sophisticated neural networks.

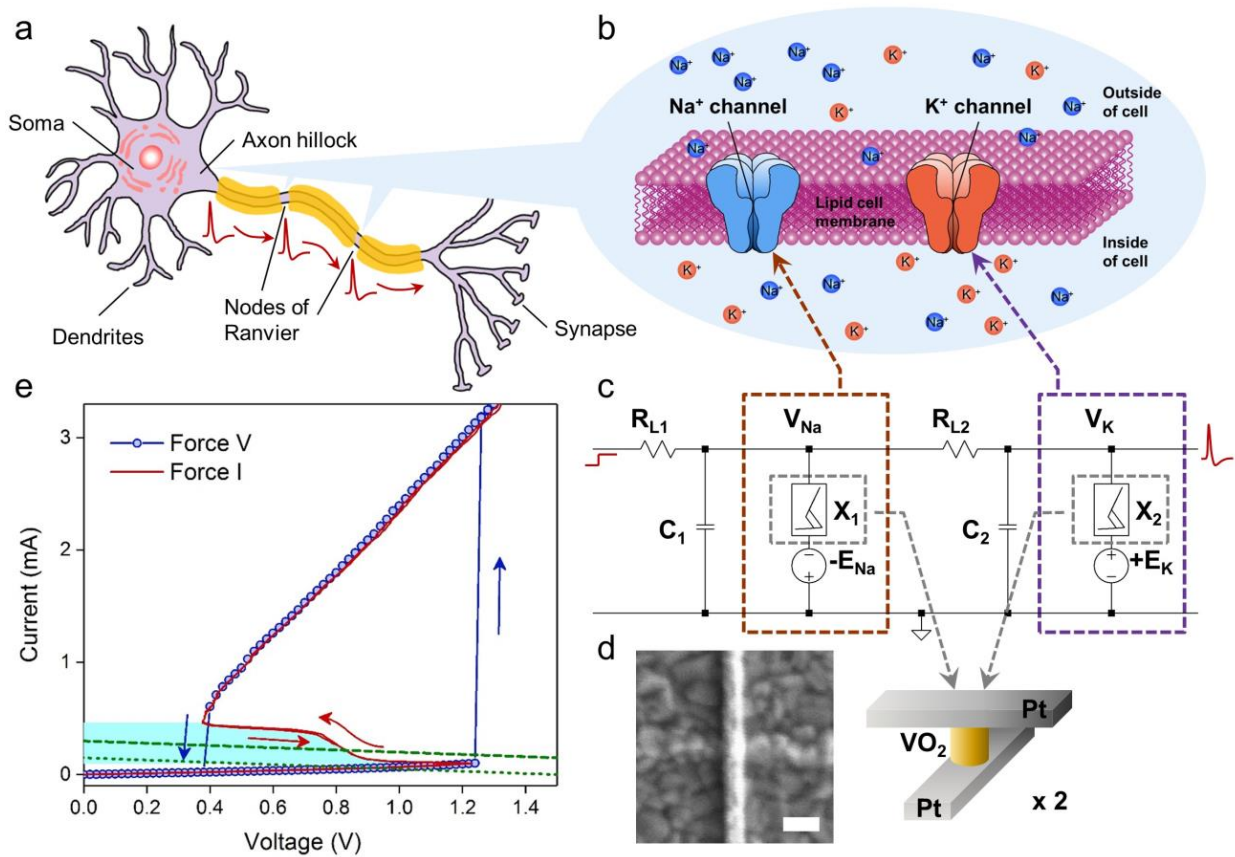

**Figure 1. Circuit diagram of a biomimetic active-memristor neuron and active memristor device characteristics. a**, Schematic structure of a biological neuron, showing that an action potential is fired near the axon hillock (under sufficient input stimulus) and propagates along the cell axon towards the output synapses. **b**, Mechanism of voltage-gated $Na^+$ and $K^+$ ion flows across the cell membrane that accounts for the action potential generation and repetition across the nodes of Ranvier (myelin-sheath gaps). A similar mechanism exists in neurons that lack a myelin sheath. **c**, Basic circuit topology of a two-channel active-memristor neuron to emulate the neuronal dynamics in **b**. A voltage-gated $Na^+$ ($K^+$) channel is emulated by a negatively (positively) d.c. biased active memristor device, which is closely coupled with a local membrane capacitor $C_1$ ($C_2$) and a series load resistor $R_{L1}$ ($R_{L2}$). **d**, Schematic structure and a scanning electron micrograph of a typical $VO_2$ active memristor nano-crossbar device ($X_1$ or $X_2$ in **c**). Scale bar: 100 nm. **e**, Typical two-terminal quasi d.c. voltage-controlled (force V) and current-controlled (force I) I-V characteristics of a $VO_2$ active memristor device. A wide hysteresis loop exists in the voltage-controlled mode due to the Mott transitions (blue arrows). The same Mott transitions are manifested by an "S" shaped negative differential resistance (NDR) regime (highlighted by cyan color) with a much narrower hysteresis (red arrows) in the current-controlled mode. In its resting state, the resistor load line for memristor $X_1$ (or $X_2$) intersects with its I-V loci outside the NDR regime (green dotted line). An input current or voltage stimulus can shift the load line into the NDR regime (green dashed line line) and elicit an action potential generation (spiking).

The neuron circuit shown in Fig. 1c continues to fire a train of evenly spaced spikes, or it fires periodic bursts of spikes when stimulated by a steady d.c. current input. These characteristics are known as part of the spiking behaviors for tonically active neurons (TANs). Phasically active neurons (PANs), on the contrary, may fire only a single spike at the onset of the steady d.c. current input due to transient dynamics, and remain quiescent afterwards as the system reaches the steady state. Phasic firing is known as Class 3 excitability[31]. Both TANs and PANs play important roles in the central nervous system. In a number of brain areas such as cortex, striatum, and midbrain, PANs act as differentiators or slope-detectors and are involved in a wide range of processes including motor control, coincidence detection in the auditory brainstem, cognition, and reward-related learning[32,33]. However, there is yet no demonstration of phasic spiking behaviors in active memristor neurons. We found that phasic spiking behaviors can be realized simply by replacing the load resistor $R_{L1}$ with a capacitor $C_{in}$, or by inserting a capacitor $C_{in}$ before $R_{L1}$ in the tonic neuron circuit. In other words, the main difference between TANs and PANs is that TANs have resistively coupled dendritic inputs and PANs have capacitively-coupled dendritic inputs. Otherwise both types of neurons share the same circuit topology. If a capacitor $C_{in}$ is placed in parallel with the load resistor $R_{L1}$, the circuit turns into a mixed-mode neuron, and fires a phasic burst followed by a train of tonic spiking when stimulated by a steady d.c. current input. In biological neurons, Class 3 phasic behavior is attributed to a subthreshold $K^+$ current, acting as a dynamic negative feedback to preclude spiking if the input rises too slowly, and the neuron shows no bifurcation to repetitive spiking no matter how strong the input is[34].

Fig. 2 shows the three $VO_2$ active memristor prototype neuron circuits and the lists of their experimentally demonstrated biological neuron spiking behaviors. In the circuit diagrams, $R_{e1}$ and $R_{e2}$ are parasitic series resistance of metal electrodes in crossbar devices with typical values of several hundred Ohms. Experimental circuit parameters are listed in Supplementary Table 3. Four basic biological neuron spiking behaviors, including all-or-nothing firing (See Supplementary Fig. 11), refractory period (See Supplementary Fig. 12–13), spike frequency adaptation (See Supplementary Fig. 16–17), and spike latency (See Supplementary Fig. 18), are shared properties of both tonic and phasic neurons. Besides these shared spiking behaviors, In TANs, ten unique spiking behaviors are observed, including tonic spiking (See Supplementary Fig. 14), tonic bursting (See Supplementary Fig. 15), Class 1 excitability (See Fig. 4c), Class 2 excitability (See Fig. 4b), subthreshold oscillations (See Supplementary Fig. 19), integrator (See Supplementary Fig. 20), bistability (See Supplementary Fig. 21), inhibition-induced spiking (See Supplementary Fig. 22), inhibition-induced bursting (See Supplementary Fig. 23), and excitation block (See Supplementary Fig. 24). In PANs, eight unique spiking behaviors are observed, including phasic spiking, i.e. Class 3 excitability (See Supplementary Fig. 26), phasic bursting (See Supplementary Fig. 27), rebound spike (See Supplementary Fig. 28–30), rebound burst (See Supplementary Fig. 31), resonator (See Supplementary Fig. 25), threshold variability (See Supplementary Fig. 32), depolarizing after-potential (See Supplementary Fig. 33), and accommodation (See Supplementary Fig. 34). Together with the mixed-mode spiking behavior observed in mixed-mode neurons (See Supplementary Fig. 35), we have observed 23 types of known biological neuron spiking behaviors.

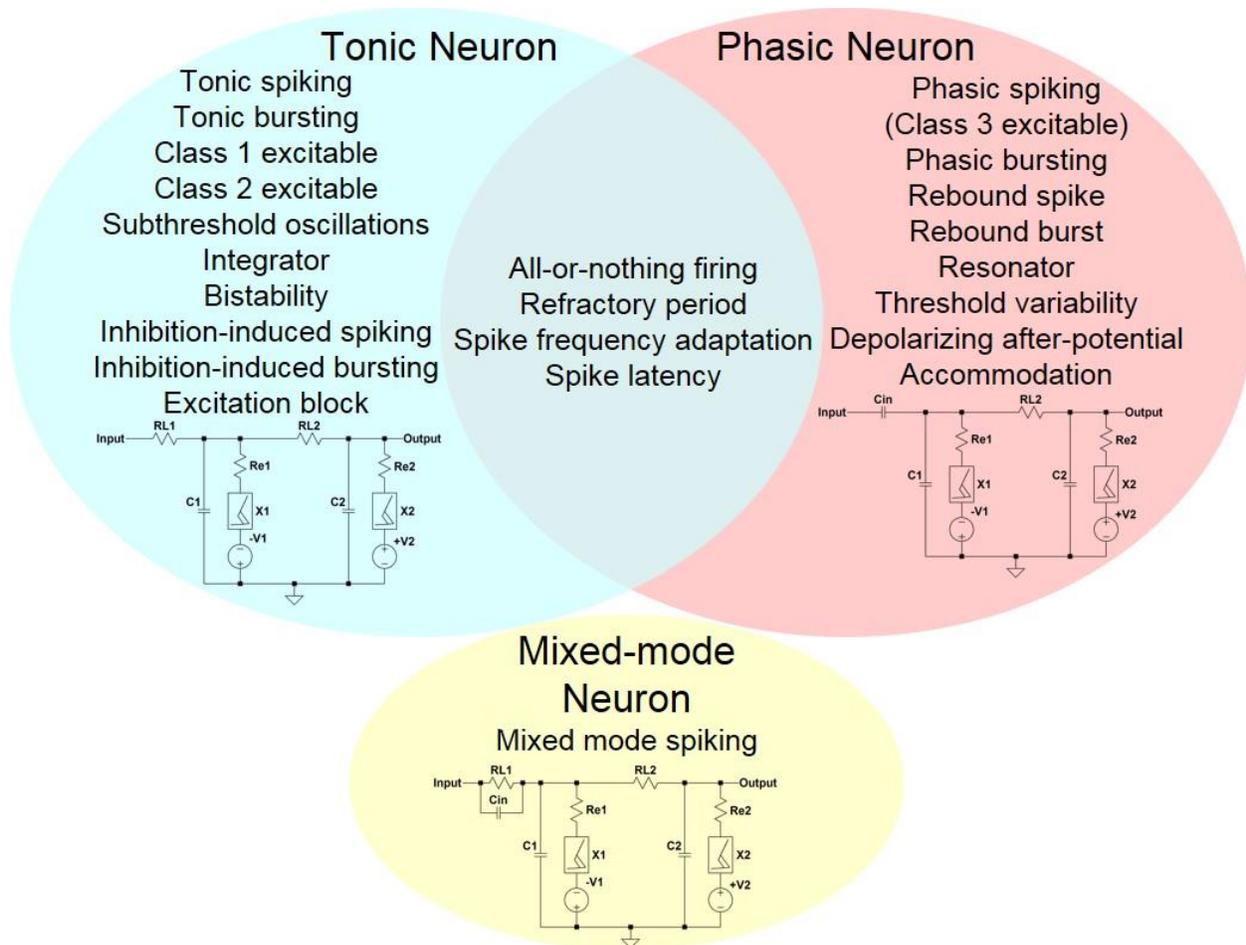

**Figure 2. Three VO$_2$ active memristor prototype neuron circuits and the lists of their experimentally demonstrated biological neuron spiking behaviors. a**, Tonic excitatory neurons, with a resistive coupling to dendritic inputs, show tonic spiking (Supplementary Fig. 14), tonic bursting (Supplementary Fig. 15), Class 1 excitable (Fig. 4c), Class 2 excitable (Fig. 4b), subthreshold oscillations (Supplementary Fig. 19), integrator (Supplementary Fig. 20), bistability (Supplementary Fig. 21), inhibition-induced spiking (Supplementary Fig. 22), inhibition-induced bursting (Supplementary Fig. 23), and excitation block (Supplementary Fig. 24). **b**, Phasic excitatory neurons, with a capacitive coupling to dendritic inputs, show phasic spiking, i.e. Class 3 excitable (Supplementary Fig. 26), phasic bursting (Supplementary Fig. 27), rebound spike (Supplementary Fig. 28–30), rebound burst (Supplementary Fig. 31), resonator (Supplementary Fig. 25), threshold variability (Supplementary Fig. 32), depolarizing after-potential (Supplementary Fig. 33), and accommodation (Supplementary Fig. 34). Other biological neuron spiking behaviors, including all-or-nothing firing (Supplementary Fig. 11), refractory period (Supplementary Fig. 12–13), spike frequency adaptation (Supplementary Fig. 16–17), and spike latency (Supplementary Fig. 18), are shared properties of both tonic and phasic neurons. **c**, mixed-mode neurons, with both resistive and capacitive couplings (R, C in parallel) to dendritic inputs, show mixed mode spiking (Supplementary Fig. 35) behavior.

Fig. 3 summarizes the 23 experimentally demonstrated spiking behaviors in VO$_2$ neurons. All the behaviors are measured from a single tonic, phasic, or mixed-mode neuron circuit that consist of only 2 VO$_2$ active memristors and 4 or 5 passive R, C elements. The VO$_2$ memristors are homogeneous in the sense that they have the same size (100 × 100 nm$^2$) and are fabricated on the same wafer, with small spreading in switching characteristics. The heterogeneity in spiking dynamics is achieved by controllable circuit parameters, i.e. values of R, C elements or input-stage impedance. One can also control the area ratio of the two VO$_2$ devices to achieve asymmetry in the emulated Na$^+$ and K$^+$ ion channels. In comparison, a CMOS artificial neuron constructed with nearly 1300 logic gates (each gate uses a minimum of two transistors) replicated 11 biological neuron behaviors using single neurons, and another 9 behaviors using 2 or 3 neurons[35]. The stark contrast in active device counts between memristor neurons and CMOS neurons to achieve similar level of biological fidelity is a manifestation of the importance of native biomimeticity. Controllability of neuron spiking characteristics is important for network design and neural coding. Spike timing properties, e.g. tonic spike frequency (See Fig. 4c), tonic burst frequency and number of spikes per burst (See Supplementary Fig. 15k, 15l), and spike latency (See Supplementary Fig. 18c), can be directly controlled by the capacitor values or the strength of input stimuli.

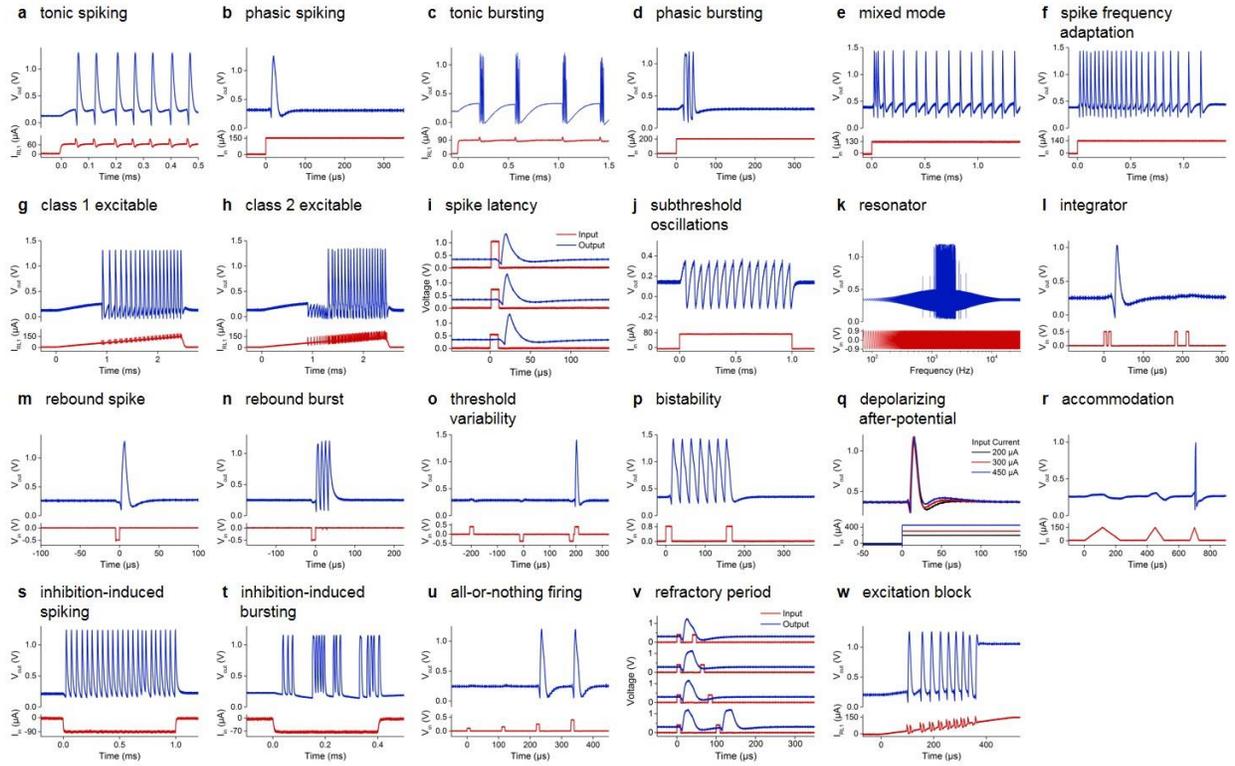

**Figure 3. The 23 biological neuron spiking behaviors experimentally demonstrated in single VO$_2$ active memristor neurons**. **a**, Tonic spiking. **b**, Phasic spiking. **c**, Tonic bursting. **d**, Phasic bursting. **e**, Mixed mode, **f**, Spike frequency adaptation. **g**, Class 1 excitable. **h**, Class 2 excitable. **i**, Spike latency. **j**, Subthreshold oscillations. **k**, Resonator. **l**, Integrator. **m**, Rebound spike. **n**, Rebound burst. **o**, Threshold variability. **p**, Bistability. **q**, Depolarizing after-potential. **r**, Accommodation. **s**, Inhibition-induced spiking. **t**, Inhibition-induced bursting. **u**, All-or-nothing firing. **v**, Refractory period. **w**, Excitation block. All the behaviors are measured from a single tonic, phasic, or mixed-mode neuron circuit that consist of only 2 VO$_2$ active memristors and 4 or 5 passive R, C elements. For more details, see Supplementary Fig. 11–35.

In Hodgkin's classification, there are three basic classes of neuron excitability that can be discerned by spiking patterns: Class 1, Class 2, and Class 3[31]. The nonlinear dynamical mechanism responsible for each class of excitability is reasonably well understood. Class 3 excitability, or phasic spiking, has been discussed above. A closer look at tonic neurons found that they can be converted between being Class 1 and Class 2 excitable, simply by adjusting the $Na^+$ and $K^+$ membrane time constants. In Fig. 4, spike patterns of a tonic $VO_2$ neuron subject to a linearly ramped input current are recorded at different values of $C_1$ and $C_2$ membrane capacitors. Other circuit parameters are left unchanged and the same value is used for the two load resistors $R_{L1}$ and $R_{L2}$. Therefore the membrane time constants are determined by $C_1$ and $C_2$. For the cases of $C_2 < C_1$ (fast $K^+$, slow $Na^+$), the neuron exhibits Class 1 excitability, and shows tonic spiking (if $0.35C_1 < C_2 < C_1$) or tonic bursting (if $C_2 < 0.35C_1$). In Class 1 regime, the observed spike onset threshold and initial frequency are relatively low. Theoretically, the initial frequency can be arbitrarily low. The spike frequency increases with the strength of input current with a pronounced slope. For the cases of $C_2 > C_1$ (slow $K^+$, fast $Na^+$), the neuron exhibits Class 2 excitability with much larger spike onset thresholds. The spike frequency is relatively constant and insensitive to changes in the input strength. In Class 2 regime, spiking is oftentimes preluded by subthreshold oscillations with sawtooth-shaped waveforms, indicating that it is the relaxation oscillation of the $K^+$ channel. In canonical models, Class 1 and Class 2 excitabilities belong to different bifurcations from stable steady state to periodic (spiking) behavior as the stimulus parameter is varied[36]. There are 4 to 6 possible types of bifurcations for each class[37]. To find out the specific bifurcations responsible for the observed classes of excitabilities, a nullcline analysis mapping the four-dimensional nonlinear system into the $V_{Na}$– $V_K$ phase plane without losing part of the dynamics is needed, in a way similar to FitzHugh-Nagumo dimensionality reduction of the HH model[25,38]. Since we experimentally observed excitation block in a Class 2 tonic $VO_2$ neuron (See Supplementary Fig. 24), it is possible that a supercritical Andronov-Hopf bifurcation drives the observed Class 2 excitability[37].

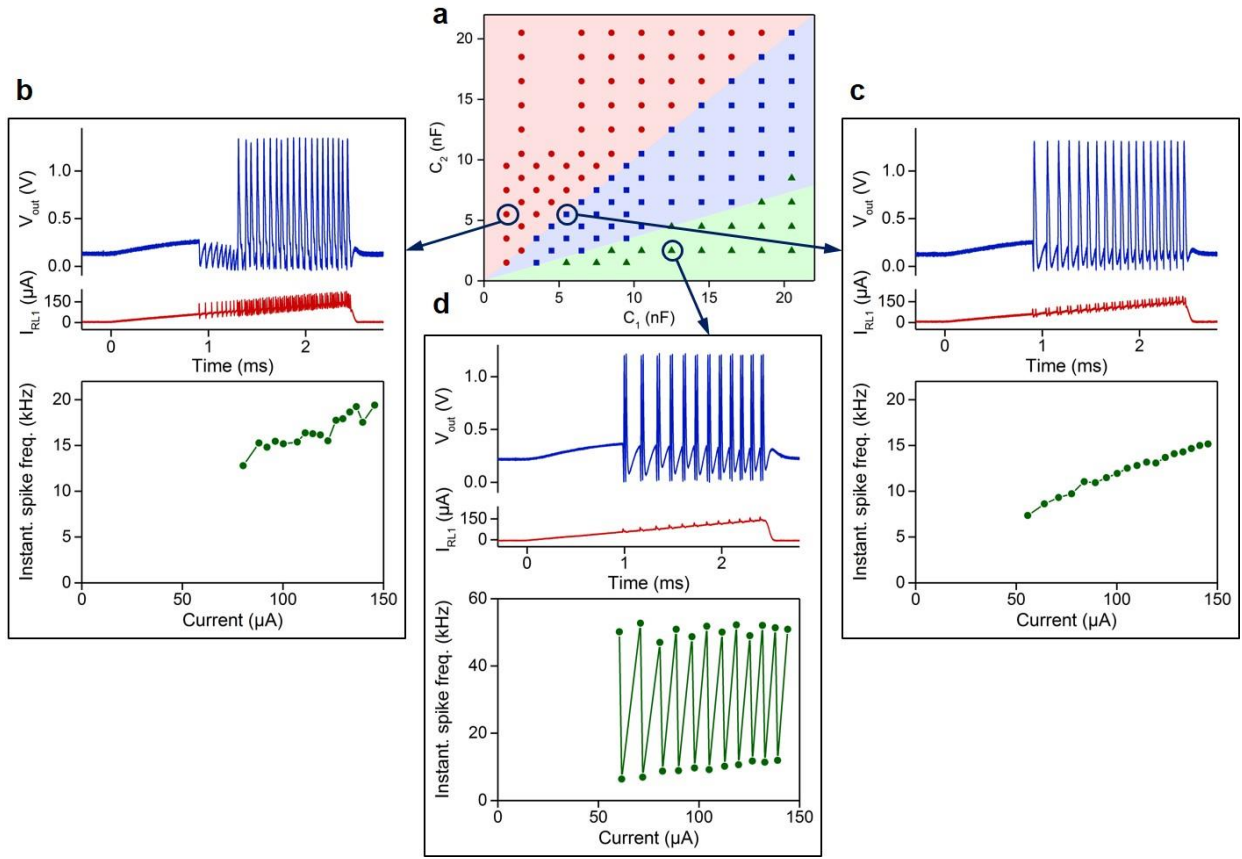

**Figure 4. Capacitance-dependent operating regimes in a tonic $VO_2$ active memristor neuron**. **a**, Diagram of operating regime determined by the values of $C_1$ and $C_2$ membrane capacitors. When $C_2 > C_1$, the neuron exhibits Class 2 excitable spiking and subthreshold oscillations (see **b**). When $C_2 < C_1$, the neuron exhibits Class 1 excitable spiking (see **c**). When $C_2 < 0.35 C_1$, the neuron exhibits Class 1 excitable bursting (see **d**). Various combinations of $C_1$ and $C_2$ are sampled (colored dots) by measuring time dependence of neuron output with a linearly ramped input current. **b**–**d**, Typical neuron input and output vs. time (top panels), and the current-dependence of instantaneous spike frequency (bottom panels) sampled from Class 2 excitable spiking, Class 1 excitable spiking, and Class 1 excitable bursting regimes, respectively.

Finally, we applied the classic stochastic spike train analysis, i.e. the joint interspike interval (JISI) analysis, to study stochasticity and correlation in spike patterns. The results show that $VO_2$ memristor neurons exhibit input-noise sensitive stochastically phase locked firing, aka skipping, in a manner similar to biological neurons[39,40]. In a first-order JISI analysis, the relationship between consecutive spike firings is inspected by analyzing a 3-spike pattern, which includes two interspike intervals (ISIs). An ISI is defined as the time difference between consecutive spikes $\tau_n = t_n - t_{n-1}$, where $t_n$ is the time of occurrence for the $n^{th}$ spike. The scatter plot of ($\tau_n$, $\tau_{n+1}$) pairs, which is referred to as a return map (Poincaré map), is used to reveal the correlation between consecutive first-order ISIs. White noise signals with different amplitudes are superimposed on a steady current clamp (82.5 µA) input for a tonic $VO_2$ neuron, and the excited spike trains within a 35 ms time duration are recorded. Fig. 5 shows the measured spike patterns (only the initial sections of ~3 ms are shown for clarity) and JISI return maps at peak-to-peak white noise amplitudes in the range of 5 µA to 50 µA. When the input noise is low (Fig. 5a), the neuron exhibits regular tonic spiking with ISIs tightly clustered around a predominant fundamental value (ISI median = 29.3 µs). As the input noise increases (Fig. 5b), sporadic drop-outs in firing start to emerge, while most of the firing is still clustered around the fundamental ISI. At even higher noise levels (Fig. 5c, 5d), the neuron exhibits irregular spiking with many drop-outs, resulting in widely scattered JISI pairs in return maps. The JISI pairs cluster around regular grids roughly at multiples of the fundamental ISI, which is also evident in the ISI histograms. Similar arrhythmic firings, or skipping, have been observed in biological neurons such as thermosensitive mammalian cold receptors[41]. These receptors show irregular spikes that are phase-locked to an underlying periodic oscillation, with a random integer number of oscillation cycles skipped between spikes. The firing irregularity increases with temperatures, consistent with our case of elevated input noise. The fundamental ISI, however, is robust against noise and does not shift or vanish. When the firing is missed, the neuron still exhibits subthreshold oscillations with sawtooth waveforms. Scatter recurrence plots of adjacent spike amplitudes (See Supplementary Fig. 36 a–f) show that spike amplitude also develops irregularity and skewness as the input noise rises. The mean spike amplitude first decreases quickly with the input noise, then partially recovers at input noise higher than ~20 µApp. A similar trend is seen in the skewness of spike amplitude distribution (See Supplementary Fig. 36g).

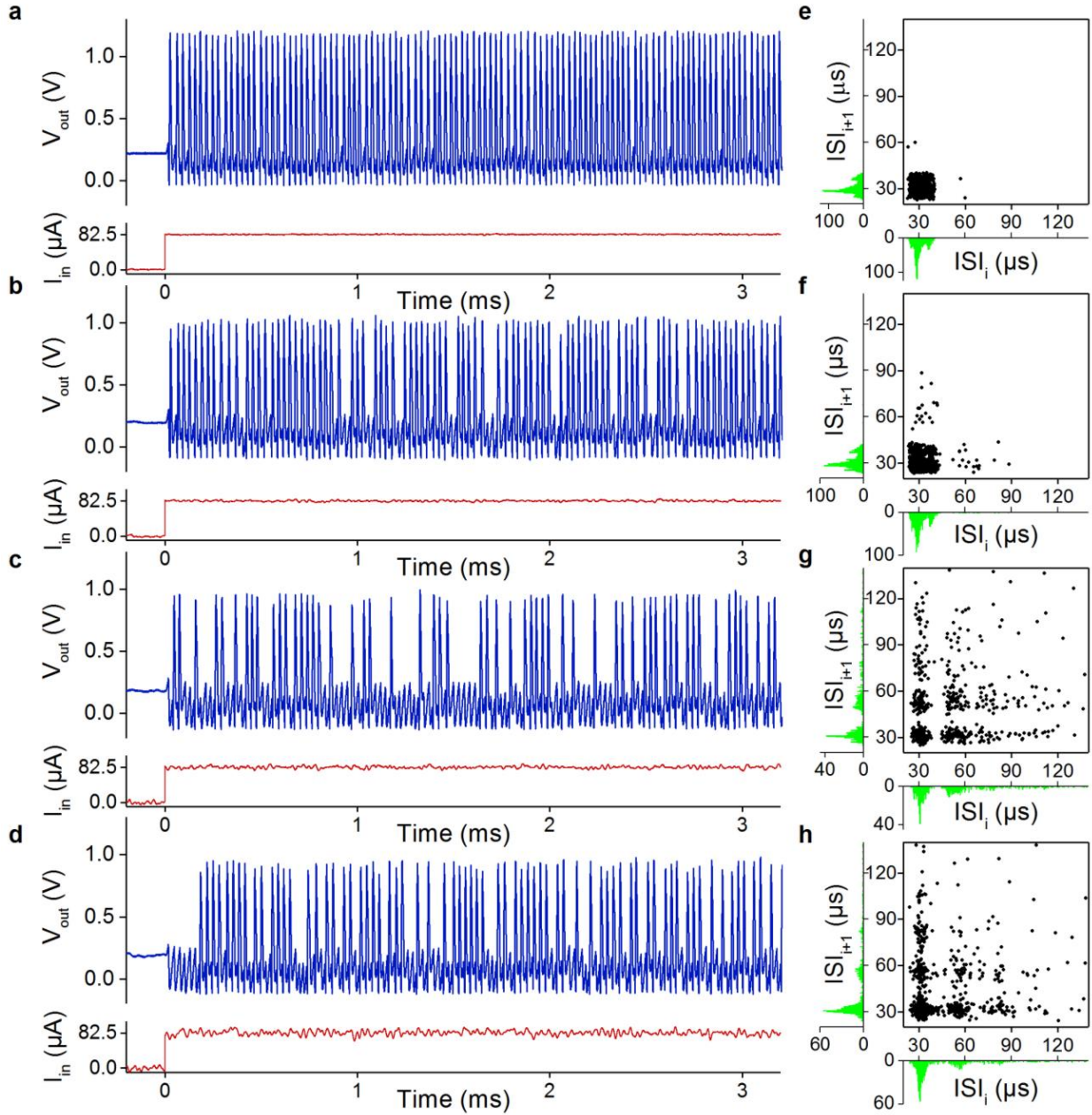

**Figure 5. Stochastically phase-locked firing, or skipping, in a tonic VO₂ active memristor neuron**. **a–d**, Tonic spike trains excited by an input d.c. current of 82.5 µA amplitude and 35 ms duration. For clarity, only the initial sections of ~3ms duration are displayed. White noise signals with 5 µA, 15 µA, 25 µA, and 50 µA peak-to-peak values, respectively, are added to the current input to study its impact on the firing pattern and the correlation between consecutive interspike intervals (ISI). **e–h**, Joint interspike interval scatter plots (*aka* return maps) of the spike trains shown in **a–d**. Also shown are the histograms of the ISI distributions. The numbers of spikes used to generate the JISI plots are 1149, 1113, 620, and 754, respectively.

**Discussion**

For memristor neurons to be biologically competitive, their EE needs to be >$10^{13}$ spikes/J (energy use <0.1 pJ/spike), and their area needs to be <100 µm$^2$ (See Supplementary Fig. 2). We used SPICE simulations to analyze the dynamic and static power scaling of tonic VO$_2$ neurons (See Supplementary Figs. 2, 37, 38, and Note 5). Our simulations show that at 1 fF/µm$^2$ specific membrane capacitance (a value easily attainable by using common dielectrics), VO$_2$ neurons show superior EE-area scaling than the best-case simulated HH cells at neuron sizes smaller than 70 µm$^2$, and can surpass the estimated human brain EE of 1.8x$10^{14}$ spike/J (or 5.6 fJ/spike energy use) at neuron sizes smaller than 3 µm$^2$. The results show that it is feasible for VO$_2$ neurons to achieve biologically-competitive EE and area.

In our case, a variety of neuronal dynamics is achieved by customizing the passive R, C elements (and circuit topology) without the need of varying VO$_2$ device parameters. This simplifies the IC design and fabrication. To achieve adaptivity in neuronal dynamics, such as evolution from Class 1 to Class 2, one may introduce passive memristors or memcapacitors[42] in lieu of fixed R, C elements, if integration of memristor and memcapacitor technologies on the same substrate can be resolved.

Very recently, unsupervised all-memristor learning and pattern classification have been demonstrated using single-stage memristor neurons[30]. We have simulated a simple one-neuron one-synapse circuit using a VO$_2$ model neuron and a TaO$_x$ passive memristor model[3] (See Supplementary Fig. 39). Without needing to adjust the VO$_2$ or TaO$_x$ model parameters, simulated TaO$_x$ synaptic weight (conductance) can be continuously increased or decreased by the spikes sent from the VO$_2$ neuron. This shows the feasibility of analog potentiation and depression as the precursors for biologically-plausible STDP learning. All-memristor neural networks with unsupervised causal STDP learning may provide compelling solutions to computational hard problems such as Bayesian reasoning[43] in highly parallel and energy-efficient fashion.

A potential issue for VO$_2$ neurons is that the Mott transition at near 67 °C posts a stringent requirement for core-level and chip-level thermal management, especially if a CMOS co-processor is involved. This risk can be mitigated by introducing a dopant that may lift the $T_C$ higher. $T_C$ ~96 °C has been reached with 5.9 at. % Ge doping in sputtered VO$_2$ films[44].

Finally, since both active memristor neurons and passive memristor synapses are fabricated from deposited thin film structures, it is feasible to vertically stack repeated pairs of memristive neurosynaptic cores to directly map to the brain cortical layers. A possible fabrication procedure for stackable integrated memristor neuron is shown in Supplementary Fig. 40. The proposed integrated neuron only requires up to three layers of interconnect metals. Passive memristor synapse array can be directly stacked on top. Such a pseudo-three-dimensional (two-and-a-half-dimensional) connectivity cannot be easily achieved using conventional CMOS technology.

## Methods
**VO$_2$ Device fabrication.** The VO$_2$ active memristor devices for the experimental demonstration were fabricated in house using electron-beam lithography, thin-film deposition and liftoff. Bottom electrodes of 50-600 nm width and 30 nm thickness (5nm Ti/25nm Pt) were patterned on silicon nitride covered silicon substrates with a liftoff process. This was followed by a blanket deposition of 100nm-thick polycrystalline VO$_2$ films deposited by reactive sputtering. Finally, top electrodes of 50-600 nm width and 75 nm thickness (5nm Ti/70nm Pt) were patterned with a liftoff process perpendicular to the bottom electrode to form the 50 x 50 nm$^2$ to 600 x 600 nm$^2$ metal/VO$_2$/metal crossbar junction. The nearly-pure monoclinic VO$_2$ phase in the sputtered oxide films is confirmed by comprehensive structural and compositional characterizations, including grazing incidence X-ray diffraction, X-ray photoemission spectroscopy, Rutherford backscattering spectroscopy, and secondary ion mass spectroscopy (See Supplementary Fig. 4). High-resolution transmission electron microscopy reveals a sharp interface between the columnar $VO_2$ grains and the amorphous $SiN_x$ substrate without sign of interface roughening or interfacial layer. The monoclinic $VO_2$ phase is further confirmed by selected area electron diffraction (See Supplementary Fig. 5). In contrast to Ref. 5, the as-deposited $VO_2$ films do not require electroforming, and all the tested devices showed upfront Mott-transition-induced resistive switching and negative differential resistance in their very first current-voltage sweep, as shown in Supplementary Fig. 7.

**Electrical characterization.** Electrical characterization of $VO_2$ active memristor devices and $VO_2$ neuron circuits was carried out using a probe station equipped with four source measure units, an oscilloscope, an arbitrary waveform generator, and a current-to-voltage converter (stimulus isolator). The oscilloscope voltage probes have 10MΩ and <4pF input impedance. The discrete neuron circuits were constructed by connecting the two $VO_2$ memristors on the same wafer to external resistors and capacitors through coaxial cables. Electrical characterizations of the completed $VO_2$ devices, summarized in Supplementary Fig. 6, showed device metrics favorable for large-scale neuron ICs. Electroforming-free and volatile resistive switching are observed for almost all of the as-grown devices, as illustrated by highly uniform switching I-V traces from devices located across all the reticles (See Supplementary Fig. 7). The coefficient of variation in switching threshold voltage is 7% to 13% for device sizes from 50nm to 600nm (See Supplementary Fig. 8). The Mott transition mechanism is supported by the temperature dependence of the zero-bias conductance measured with the wafer mounted on a temperature-controlled heater stage. It shows a thermally activated transport in the insulating state at T < 60 °C, with a single activation energy of ~0.2 eV that is close to reported values[45]. At T > 60 °C, the conductance surges up as the material turns from an insulator to a metal. Typical device yield, reproducible across samples and deposition sessions, is in the range of 98–100 % for the 576 devices (36 reticles and 16 devices/reticle) fabricated on a 3-inch $SiN_x$/Si wafer, for crossbar devices having junction area from 50 x 50 $nm^2$ to 600 x 600 $nm^2$. The switching voltage threshold, ranging from 0.4 V to 1.3 V for all the devices tested so far, is size-dependent and tunable by the $VO_2$ film process conditions. It also scales with the film thickness in theory, which is not yet studied experimentally. Threshold voltages of ~0.5 V or lower is competitive if compared to the supply voltage in advanced CMOS transistors. The robustness in resistive switching is demonstrated by a switching endurance of >26.6 million pulsed-mode on/off switching cycles, without discernible change in the device I-V after the endurance test. The actual endurance number is unknown, but could be several decades higher than the instrumentation-limited measured number.

**SPICE simulations.** The SPICE model used to simulate the VO$_2$ switching dynamics and neuron spiking behaviors is based on the same mathematical equations outlined in Ref. 21. See Supplementary Table 2 in Supplementary Information for the values used for the VO$_2$ material parameters. All the simulated neuron behaviors used the same VO$_2$ device model with a cylindrical-shaped VO$_2$ conduction channel of 56 nm in radius and 100 nm in length to match the actual VO$_2$ crystal volume in 100 × 100 nm$^2$ sized and 100 nm-thick nano-crossbar devices used in the experiments, and only varied the values of R, C elements. Series electrode resistance of 150-500 Ω, and parallel VO$_2$ channel leakage resistance of 13 to 17 kΩ were included in simulations to take into account their effects on the voltage drop across the memristors and the standby current in the insulating phase.

**Data availability.** The data that support the plots within this paper and other findings of this study are available from the corresponding author upon reasonable request.

**Acknowledgements**

This work was supported by HRL Laboratories, LLC. We acknowledge P. A. King and S. J. Kim for fabrication and electrical test support, and T. C. Oh for electrical tests at the early stage of this project.

**Author contributions**

W.Y. conceived and simulated the devices and circuits, designed the experiments and supervised the project. W.Y., S.K.L., X.B., J.A.C., E.A.F. fabricated the devices. W.Y. and K.K.T. designed the electrical test setups and carried tests. W.Y. wrote the manuscript. K.K.T. contributed to analysis of the results. All authors commented on the manuscript.

**Competing interests**

The authors declare no competing financial interests.

**Additional information**

**Supplementary information** is available for this paper at *URL*

**Reprints and permissions information** is available at www.nature.com/reprints.

**Correspondence and requests for materials** should be addressed to W.Y.

**References**

1   Xia, Q. et al. Memristor-CMOS hybrid integrated circuits for reconfigurable logic. *Nano Lett* **9**, 3640-3645 (2009).

2   Khiat, A., Ayliffe, P. & Prodromakis, T. High density crossbar arrays with sub- 15 nm single cells via liftoff process only. *Sci. Rep.* **6**, 32614 (2016).
3   Kim, S., Du, C., Sheridan, P., Ma, W., Choi, S. & Lu, W. D. Experimental demonstration of a second-order memristor and its ability to biorealistically implement synaptic plasticity. *Nano Lett.* **15**, 2203–2211 (2015).
4   Wang, Z. et al. Memristors with diffusive dynamics as synaptic emulators for neuromorphic computing. *Nat. Mater.* **16**, 101-108 (2016).
5   Pickett, M. D., Medeiros-Ribeiro, G. & Williams, R. S. A scalable neuristor built with Mott memristors. *Nat. Mater.* **12**, 114-117 (2013).
6   Lim, H. et al. Reliability of neuronal information conveyed by unreliable neuristor-based leaky integrate-and-fire neurons: a model study. *Sci. Rep.* **5**, 09776 (2015).
7   Izhikevich, E. M. Which model to use for cortical spiking neurons? *IEEE Trans. Neural Netw.* **15**, 1063-1070 (2004).
8   Chua, L. O., Sbitnev, V. & Kim, H. Hodgkin–Huxley axon is made of memristors. *Int. J. Bifur. Chaos* **22**, 1230011 (2012).
9   Burr, G. W. et al. Experimental demonstration and tolerancing of a large-scale neural network (165000 synapses) using phase-change memory as the synaptic weight element. *IEEE Trans. Electron Dev.* **62**, 3498-3507 (2015).
10  Prezioso, M. et al. Training and operation of an integrated neuromorphic network based on metal-oxide memristors. *Nature* **521**, 61-64 (2015).
11  Eryilmaz, S. B. et al. Brain-like associative learning using a nanoscale non-volatile phase change synaptic device array. *Front. Neurosci.* **8**, 205 (2014).
12  Choi, S., Shin, J. H., Lee, J., Sheridan, P. & Lu, W. D. Experimental demonstration of feature extraction and dimensionality reduction using memristor networks. *Nano Lett.* **17**, 3113–3118 (2017).
13  Tuma, T., Pantazi, A., Le Gallo, M., Sebastian, A. & Eleftheriou, E. Stochastic phase-change neurons. *Nat. Nanotech.* **11**, 693-699 (2016).
14  Chua, L. O. & Kang, S. M. Memristive devices and systems. *Proc. IEEE* **64**, 209–223 (1976).
15  Chua, L. O., Sbitnev, V. & Kim, H. Neurons are poised near the edge of chaos. *Int. J. Bifurcation Chaos* **22**, 1250098 (2012).
16  Ascoli, A., Tetzla, R., Chua, L. O., Yi, W. & Williams, R. S. Memristor emulators: a note on modeling. in *Advances in Memristors, Memristive Devices and Systems* (Springer International Publishing AG, 2017).
17  Alexandrov, A. S., Bratkovsky, A. M., Bridle, B., Savel'ev, S. E. & Strukov, D. B. *Appl. Phys. Lett.* **99**, 202104 (2011).
18  Ovshinsky, S. R. Reversible electrical switching phenomena in disordered structures. *Phys. Rev. Lett.* **21**, 1450 (1968).
19  Czubatyj, W. & Hudgens, S. J. Thin-film ovonic threshold switch: its operation and application in modern integrated circuits. *Electron. Mater. Lett.* **8**, 157-167 (2012).
20  Ielmini, D. Threshold switching mechanism by high-field energy gain in the hopping transport of chalcogenide glasses. *Phys. Rev. B* **78**, 035308 (2008).
21  Pickett, M. D. & Williams, R. S. Sub-100 fJ and sub-nanosecond thermally driven threshold switching in niobium oxide crosspoint nanodevices. *Nanotechnol.* **23**, 215202 (2012).
22  Yi, W., Oh, T. C., Crowell, J. A., Flores, E. A. & King, P. A. Low-voltage threshold switch devices with current-controlled negative differential resistance based on electroformed vanadium oxide layer. US Patent Application No. 15/417,049 (2017).
23  Crane, H. D. The neuristor. *IRE Trans. Elect. Comput.* **9**, 370-371 (1960).
24  Crane, H. D. Neuristor - A novel device and system concept. *Proc. IRE* **50**, 2048-2060 (1962).

25    Nagumo, J., Arimoto, S. & Yoshizawa, S. An active pulse transmission line simulating nerve axon. *Proc. IRE* **50**, 2061-2070 (1962).
26    Wilamowski, B. M., Czarnul, Z. & Bialko, M. Novel inductorless neuristor line. *Electron. Lett.* **11**, 355-356 (1975).
27    Farhat, N. H. & Eldefrawy, M. H. Bifurcating neuron: characterization and dynamics. *Proc. SPIE* **1773**, 23-35 (1992).
28    Moon, K. et al. High density neuromorphic system with Mo/$Pr_{0.7}Ca_{0.3}MnO_3$ synapse and $NbO_2$ IMT oscillator neuron. *2015 IEEE Int. Electron Dev. Meeting* https://doi.org/10.1109/IEDM.2015.7409721 (2015).
29    Stoliar, P. et al. A leaky-integrate-and-fire neuron analog realized with a Mott insulator. *Adv. Funct. Mater.* **27**, 1604740 (2017).
30    Wang, Z. et al. Fully memristive neural networks for pattern classification with unsupervised learning. *Nat. Electron.* **1**, 137–145 (2018).
31    Hodgkin, A. L. The local electric changes associated with repetitive action in a non-medullated axon. *J. Physiol.* **15**, 165–181 (1948).
32    Schultz, W. Predictive reward signal of dopamine neurons. *J. Neurophysiol.* **80**, 1-27 (1998).
33    Meng, X., Huguet, G. & Rinzel J. Type III excitability, slope sensitivity and coincidence detection. *Discrete Contin. Dyn. Syst. A* **32**, 2729–2757 (2012).
34    Rothman, J. S. M., P. B. The roles potassium currents play in regulating the electrical activity of ventral cochlear nucleus neurons. *J. Neurophysiol.* **89**, 3097–3113 (2003).
35    Cassidy, A. S. et al. Cognitive computing building block: A versatile and efficient digital neuron model for neurosynaptic cores. *2013 IEEE Int. J. Conf. Neural Netw.* https://doi.org/10.1109/IJCNN.2013.6707077 (2013).
36    Rinzel, J. E., B. Analysis of neural excitability and oscillations. in *Methods in neuronal modeling* (MIT Press, 1989).
37    Izhikevich, E. M. Neural excitability, spiking and bursting. *Int. J. Bifur. Chaos* **10**, 1171–1266 (2000).
38    FitzHugh, R. Impulses and physiological states in theoretical models of nerve membrane. *Biophys J.* **1**, 445–466 (1961).
39    Segundo, J. P., Altshuler, E., Stiber, M. & Garfinkel, A. Periodic inhibition of living pacemaker neurons. I. Locked, intermittent, messy and hopping behaviors. *Int. J. Bifur. Chaos* **1**, 549-581 (1991).
40    Fitzurka, M. A. & Tam, D. C. A joint interspike interval difference stochastic spike train analysis: detecting local trends in the temporal firing patterns of single neurons. *Biol. Cybern.* **80**, 309-326 (1999).
41    Longtin, A. H., K. Encoding with bursting, subthreshold oscillations, and noise in mammalian cold receptors. *Neural Computation* **8**, 215-255 (1996).
42    Di Ventra, M., Pershin, Y. V. & Chua, L. O. Circuit elements with memory: memristors, memcapacitors, and meminductors. *Proc. IEEE* **97**, 1717-1724 (2009).
43    Dagum, P. & Luby, M. Approximating probabilistic inference in Bayesian belief networks is NP-hard. *Artificial intelligence* **60**, 141–153 (1993).
44    Krammer, A. et al. Elevated transition temperature in Ge doped $VO_2$ thin films. *J. Appl. Phys.* **122**, 045304 (2017).
45    Beaumont, A., Leroy, J., Orlianges, J.-C. & Crunteanu, A. Current-induced electrical self-oscillations across out-of-plane threshold switches based on $VO_2$ layers integrated in crossbars geometry. *J. Appl. Phys.* **115**, 154502 (2014).

**Supplementary Figures**

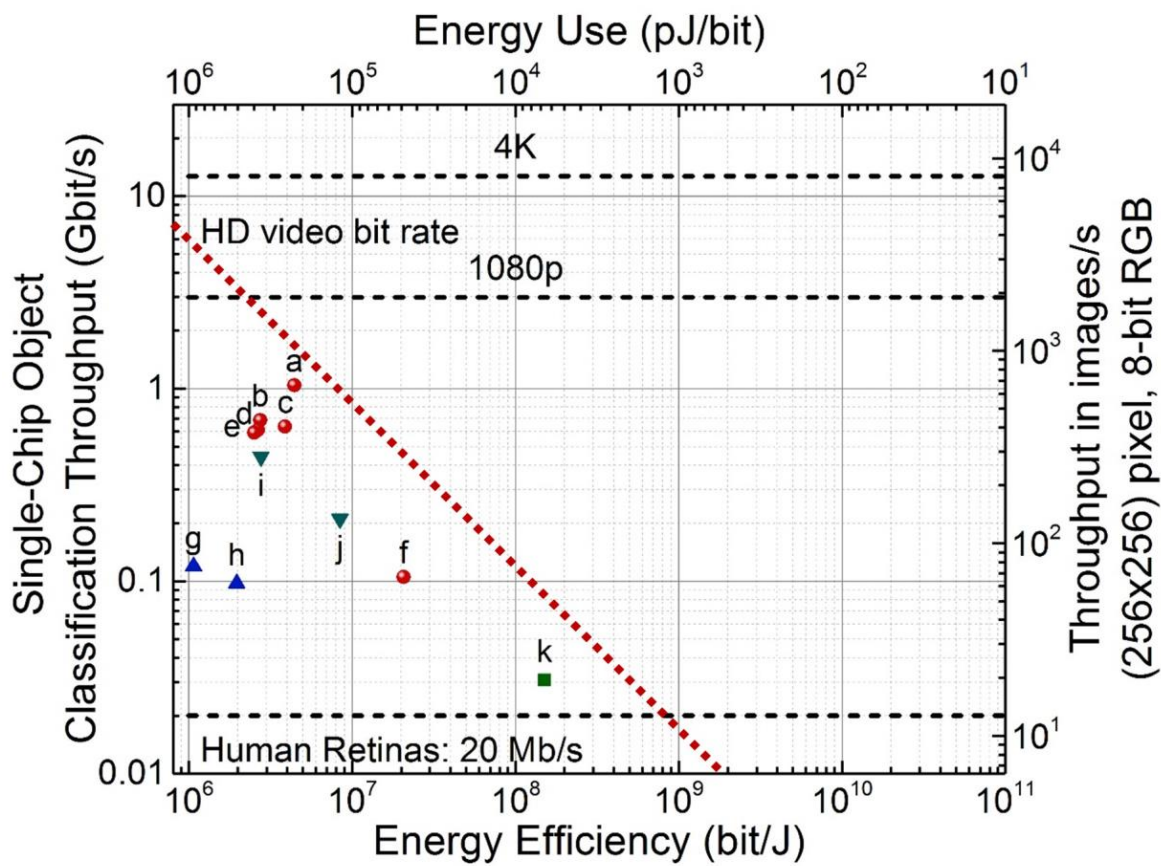

**Supplementary Figure 1. Benchmarks of single-chip image object classification performance and energy efficiency (EE) in silicon GPU (data a-f), CPU (data g, h), ASIC (data i, j), and neuromorphic (NMC) (data k) processors,** all running deep convolutional neural network (CNN) algorithms for inference on standard image sets. A record EE of 6.7 nJ/bit was realized in a silicon NMC processor (TrueNorth, data k)[1], which is one decade better than the record EE of 48.4 nJ/bit for GPUs (Nvidia Tegra X1, data f), and two decades better than the record EE of 509.7 nJ/bit for CPUs (Intel i7 6700K, data h)[2,3]. However, the throughput (in bit/s) of a silicon NMC chip is the lowest among the three categories, reaching only 31 Mbit/s, barely faster than a human operator (~20 Mbit/s). Best-case performance of ASIC (TPUv2, data i)[4] is comparable to that of GPUs. Dotted line is a speculated empirical boundary of the chip-level inference throughput·EE product for the surveyed technologies, i.e. the so-called "performance-efficiency dilemma", possibly limited by the silicon CMOS device physics. The bit rates for 60Hz 1080p or 4K video are 2.99 Gb/s and 12.7 Gb/s, respectively. For a GPU to deliver inference throughput at 1080p HD bit rate, the extrapolated chip-level power consumption is ~3.8 kW. Standard AlexNet architecture and non-batched ImageNet image set (8-bit RGB images with 65536 pixels) are used in GPU and CPU benchmarking. A custom CNN architecture and CIFAR10 image set (8-bit RGB images with 1024 pixels) is used in NMC benchmarking. The reported inference throughput values in frame/s were first converted to pixel/s then to bit/s by the formula of (bit count) = (pixel count) × (color channel) × (color depth) = pixel count × 24.

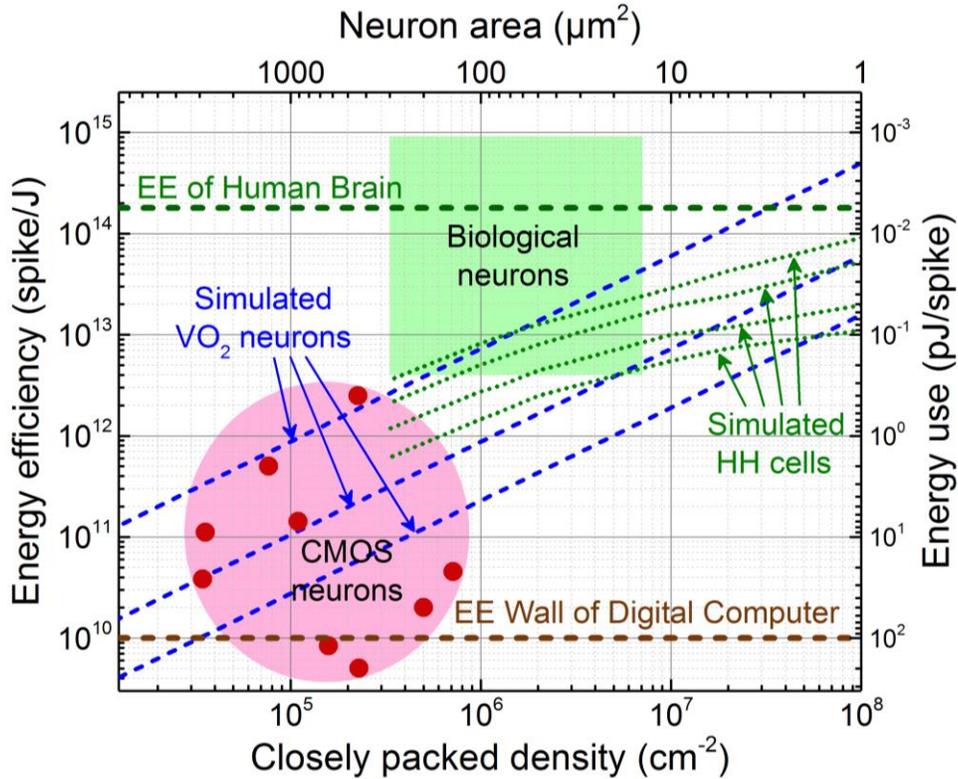

**Supplementary Figure 2. The scaling of neuron energy efficiency (EE, in number of spikes per Joule) vs. neuron area.** Data points encircled by the magenta ellipse are from silicon CMOS neurons published in 2008-2014[5-13], with the lowest reported energy use at 0.4 pJ/spike[9]. The estimated domain of biological neurons is outlined by the green rectangle. Simulated Hodgkin–Huxley (HH) cells (green dotted lines, simulated at ion channel density of 0.5, 1, 2, and 4 µm$^{-2}$ respectively, from top to bottom) illustrate trends of higher EE for smaller neurons, and higher EE for lower ion channel densities[14]. Simulated VO$_2$ memristor neurons (blue dashed lines, at specific membrane capacitance of 1, 10, and 43 fF/µm$^2$, from top to bottom) show a similar trend of higher EE for smaller neurons (but with a higher slope of change), and higher EE for smaller specific membrane capacitance. This is understandable, since the dynamic spiking energy is proportional to the membrane capacitance (see Supplementary Fig. 37). At the same neuron (and capacitor) area, lower specific membrane capacitance translates into lower dynamic spiking energy and hence higher EE. At 1 fF/µm$^2$ specific capacitance (the topmost blue line), VO$_2$ neurons show superior EE-area scaling than the best-case HH cells at neuron sizes smaller than 70 µm$^2$, and can surpass the estimated human brain EE (horizontal green dashed line) of $1.8 \times 10^{14}$ spike/J (or 5.6 fJ/spike energy use) at neuron sizes smaller than 3 µm$^2$. Orange line is the conceived fundamental limit of EE for digital computers[15]. For simplicity, a one-to-one conversion is assumed between the EE of multiply–accumulate operations (MAC) per Joule in a digital computer and the EE of spikes per Joule in the brain, i.e. 1 MAC/J = 1 spike/J. The unit of EE used in simulated HH cells[14] is bit/ATP, which is converted to spike/J by conversion factors of 1 ATP = $10^{-19}$ J and 1 bit = 1 spike. In simulated VO$_2$ neurons, VO$_2$ channel radius (length) is fixed at $r$ ($L$) =10 (10) nm, and the two VO$_2$ memristors contribute 2.3fJ of switching energy in each spike. It is assumed that 80 % of the total neuron area is occupied by membrane capacitors. This ratio may vary with specific designs, but adjusting it will only cause a small lateral shift in the trend lines without affecting the slope and the main conclusions.

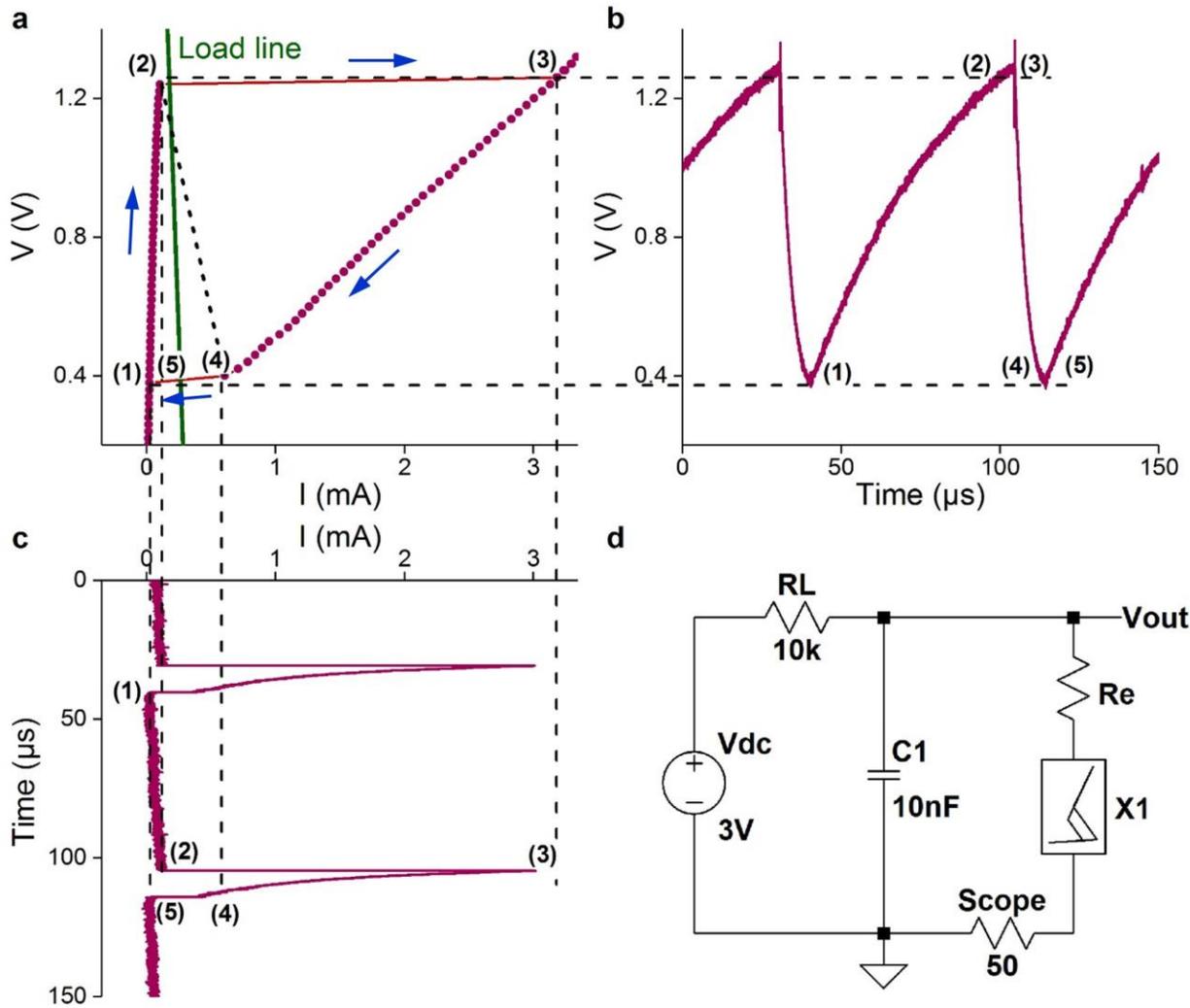

**Supplementary Figure 3. Astable oscillator characteristics measured in a VO$_2$ Pearson-Anson relaxation oscillator. a**, Two-terminal quasi-d.c. V–I characteristics (force V, measure I) of the VO$_2$ device X$_1$ in circuit **d** (including R$_e$) by sweeping the d.c. bias from 0 to 1.6 V and then back to 0. R$_e$ ($\approx$370 $\Omega$) is the metal wire resistance in the crossbar device. To enable astable oscillations, the load line (green), defined by the d.c. bias (V$_{dc}$) of 3 V and the load resistor (R$_L$) of 10 k$\Omega$ in circuit **d**, must intersect the V-I curve in its negative resistance region connecting (2) and (4). **b**, Waveform of the output voltage (V$_{out}$) in circuit **d**, showing sawtooth-shaped relaxation oscillations. **c**, Waveform of the current flowing through the VO$_2$ device X$_1$ (monitored by an oscilloscope channel with 50 $\Omega$ input resistance to ground), showing Mott transitions from (2) to (3) and from (4) to (5). The actual rise/fall time in (2)–(3) and (4)–(5) transitions are much shorter than the sample interval used (2 ns). A complete cycle of astable oscillation from (1) to (5) has four stages (see arrows). From (1) to (2): switch X$_1$ remains open, capacitor C$_1$ is charged till V$_{out}$ reaches the switching threshold of X$_1$. From (2) to (3): closing of X$_1$ causes a surge in current, but V$_{out}$ is held constant by C$_1$. From (3) to (4): capacitor C$_1$ is discharged till V$_{out}$ reaches the minimum holding threshold for X$_1$ to stay metallic. From (4) to (5): X$_1$ is reopened.

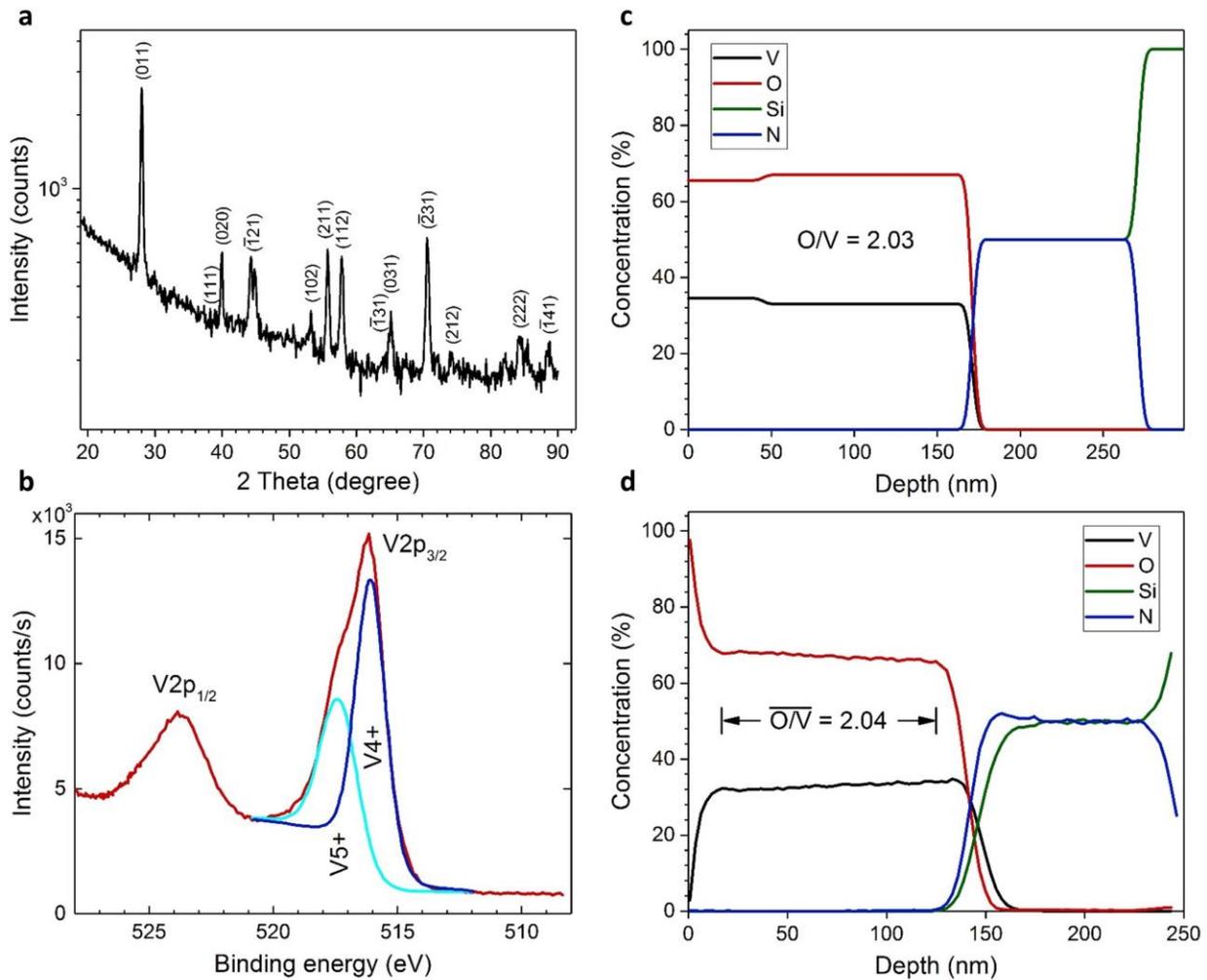

**Supplementary Figure 4. Structural and compositional characterizations of a 100nm-thick VO$_2$ film grown on SiN$_x$/Si substrate. a**, Grazing incidence X-ray diffraction (GIXRD) spectrum acquired at 1.54059 Å Cu Kα1 wavelength. Indexed lines are the results of a phase-identification analysis by using the whole pattern fitting method. The best match (at an R factor of 7.28 %) is found with a monoclinic VO$_2$ phase (space group: P2$_1$/c (14) PDF# [98-001-4290]). The relative intensities indicate some preferred orientation of the crystallites. **b**, High-resolution X-ray photoemission spectroscopy (XPS) of the V2p spectral doublet (V2p$_{1/2}$ and V2p$_{3/2}$). The V2p$_{3/2}$ peak is curve fitted in an attempt to quantify the oxidation states of V, showing a dominating V$^{4+}$ oxidation state (64 %) with the rest of it being at V$^{5+}$ state (36 %). Since XPS only detects the top few nm thickness of the film, the V$^{5+}$ state was possibly due to native oxidation after the film was exposed to air. **c**, Rutherford backscattering spectrum (RBS) showing atomic concentrations of O at (67±4) % and V at (33±1) %, or a O:V ratio of 2.03:1. The thickness in RBS is estimated by assuming a density of 7.15×10$^{22}$ atoms/cm$^3$. **d**, Secondary ion mass spectroscopy (SIMS) showing an average O:V ratio of 2.04:1 (in the depth of 40–120 nm).

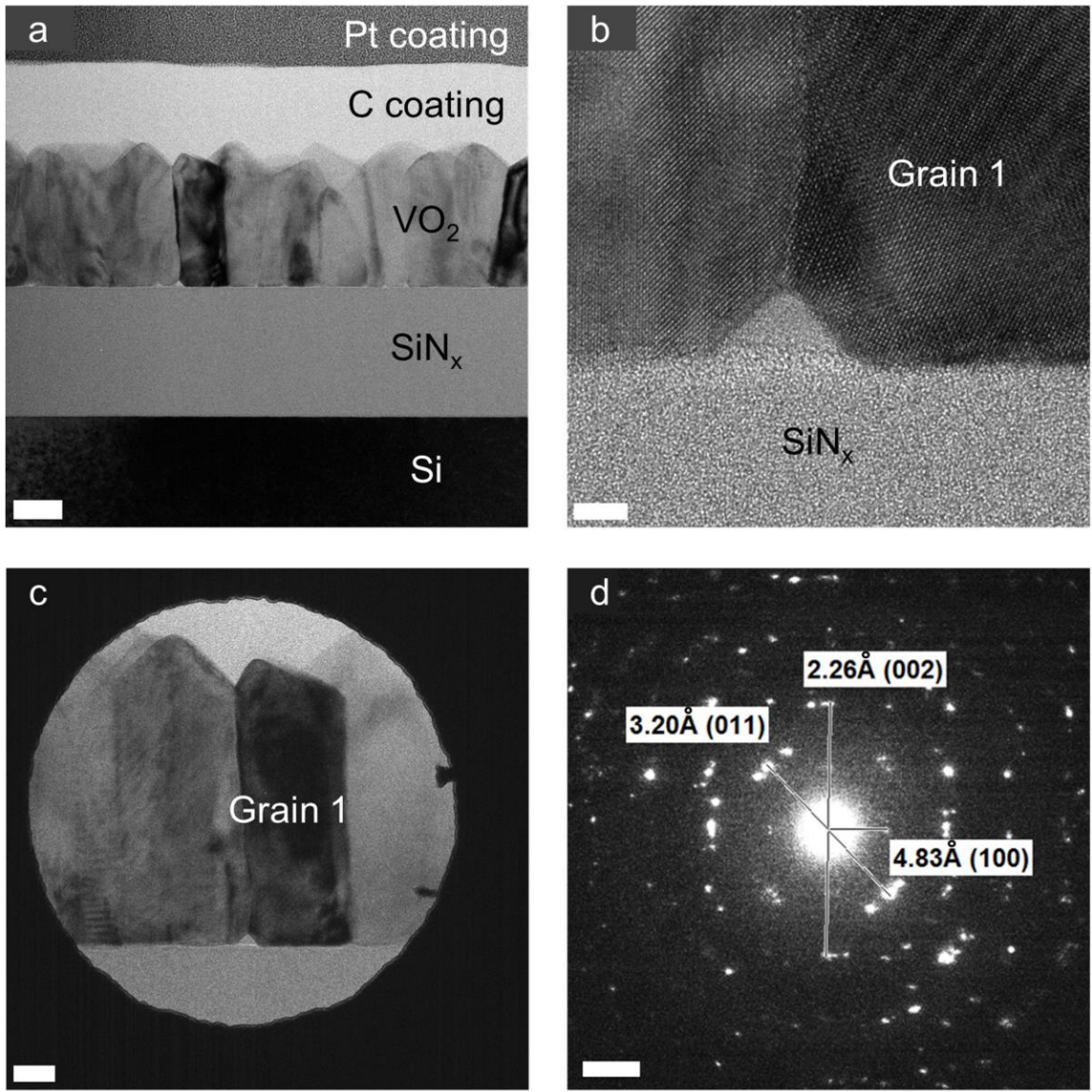

**Supplementary Figure 5. Transmission electron microscopy and electron diffraction characterizations of a 100nm-thick VO₂ film grown on SiNₓ/Si substrate. a**, Cross-sectional bright-field transmission electron microscopy (BFTEM) image of the VO₂ sample prepared by focused ion beam cutting. The polycrystalline nature of the VO₂ film with columnar grain structures is clearly resolved. The brighter and darker contrasts seen across grain boundaries are caused by electron beam diffractions by lattice planes with a slight tilt from one grain to another. Scale bar: 50 nm. **b**, High-resolution (HRTEM) image of the VO₂/SiNₓ interface. Scale bar: 4nm. **c**, BFTEM image of one particular grain ("grain 1") selected for electron diffraction study. Scale bar: 20 nm. **d**, Selected area electron diffraction (SAED) pattern of grain 1 showing diffraction spots that can be matched nearly perfectly with the d-spacing of a few low-index lattice planes, (002), (100), and (011), from a monoclinic VO₂ phase (space group: P2₁/c (14) PDF# [98-001-4290]). Scale bar: 2 nm$^{-1}$.

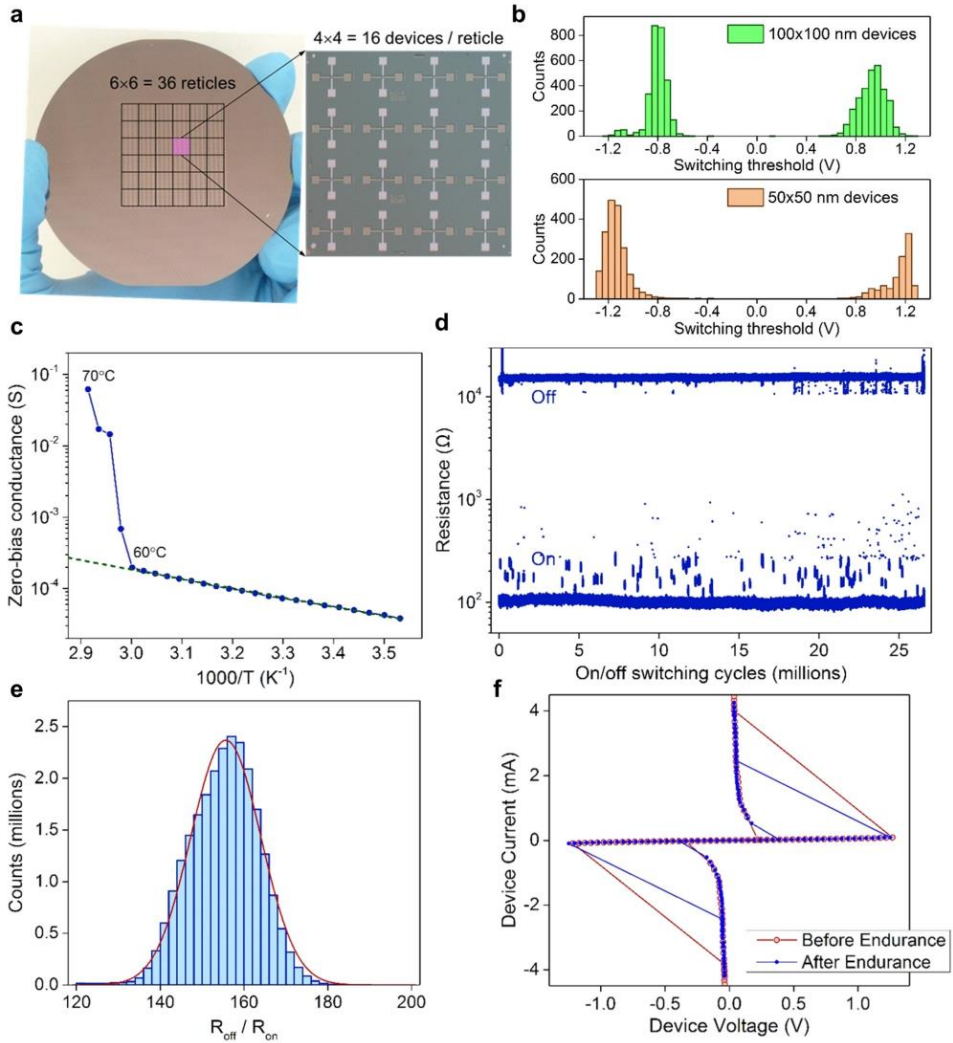

**Supplementary Figure 6. Characteristics of electroform-free VO₂ active memristor devices.**
**a**, Photo of an array of 576 VO$_2$ crossbar devices (36 reticles and 16 devices in each reticle) fabricated on a 3-inch SiN$_x$/Si substrate. **b**, Histograms of the switching threshold voltage measured from 288 $100 \times 100$ nm$^2$ VO$_2$ devices (top) and 288 $50 \times 50$ nm$^2$ devices (bottom) fabricated on the same wafer. The mean values of switching thresholds are size-dependent and tunable by the VO$_2$ film process conditions. **c**, Temperature dependence of device conductance near zero bias (50 mV) measured in a heating cycle, showing a sharp Mott transition when temperature rises above 60 °C. Thermally activated electron transport in the insulating state is shown by a least-square fit ($R^2 = 0.996$) by $\ln(G/T^2)$ vs. 1000/T (dashed line) with a single activation energy of $(0.205 \pm 0.003)$ eV. **d**, Switching endurance data of > 26.6 million pulsed-mode on/off switching cycles. All the switching events were measured without subsampling. The data show no sign of device degradation or drift in resistance values during the endurance test. **e**, Histogram of the $R_{off}/R_{on}$ resistance ratio in the endurance measurement of **d**. Red line is a Gaussian fit with $R^2 = 0.99$, having a center of $155.65 \pm 0.15$ and FWHM of $16.72 \pm 0.30$. **f**, Four-terminal quasi d.c. I–V characteristics (force V, measure I) of the device tested in **d** before and after the endurance test, showing no deterioration or drift in its switching characteristics.

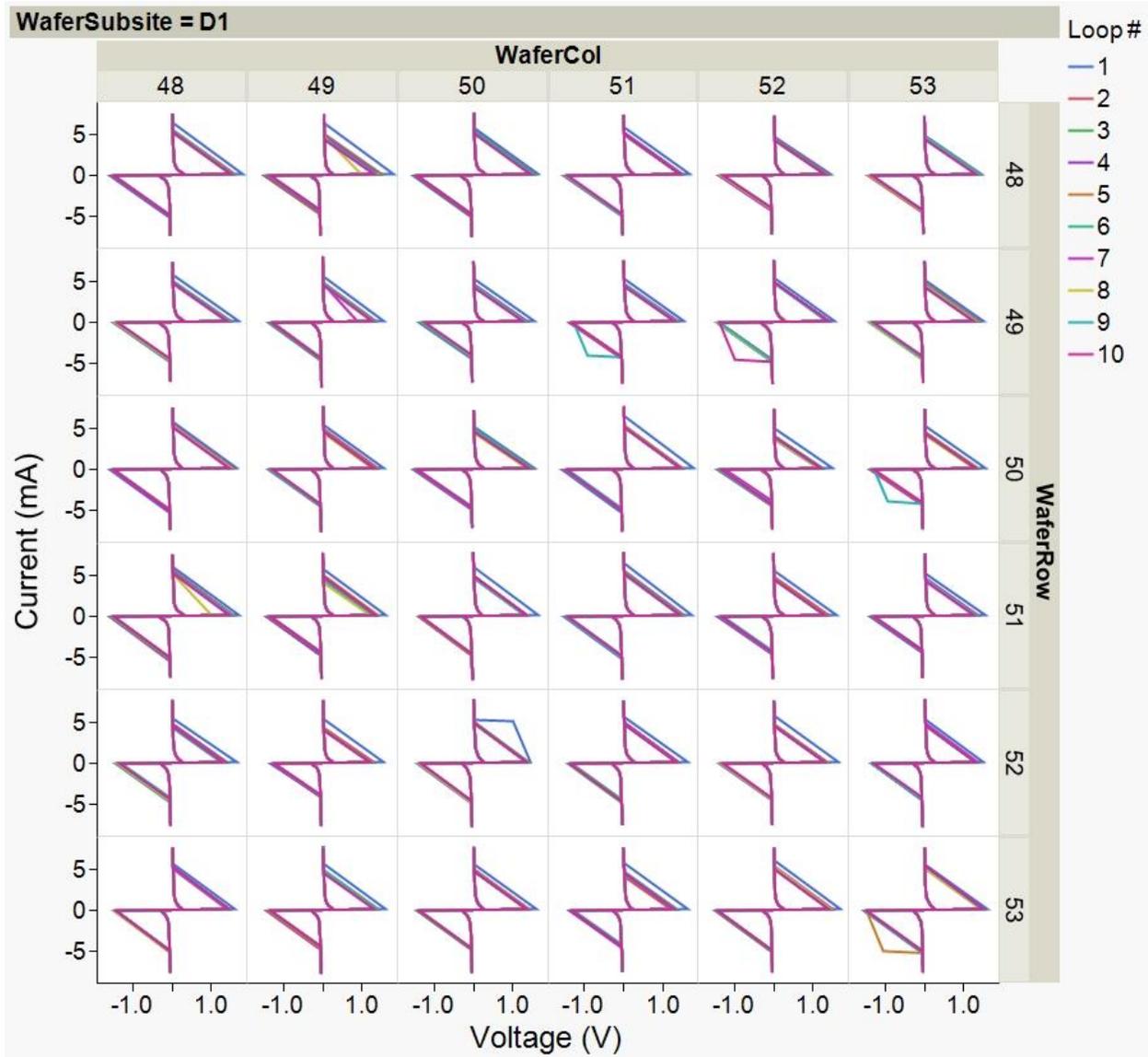

**Supplementary Figure 7. Wafer-scale uniform and electroform-free switching characteristics of VO₂ devices.** Plotted are 36 sets of four-terminal quasi-d.c. I-V traces (force V, measure I) measured from VO$_2$ crossbar devices, sampled one device per reticle from 36 reticles (labeled by the wafer row and column numbers, from 48 to 53) in the same wafer. All the devices are identified as WaferSubsite = D1, meaning that they are all located at the same relative position inside each reticle (D1 is the first device at the bottom left corner in each reticle, as shown in Supplementary Figure 6a). For each device, the force V and measure I sweeps is repeated 10 times at the same setting. Majority of the as-grown VO$_2$ devices, ~98 %, showed upfront resistive switching and NDR in the very first I-V sweep without the need of electroforming. The uniformity of switching is demonstrated in both the run-to-run repeatability and the low device-to-device variation in switching thresholds (see Supplementary Figure 8).

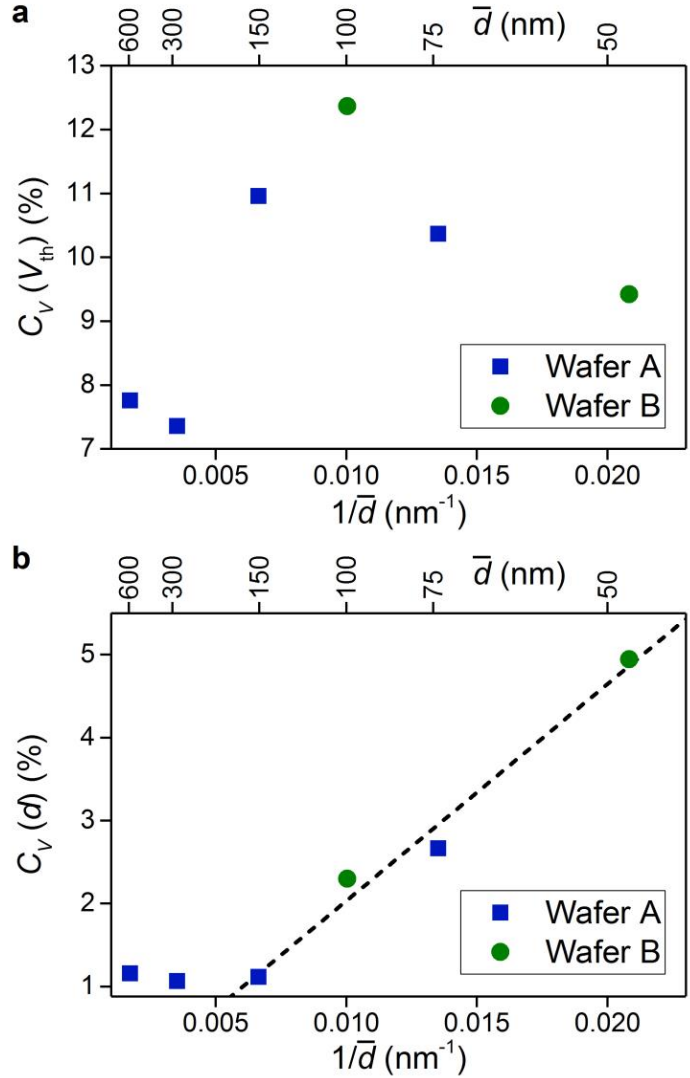

**Supplementary Figure 8. Statistical dispersions in switching threshold voltage $V_{th}$ and device size $d$ of 1152 VO$_2$ nano-crossbar devices, as measured by coefficient of variation $C_V \equiv \sigma/\mu$ (ratio of the standard deviation $\sigma$ to the mean $\mu$).** The device size $d$, as defined by the electrode linewidth, is measured by a critical dimension scanning electron microscope (CDSEM) system. **a**, $C_V$ in switching threshold $V_{th}$ vs. the mean device size $\bar{d}$ and its inverse $1/\bar{d}$. **b**, $C_V$ in device size $d$ vs. the mean device size $\bar{d}$ and its inverse $1/\bar{d}$. A total number of 1152 VO$_2$ devices, with six different designed dimensions, are measured from two 3-inch wafer samples (wafer A: 75×75 nm$^2$, 150×150 nm$^2$, 300×300 nm$^2$, and 600×600 nm$^2$; wafer B: 50×50 nm$^2$ and 100×100 nm$^2$). Each data point in wafer A is the result from 144 distinctive VO$_2$ devices with the same designed size. For each device, the force V and measure I sweeps is repeated 10 times at the same setting to obtain 20 switching events (including both positive and negative bias polarities). Each data point in wafer B is the result from 288 distinctive VO$_2$ devices in the same manner as in wafer A. At $d < 150$ nm, $C_V$ in device size starts to increase linearly with the inverse of device size due to the impact of edge roughness in ebeam lithography. In the worst-case, $C_V(d)$ is ~5% for 50×50 nm$^2$ devices. No such trend is observed in the relative variation of switching threshold, and all the $C_V(V_{th})$ data are less than 13%.

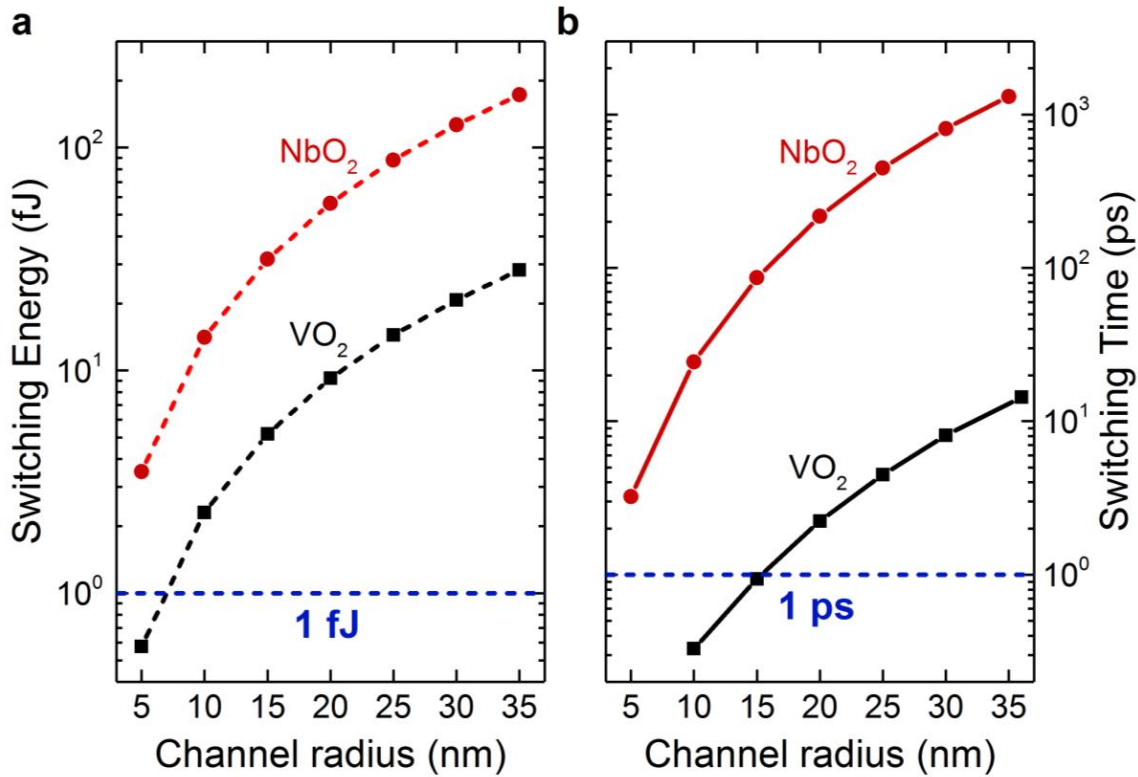

**Supplementary Figure 9. Comparison of switching energy and switching time (speed) of Mott IMT in $VO_2$ and $NbO_2$ devices with channel radius of 5–35 nm. a**, Calculated switching energy vs. channel radius for devices with a film thickness of 20 nm. In $VO_2$, due to the much smaller temperature rise needed for IMT to occur (40 K vs. 800 K), the volumetric free energy cost for IMT in $VO_2$ is only one-sixth of that for $NbO_2$ at the same crystal volume, and is less than 1 fJ at channel radii smaller than 7 nm. For details, see Supplementary Table 1. **b**, Simulated switching time vs. channel radius for devices with a film thickness of 50 nm. SPICE simulations of a $VO_2$-based Pearson-Anson relaxation oscillator (see the circuit in Supplementary Figure 3d) are used to estimate the switching time (speed) of Mott IMT from the rising edges of device current in each oscillation period[16]. The $VO_2$ channel radius is varied while all the other $VO_2$ model parameters are kept the same. Note that the switching speed is a material-dependent parameter, and is not affected by the values of $R_L$ and $C_1$ passive components ($R_L$ from 5 kΩ to 100 kΩ and $C_1$ of 22 pF were used). Simulations found that, at the same channel dimensions, Mott IMT switching in $VO_2$ is 100 times faster than in $NbO_2$, and is faster than 1 ps at channel radii smaller than 15 nm.

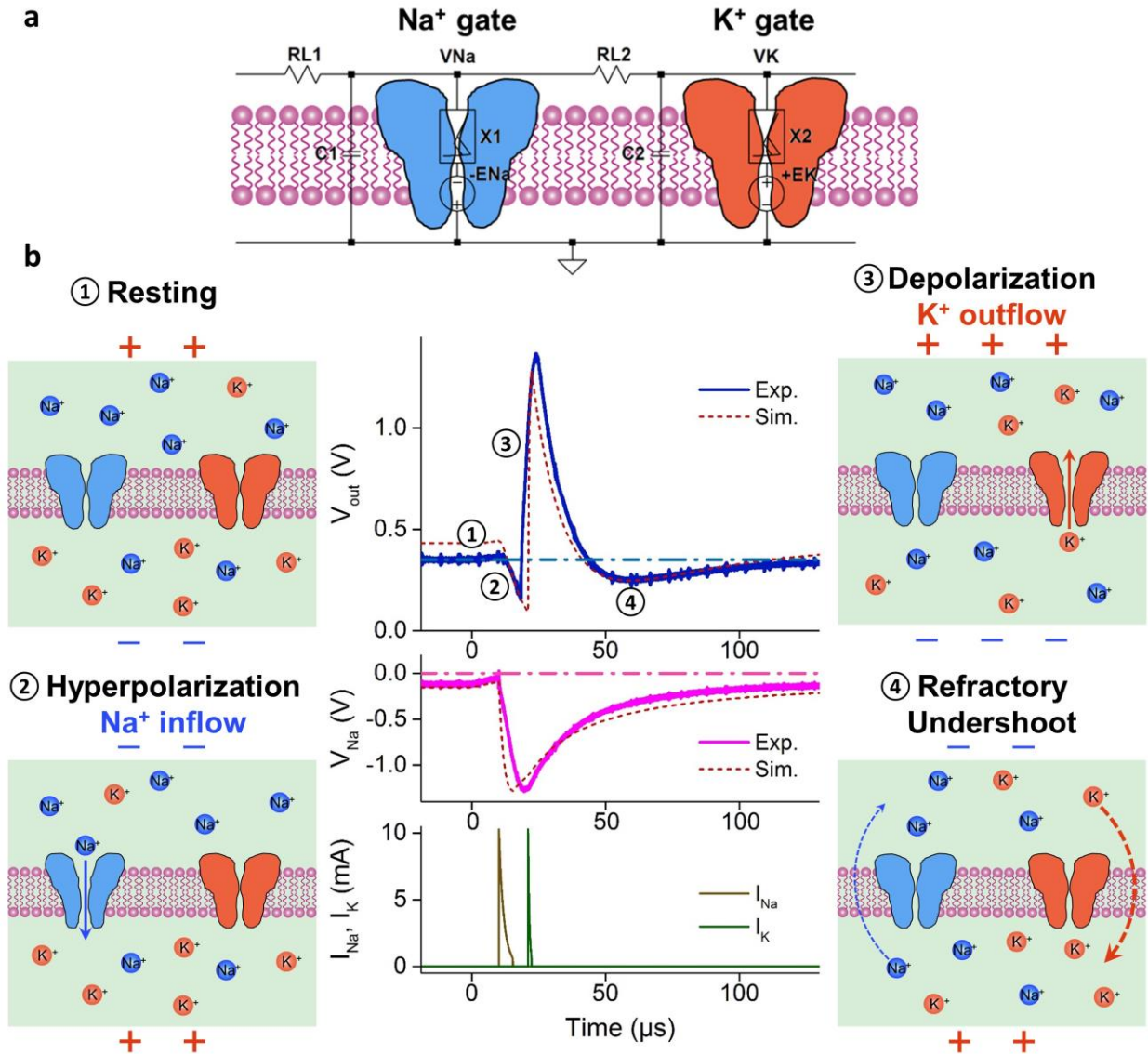

**Supplementary Figure 10. Circuit diagram and action potential generation in a VO₂ active memristor neuron**. **a**, Circuit diagram of a VO₂ memristor neuron, consisting of two resistively coupled Pearson-Anson relaxation oscillators ($R_{L1}$, $C_1$, $X_1$ and $R_{L2}$, $C_2$, $X_2$, respectively). The negatively-biased memristor $X_1$ acts as the voltage gated Na$^+$ channel, and the positively-biased memristor $X_2$ acts as the voltage-gated K$^+$ channel. Capacitors $C_1$ and $C_2$ are the corresponding membrane capacitances. **b**, Basic steps in action potential (spike) generation of a VO₂ neuron. (1) Resting state, in which both the Na$^+$ and K$^+$ channels are closed. A resting potential of 0.2–0.3V is produced by a membrane leakage current flowing through the VO₂ devices in their insulating state. (2) Hyperpolarization caused by the activation of the Na$^+$ channel, which drives the membrane potential toward negative direction. (3) Depolarization caused by the activation of the K$^+$ channel, which drives the membrane potential toward positive direction. (4) Refractory (undershoot), during which the neuron is recovering and does not respond to another stimulus. The central plots are experimental and simulated action potentials (top), the Na$^+$ channel membrane potential $V_{Na}$ (middle), and simulated Na$^+$ and K$^+$ channel currents (bottom).

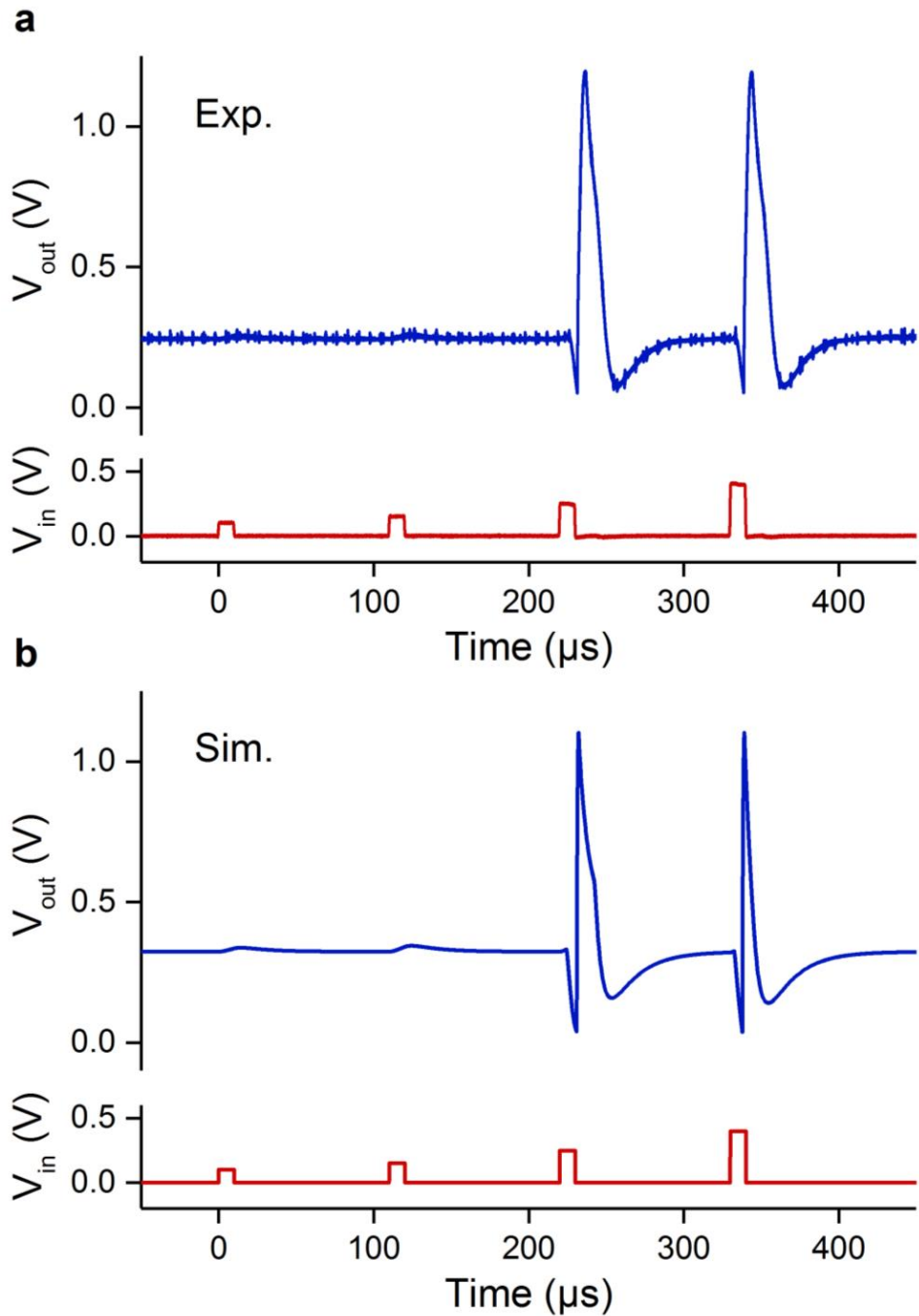

**Supplementary Figure 11. All-or-nothing firing behavior in a tonic VO₂ neuron circuit. a**, Experimentally measured all-or-nothing behavior, showing no response at subthreshold input voltage pulses of 0.1 V and 0.15 V, and spiking at suprathreshold input voltage pulses of 0.25 V and 0.4 V. In the suprathreshold regime, the shape or amplitude of neuron spikes does not change with an increase in the input voltage pulse amplitude. **b**, Simulated all-or-nothing behavior of the same tonic VO₂ neuron circuit. The pulse width in all the data shown is 10 μs.

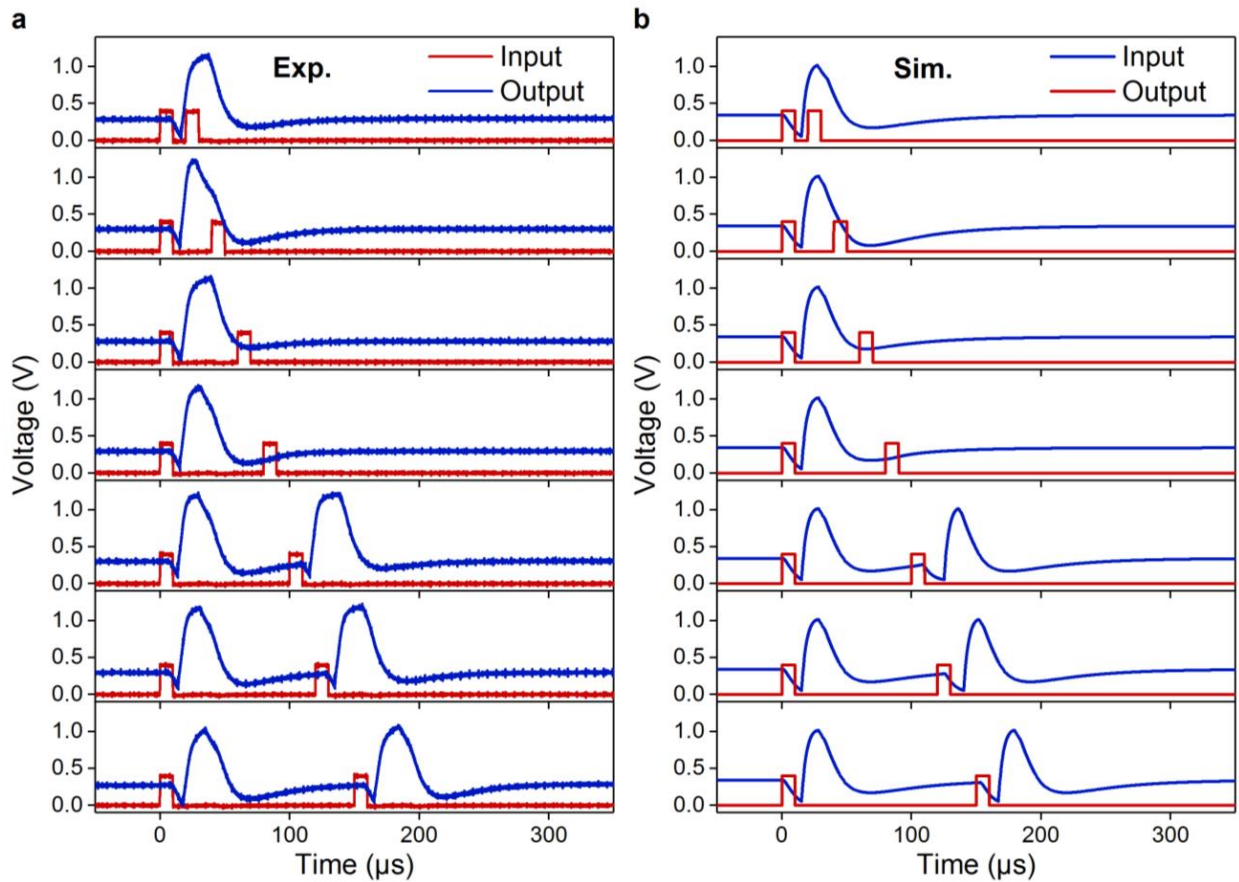

**Supplementary Figure 12. Refractory period behavior in a tonic VO$_2$ neuron circuit. a**, Experimentally measured refractory period behavior, showing a spiking in response to the first suprathreshold input voltage pulse, but no response to the second input voltage pulse if it occurred within the refractory period (panels 1 to 4 from the top). The neuron fires again if the second input voltage pulse is outside the refractory period (panels 5 to 7 from the top). From top to bottom, the periods of the input voltage pulse doublets are 20 μs, 40 μs, 60 μs, 80 μs, 100 μs, 120 μs, and 150 μs, respectively. **b**, Simulated refractory period behavior of the same tonic VO$_2$ neuron circuit. The pulse width in all the data shown is 10 μs.

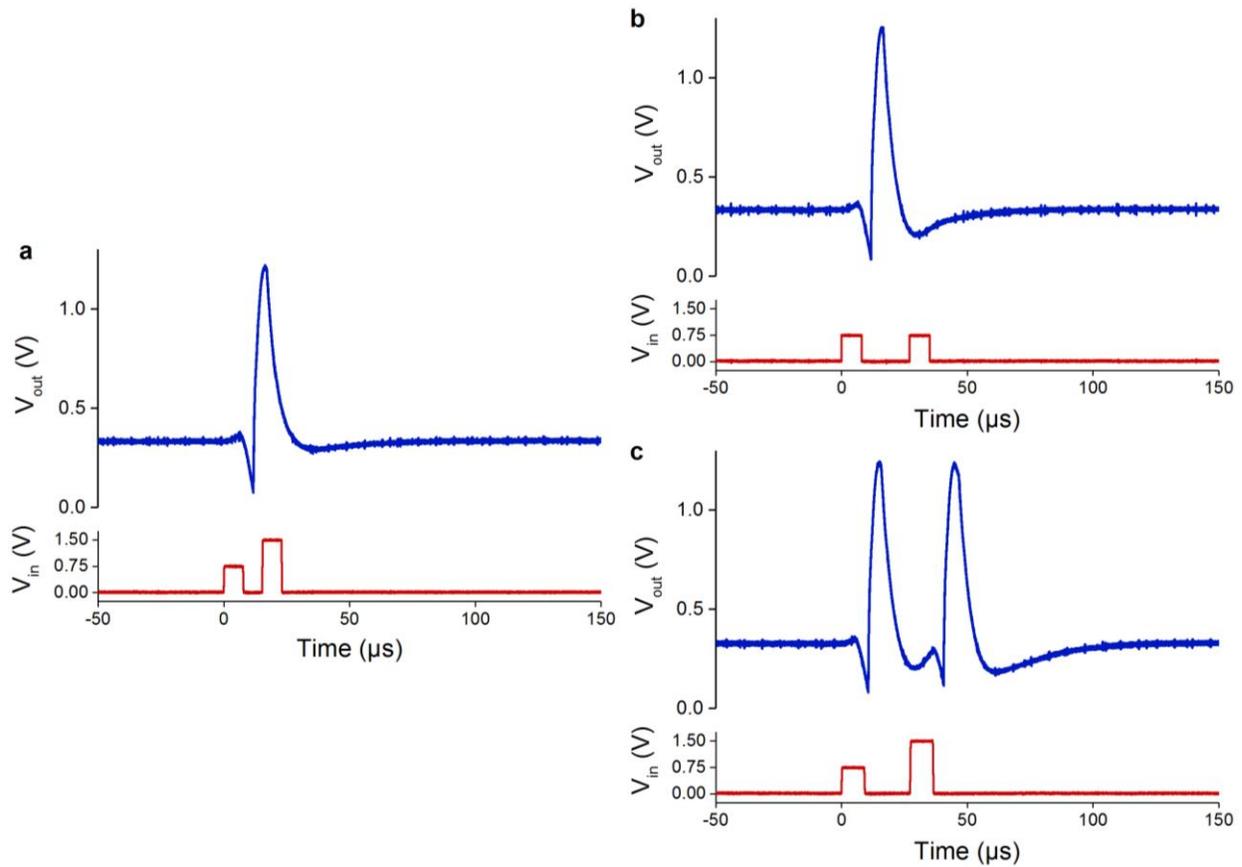

**Supplementary Figure 13. Absolute and relative refractory periods experimentally observed in a tonic VO₂ neuron circuit. a**, Absolute refractory period behavior, showing that if a second input voltage pulse is applied within the absolute refractory period, regardless of its strength (in this example 1.5 V was used, an amplitude greater than the spike amplitude), the neuron will never fire a second action potential in response. **b**, Relative refractory period behavior, showing that if the second input voltage pulse applied within the relative refractory period has the same strength as the first input pulse (in this example 0.75 V), the neuron will not fire a second action potential. **c**, Relative refractory period behavior, showing that if the second input voltage pulse applied within the relative refractory period is much stronger than the first input pulse (in this example 1.5 V vs. 0.75 V), the neuron will respond to it and fire a second action potential. The pulse width in all the data shown is 8 μs.

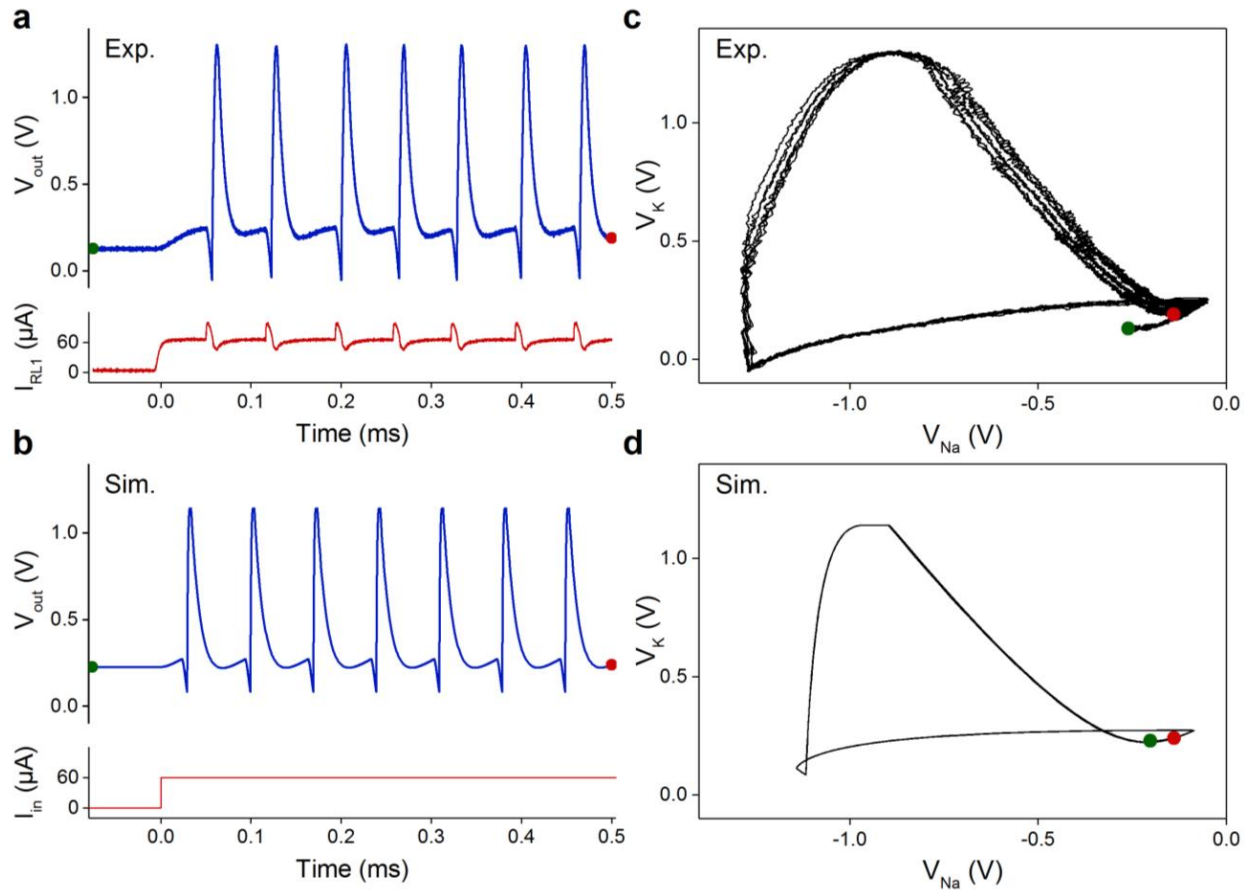

**Supplementary Figure 14. Tonic spiking behavior in a tonic VO$_2$ neuron circuit. a**, Experimentally measured tonic spiking behavior, showing that the neuron continues to fire a train of spikes in response to a d.c. input current. $I_{RL1}$, the current flowing through $R_{L1}$, was monitored by probing the voltage across it using two high-impedance (10 MΩ) oscilloscope probes. The current jitters coinciding with the output spikes are likely caused by the reflection of action potentials toward the neuron input, i.e. "back actions". **b**, Simulated tonic spiking behavior of the same tonic VO$_2$ neuron circuit in response to a d.c. input current. **c**, Experimental phase plane of the K$^+$ membrane potential $V_K$ (aka $V_{out}$) vs. the Na$^+$ membrane potential $V_{Na}$. In the phase plane representation, time-domain tonic spiking turns into a limit cycle attractor. **d**, Simulated phase plane of $V_K$ vs. $V_{Na}$, showing a limit cycle attractor similar to the experimental data in a qualitative manner. The green and red dots in **a**–**d** show the first and last plotted data points.

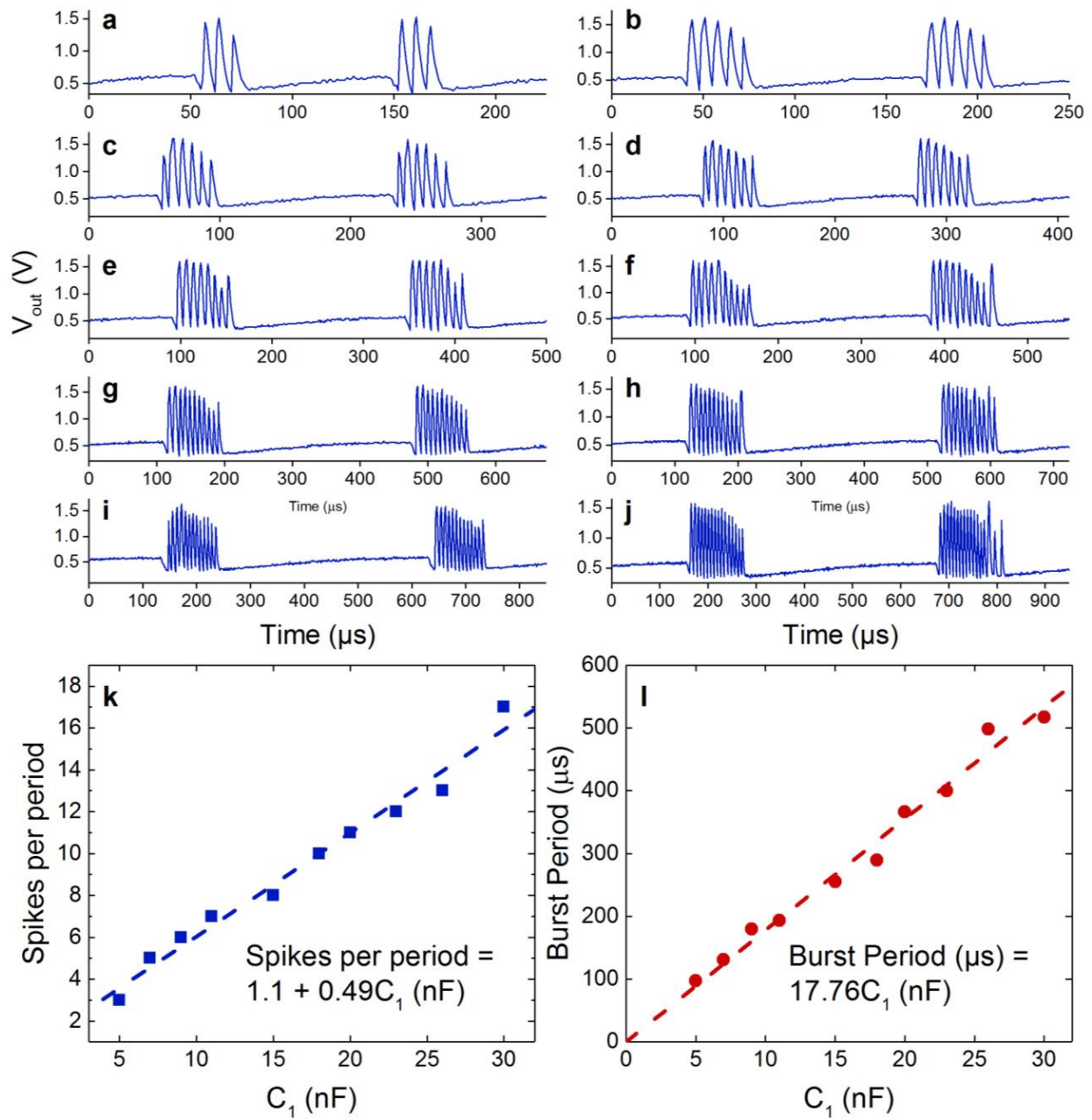

**Supplementary Figure 15. Tunable tonic bursting behavior measured in a tonic VO$_2$ neuron circuit. a–j**, Experimental tonic burst patterns measured with a fixed value of C$_2$ capacitor and an increasing value of C$_1$ capacitor (C$_1 \gg$ C$_2$ in all the cases). As C$_1$ becomes larger, both the number of spikes in each burst period and the burst period itself increase. **k**, Experimental C$_1$ dependence of the number of spike in each burst period. Dashed line is a linear fit (with R$^2$ = 0.98) which shows a linear trend that the number of spike per period = (1.1±0.5) + (0.49±0.03)·C$_1$ (nF). **l**, Experimental C$_1$ dependence of the burst period. Dashed line is a linear fit (with R$^2$ = 0.997) which shows a linear trend that the burst period (μs) = (17.76±0.33)·C$_1$ (nF). In all the data shown, C$_2 \approx$ 1 nF comes from the stray capacitance in the experimental setup.

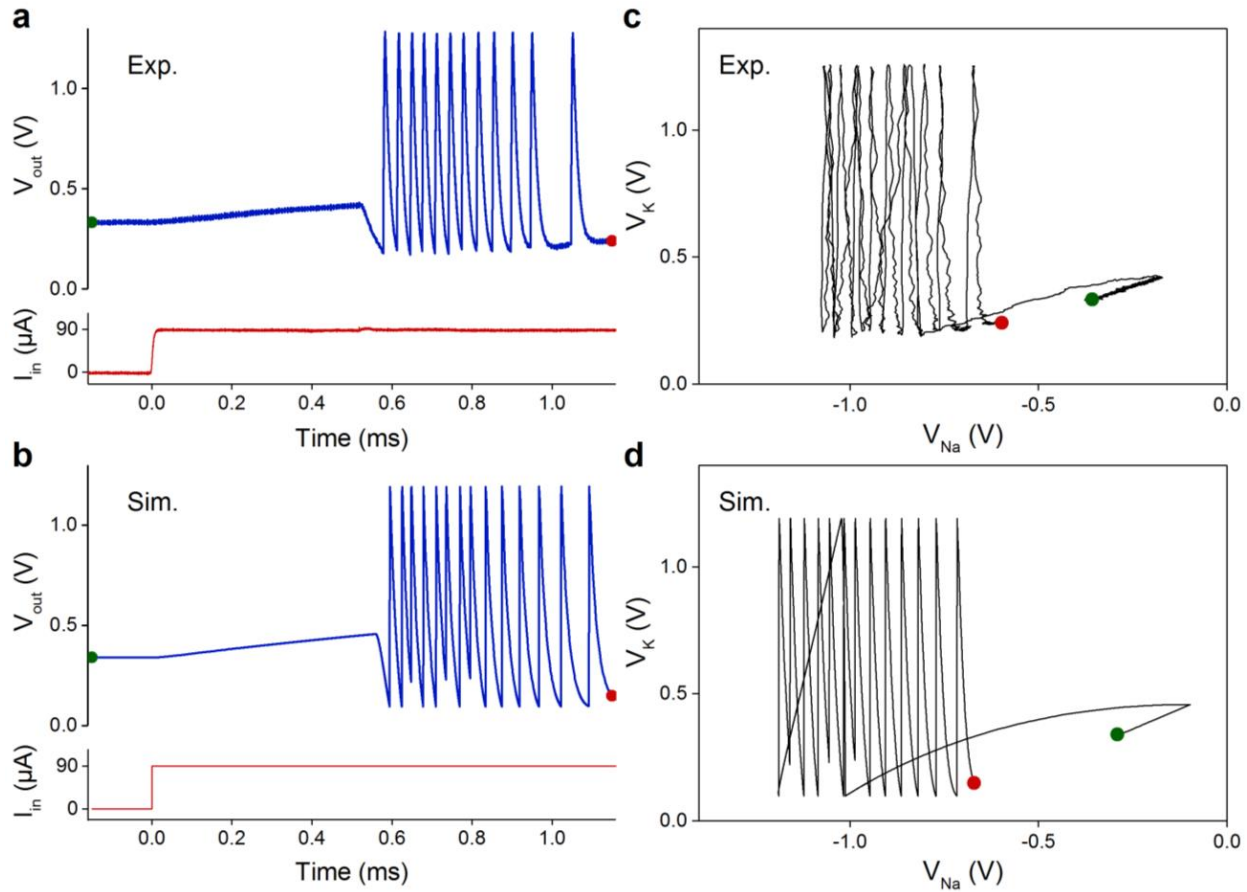

**Supplementary Figure 16. Spike frequency adaptation behavior in a tonic VO$_2$ neuron circuit. a**, Experimentally measured spike frequency adaptation in tonic burst under a sustained d.c. current stimulation. The spike frequency is relatively high at the onset of tonic bursting, and then it decreases over time, i.e., the neuron adapts. **b**, Simulated spike frequency adaptation behavior of the same tonic VO$_2$ neuron circuit in response to a d.c. input current. **c**, Experimental phase plane of the K$^+$ membrane potential $V_K$ (aka $V_{out}$) vs. the Na$^+$ membrane potential $V_{Na}$. A 100-fold down-sampling followed by 5-point adjacent-averaging was applied to the raw oscilloscope data to smooth the curve. **d**, Simulated phase plane of $V_K$ vs. $V_{Na}$, showing trajectories similar to the experimental data in a qualitative manner. The green and red dots in **a–d** show the first and last plotted data points, respectively.

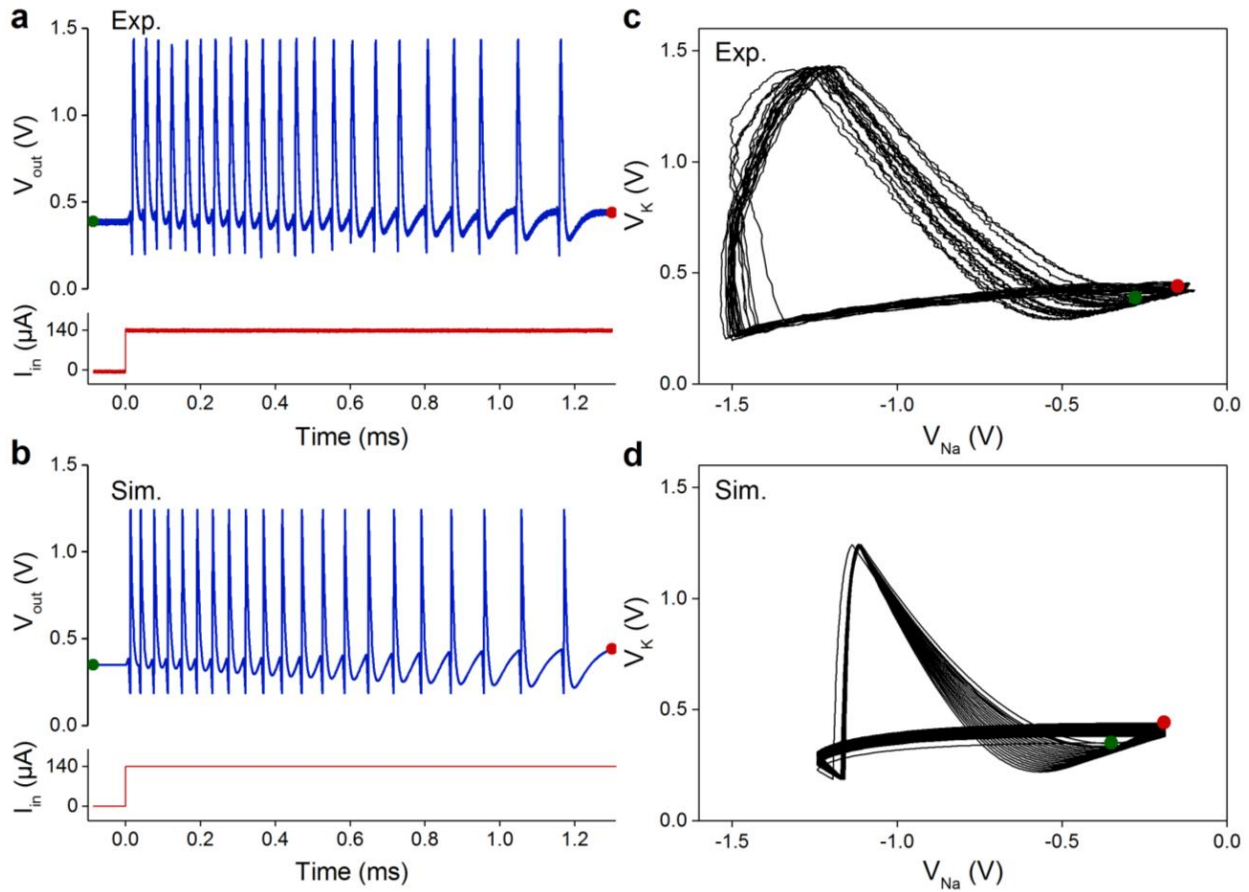

**Supplementary Figure 17. Spike frequency adaptation behavior in a phasic VO$_2$ neuron circuit. a**, Experimentally measured spike frequency adaptation in phasic burst under a sustained d.c. current stimulation. The spike frequency is relatively high at the onset of stimulation, and then it decreases over time, i.e., the neuron adapts. **b**, Simulated spike frequency adaptation behavior of the same phasic VO$_2$ neuron circuit in response to a d.c. input current. **c**, Experimental phase plane of the K$^+$ membrane potential $V_K$ (aka $V_{out}$) vs. the Na$^+$ membrane potential $V_{Na}$. A 10-fold down-sampling followed by 5-point adjacent-averaging was applied to the raw oscilloscope data to smooth the curve. **d**, Simulated phase plane of $V_K$ vs. $V_{Na}$, showing trajectories similar to the experimental data in a qualitative manner. The green and red dots in **a–d** show the first and last plotted data points.

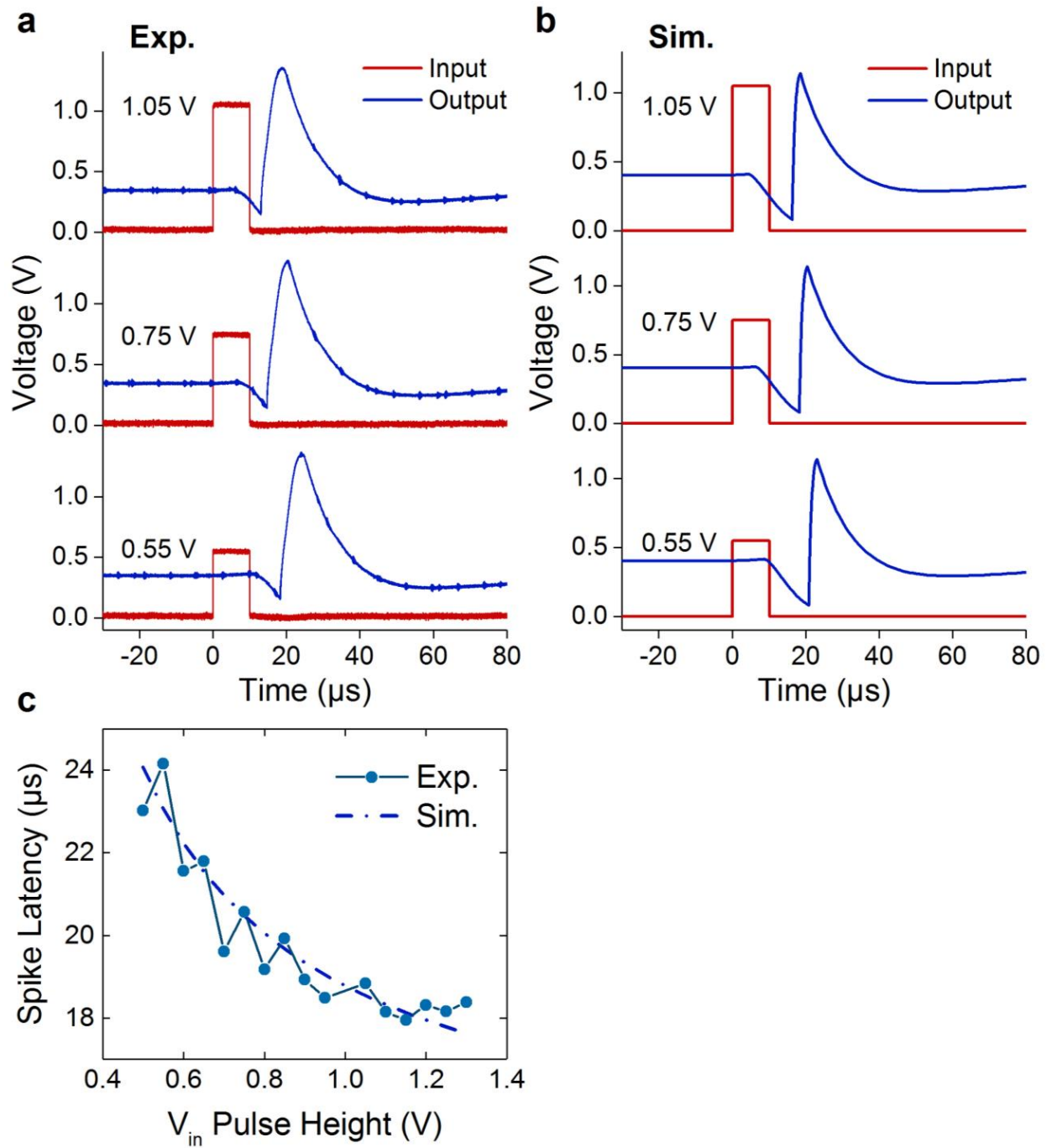

**Supplementary Figure 18. Spike latency behavior in a tonic VO₂ neuron circuit. a**, Experimental spike latencies in responding to suprathreshold 10 μs input voltage ($V_{in}$) pulses. Spike delay is longer for a relatively weak input, and it diminishes as the input gets stronger. **b**, Simulated spike latencies of the same tonic VO₂ neuron circuit. **c**, Dependences of measured and simulated spike latencies on the amplitude of the input pulse. Spike latency is arbitrarily defined as the delay between the onset of $V_{in}$ and the peak time of spiking. Simulated spike latency $\tau$ can be fitted (with $R^2 = 0.9995$) by a logarithmic formula $\tau = \tau_0 + b\ln(E - V_{in})$, where $\tau_0 = (17.29 \pm 0.02)$ μs, $b = (3.20 \pm 0.07)$ μs/ln(V), and $E = (0.382 \pm 0.007)$ V. The logarithmic dependence of spike latency on the input amplitude can be accounted for by the logarithmic formula of the capacitor discharge time in a relaxation oscillator[17].

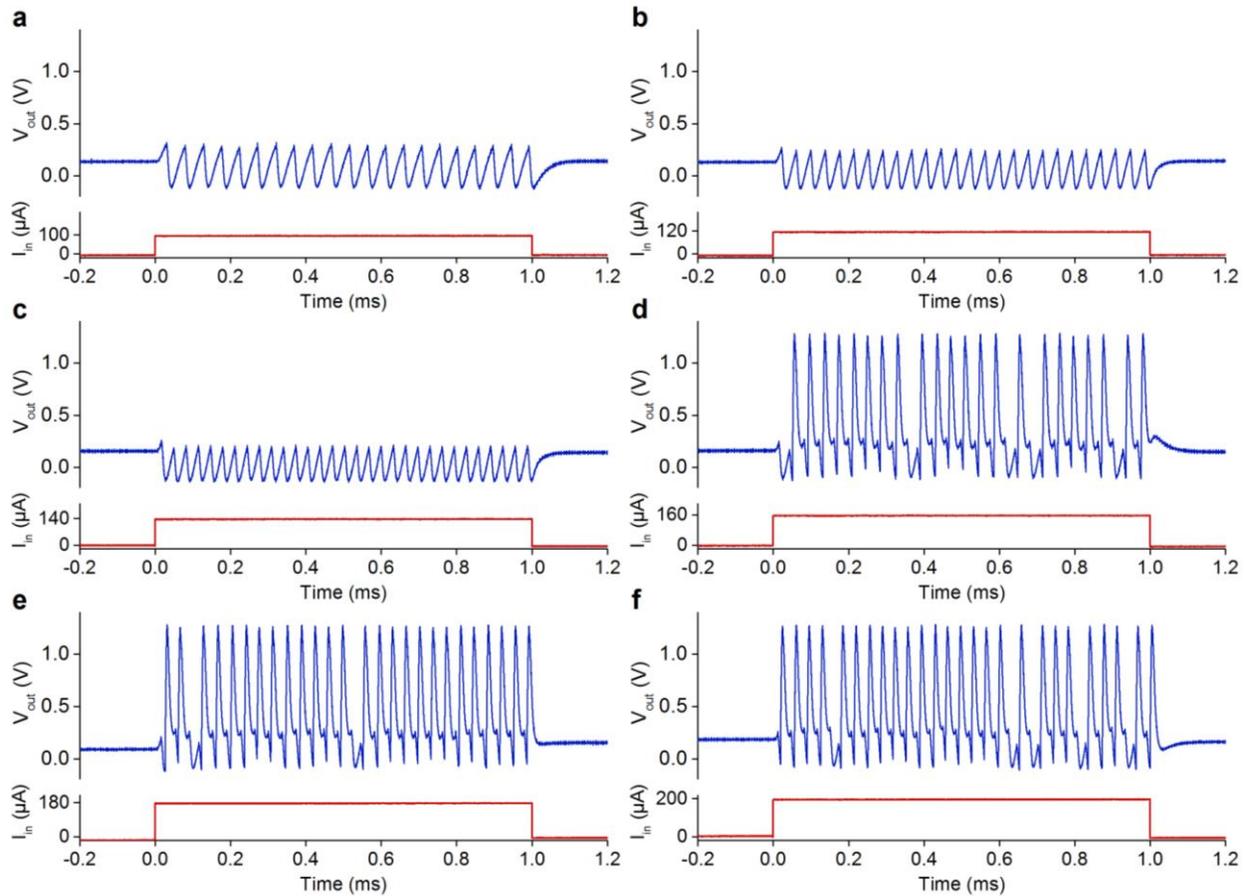

**Supplementary Figure 19. Subthreshold oscillation behavior measured in a tonic Class 2 VO₂ neuron circuit. a–c**, Subthreshold oscillations in the neuron output, i.e., the K$^+$ channel membrane potential, under a sustained d.c. current stimulation of 100 μA, 120 μA, and 140 μA, respectively. It is evident that the frequency of the oscillations increases with the input current level. **d–f**, Tonic spiking intermixed with occasional subthreshold oscillations (as demonstrated by the missing spikes) in the neuron output under a sustained d.c. current stimulation of 160 μA, 180 μA, and 200 μA, respectively. The transition from subthreshold oscillations to tonic spiking is better observed by ramping up the current stimulation. See Figure 4b in the main text for details.

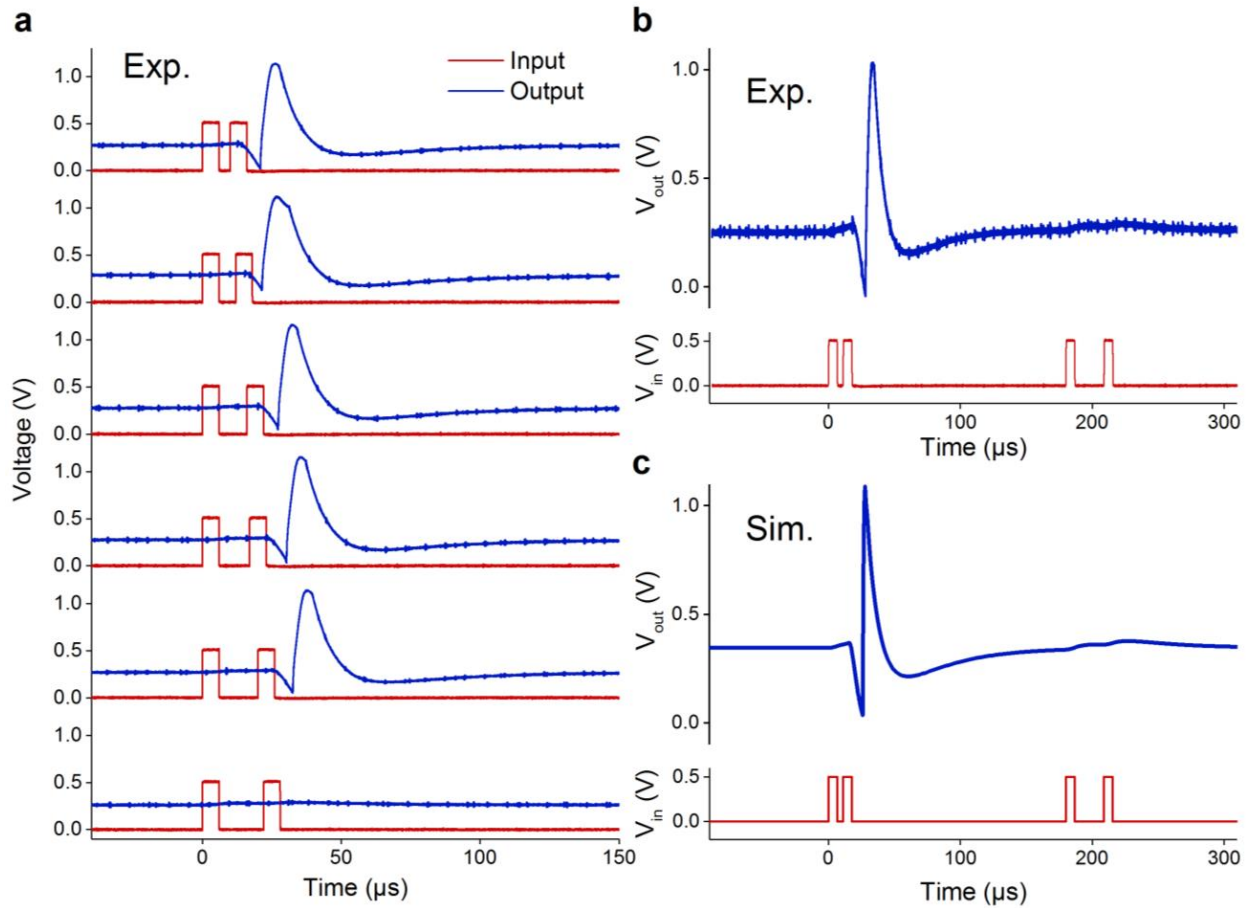

**Supplementary Figure 20. Integrator behavior in a tonic Class 1 VO$_2$ neuron circuit. a**, Experimentally measured integrator behavior, showing that the neuron fires an action potential if a doublet of two subthreshold input voltage pulses are applied with sufficiently short interval between them The pulse interval for data in panel 1–5 starting from the top is 4 µs, 6 µs, 10 µs, 11 µs, and 14 µs, respectively. The neuron does not spike if the two subthreshold input voltage pulses are too far apart from each other, i.e., the data in the bottom panel, with an interval of 16 µs. **b**, Experimental integrator behavior of the same neuron circuit as in **a**, demonstrated by applying two doublets of subthreshold input voltage pulses in the same measurement. The neuron fires a spike in response to the first input pulse doublet with a shorter interval (5 µs), but does not fire in response to the second input pulse doublet with a longer interval (23 µs). **c**, Simulated integrator behavior of the same neuron circuit as in **b**. The pulse width and amplitude in all the data shown is 6 µs and 0.5 V, respectively.

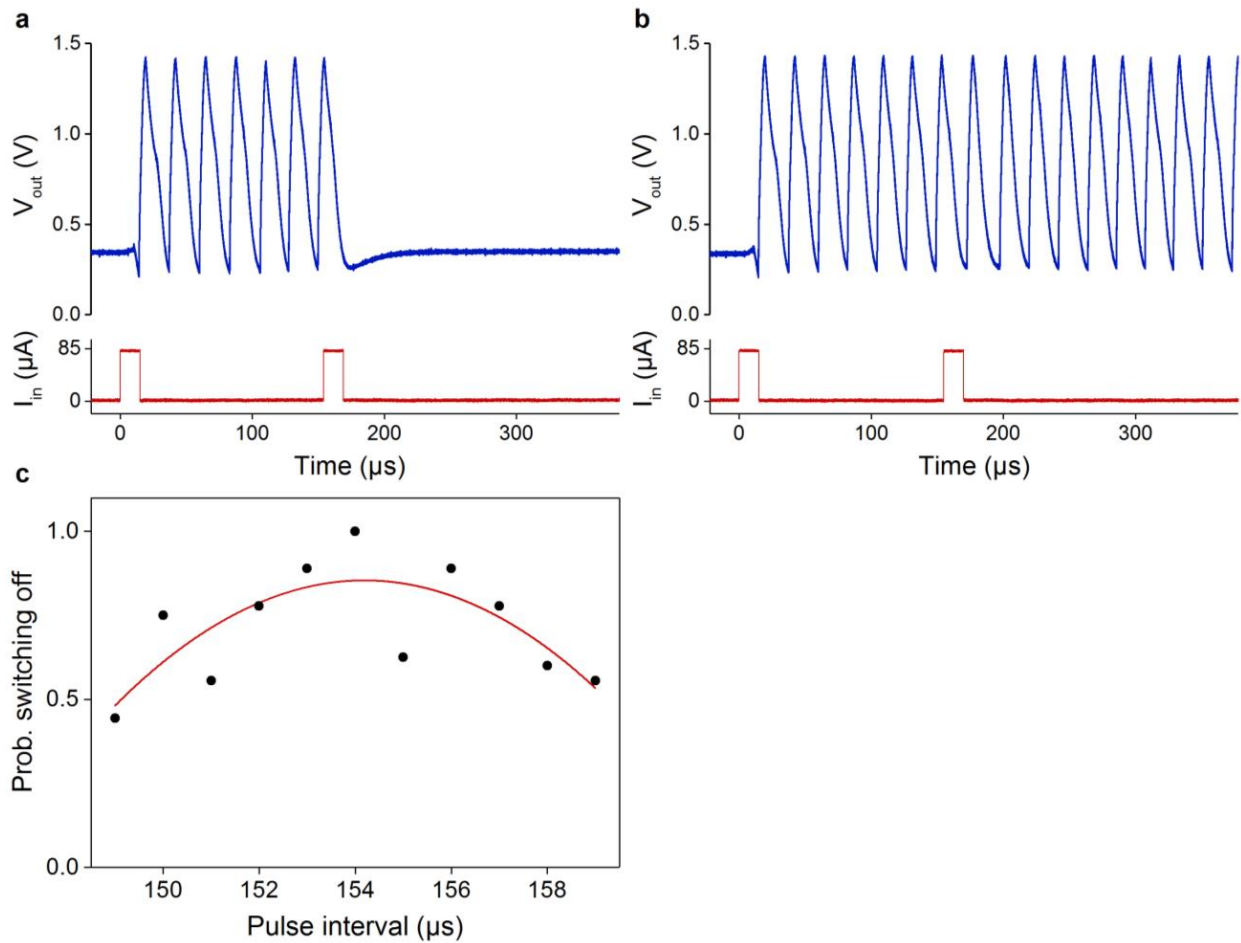

**Supplementary Figure 21. Bistability behavior measured in a tonic VO$_2$ neuron circuit. a**, The neuron is driven from the resting mode into a persistent tonic spiking mode (a self-oscillation) by applying the first input pulse stimulation. A second input pulse arriving at an interval of 154 μs successfully switches the neuron from tonic spiking back to the resting mode. **b**, A second input pulse arriving at an interval of 155 μs fails to switch the neuron from tonic spiking back to the resting mode. In the measurements, 0.85 V and 15 μs wide voltage pulses sent from an arbitrary waveform generator (AWG) were converted into 85 μA input current pulses ($I_{in}$) using a stimulus isolator with a gain of 0.1mA/V. Input current was not monitored because the load resistor $R_{L1}$ was set to be zero to enable bistability. The plotted $I_{in}$ waveforms are calculated from the monitored AWG voltage waveforms. **c**, Probability (success rate) of the second input pulse switching off the self-oscillation vs. the pulse interval. Red line is a second-order polynomial fit. Each data point represents the statistics from 8 to 10 such attempts. At an interval of 154 μs, the success rate is 100 %, or the neuron self-oscillation was switched off in 10 out of 10 attempts. At an interval of 155 μs, the success rate dropped to 62.5%, or 5 out of 8 attempts. Despite the scattering of data points, it is evident that the success rate peaks at around 154 μs interval, and it drops off as the interval is detuned away. This observation is consistent with the interpretation that the input must arrive at an appropriate phase of oscillation for it to switch the neuron from tonic spiking back to the resting mode. Large scattering in the success rates may be explained by the stochastic onset of tonic spiking, which shifts the phase of oscillation randomly with respect to the fixed intervals used in measurements.

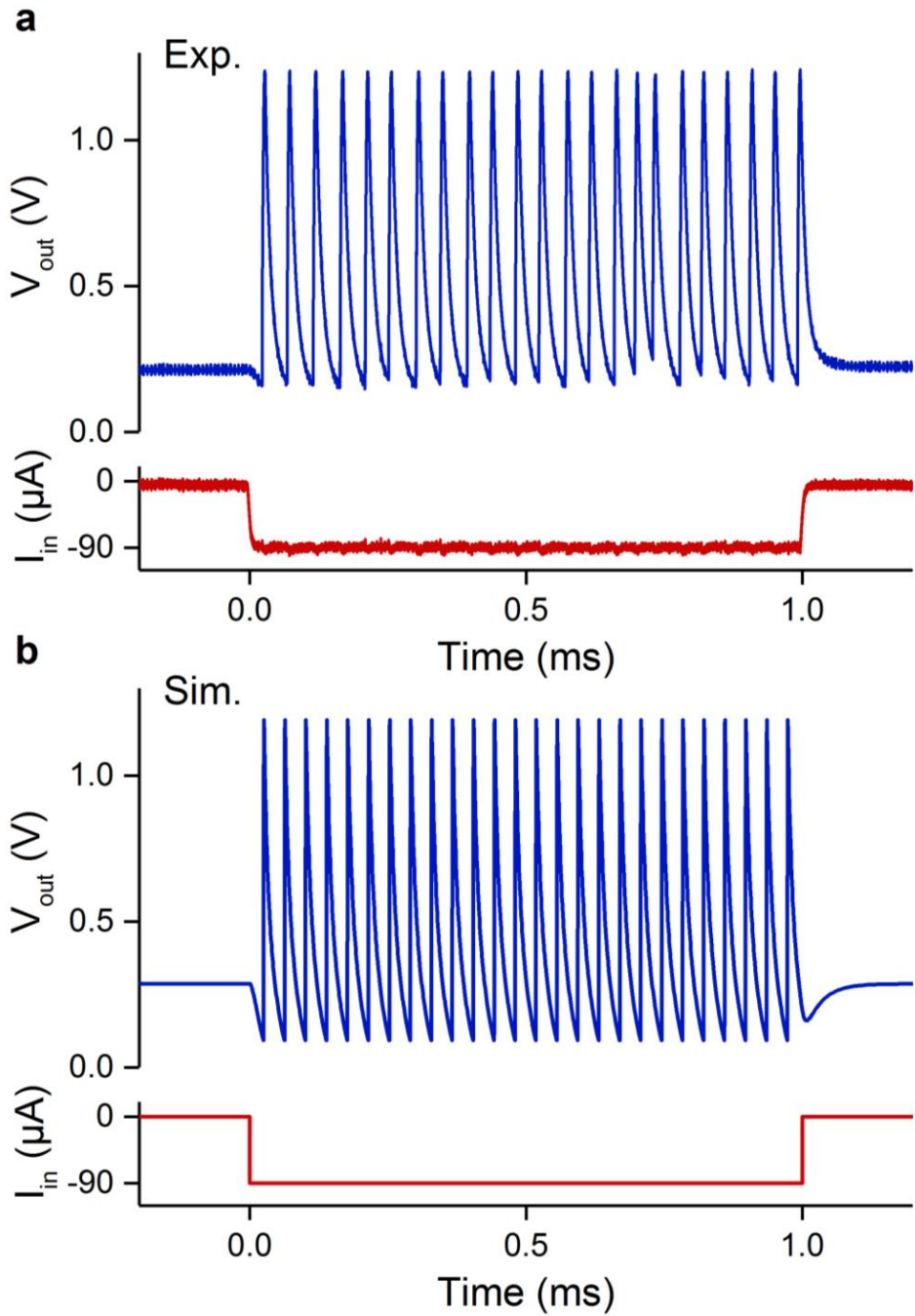

**Supplementary Figure 22. Inhibition-induced spiking (IIS) behavior in a tonic $VO_2$ neuron circuit. a**, Experimentally measured inhibition-induced spiking behavior, showing that the neuron is quiescent (at rest) when there is no input current, but fires a tonic spike train when it is hyperpolarized by an inhibitory (negative) input current of -90 µA. **b**, Simulated inhibition-induced spiking behavior of the same tonic $VO_2$ neuron circuit. In biology, many thalamo-cortical neurons exhibit the IIS feature. The mechanism was attributed to inhibitory-input induced activation of the h-current and deactivation of the $Ca^{2+}$ T-current[18].

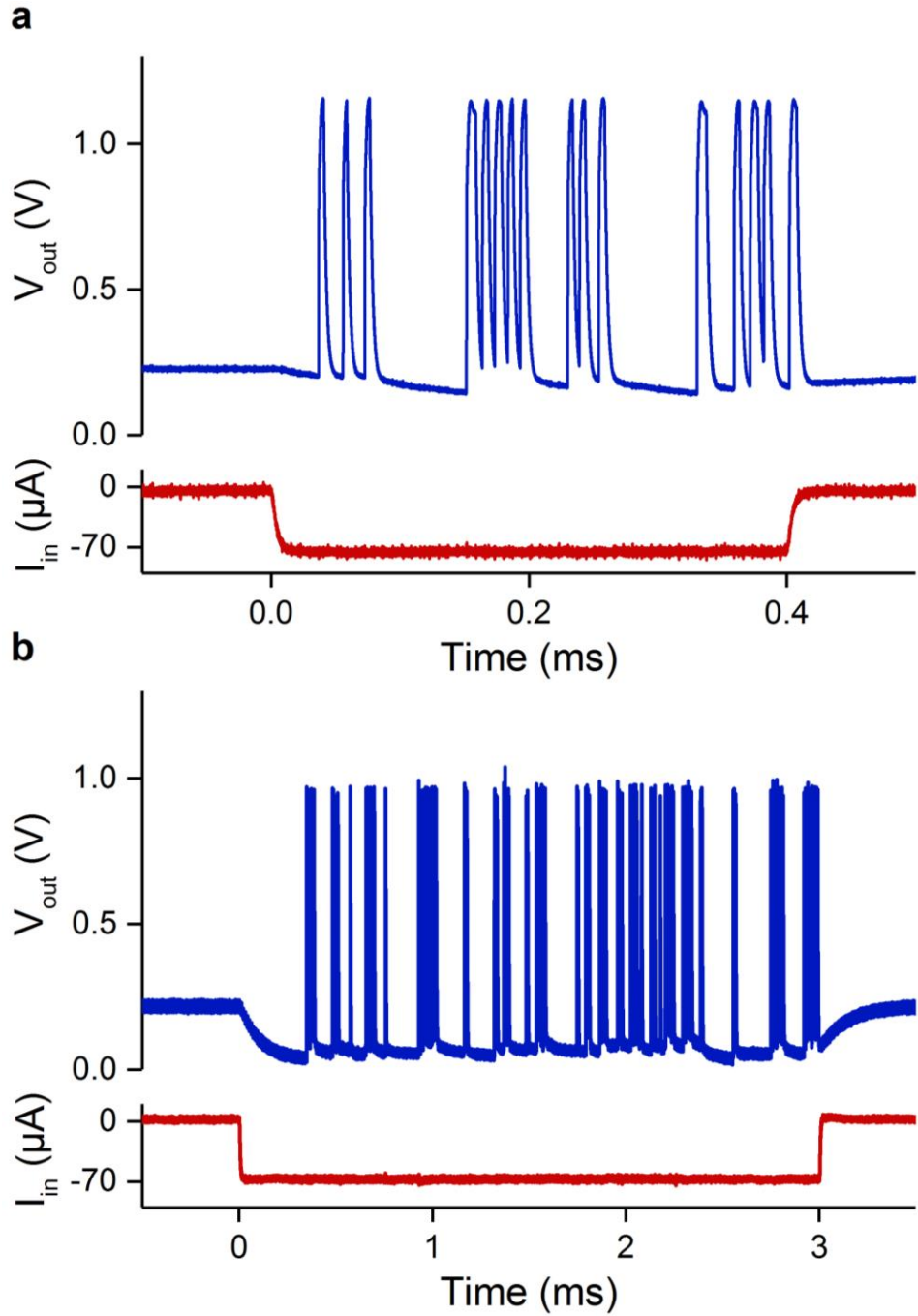

**Supplementary Figure 23. Experimental inhibition-induced bursting (IIB) behaviors in tonic VO$_2$ neuron circuits.** The neuron is quiescent (at rest) when there is no input current, but fires irregular bursts of spikes when it is hyperpolarized by an inhibitory (negative) input current of -70 µA. Similar to the case of tonic bursting induced by excitatory (positive) inputs, IIB requires a much slower Na$^+$ channel than the K$^+$ channel, or $C_1 \gg C_2$. **a**, IIB measured from a tonic VO$_2$ neuron circuit with the discrete membrane capacitors set at $C_1 = 35$ nF and $C_2 = 0$ nF. **b**, IIB measured from another tonic VO$_2$ neuron circuit with $C_1 = 21$ nF and $C_2 = 0$ nF. In the test setup, for each discrete capacitor there also exists a stray capacitance of ~1 nF, mostly contributed by the cables. Other circuit parameters can be found in Supplementary Table 3.

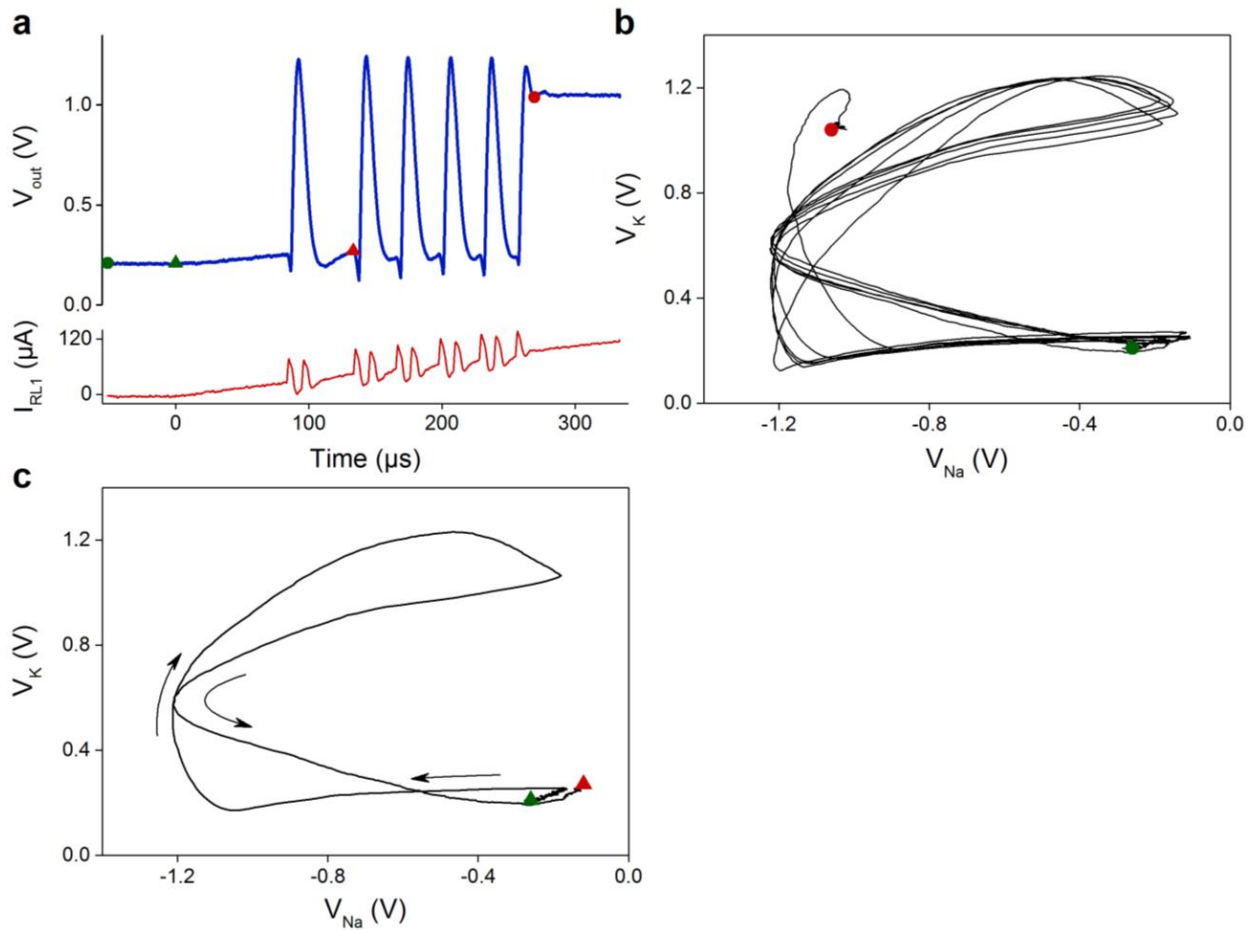

**Supplementary Figure 24. Experimental excitation block behavior in a Class 2 tonic VO₂ neuron circuit. a**, Experimentally measured excitation block behavior, showing that the neuron fires tonic spikes when the input current ramps above the spiking threshold at 25 µA. As the stimulus current further increases beyond 95 µA, the neuron suddenly ceases to spike and the output is locked to an elevated value (1.05 V). **b**, Experimental phase plane of the K⁺ membrane potential $V_K$ (aka $V_{out}$) vs. the Na⁺ membrane potential $V_{Na}$ for the data shown in **a**. The green and red dots in **a** and **b** show the onset of current ramp and the onset of excitation block. The $V_K$–$V_{Na}$ trajectory shows characteristics of a distorted letter "B"-shaped limit cycle attractor. Each loop of the trajectory corresponds to a complete cycle of action potential generation in **a**. Further increase of the stimulus current beyond 95 µA causes the disappearance of the limit cycle attractor and the locking of neuron state (the red dot). A 100-fold down-sampling followed by 16-point adjacent-averaging was applied to the raw oscilloscope data to smooth the curve. **c**, Experimental phase plane of $V_K$ vs. $V_{Na}$ for the time duration of the first tonic spike in **a** (marked by green and red triangles). The loop of letter "B"-shaped trajectory is marked with arrows. In theory, excitation block is attributed to the conversion from a spiking limit cycle to a supercritical Andronov-Hopf bifurcation phenomenon and is explained in FitzHugh-Nagumo model by the phase plane approach, which shows a stimulus-induced shift of the equilibrium through stable-unstable-stable branches of the N-shaped nullcline[19].

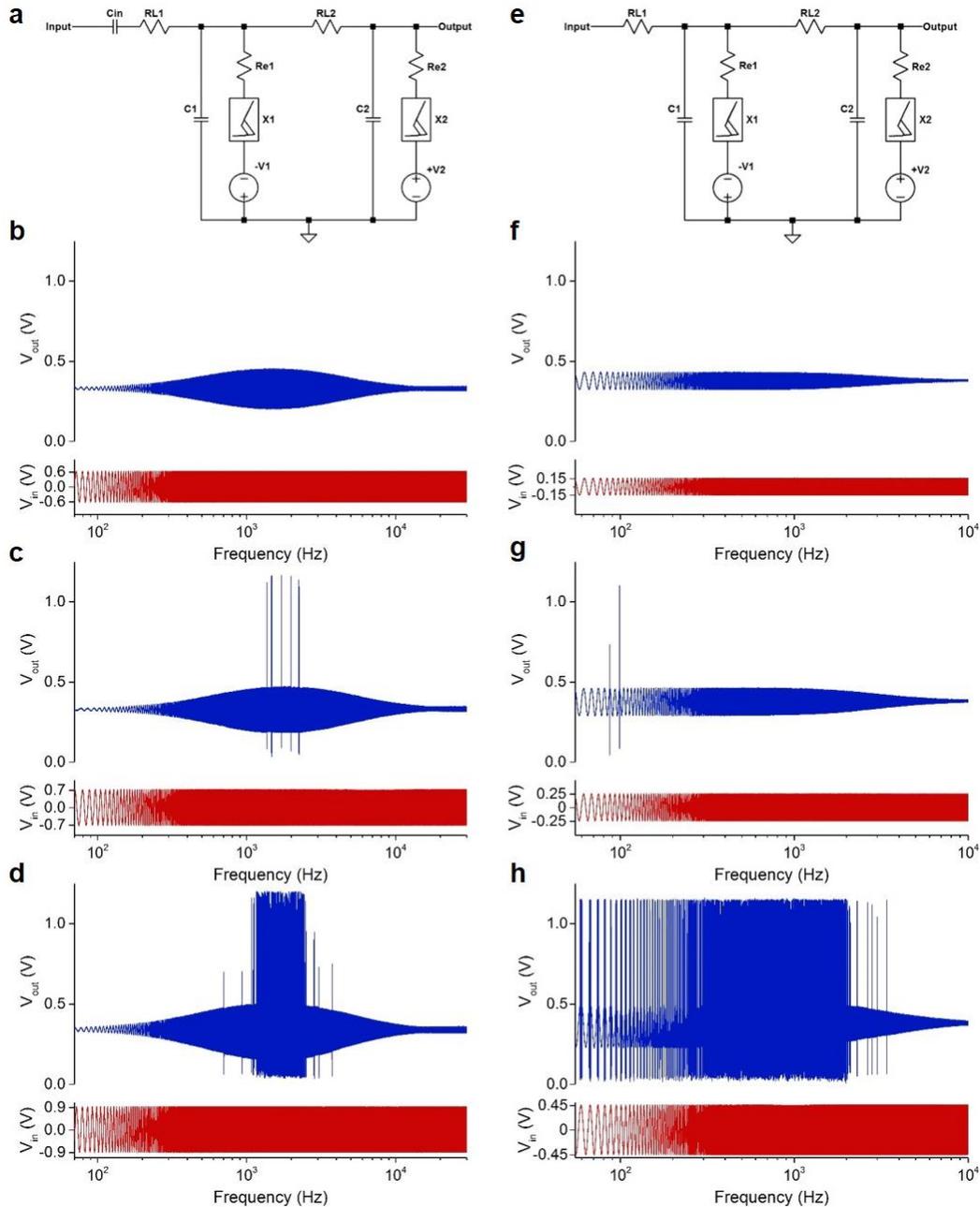

**Supplementary Figure 25. Experimental resonator behavior in a phasic VO$_2$ neuron (a) as compared with integrator behavior in a tonic VO$_2$ neuron (b). a**, The phasic neuron circuit with a capacitively-coupled input. **b**, Oscilloscope-captured phasic neuron response to a subthreshold (0.6 V) frequency-sweeping sinusoidal voltage input (ZAP sweep[20]), showing a pass band centered around ~17 kHz. **c**, Sporadic spikes occur near ~17 kHz as the ZAP sweep amplitude increased to 0.7 V. **d**, Intensified spiking near ~17 kHz as the ZAP sweep amplitude further increased to 0.9 V. **e**, The tonic neuron circuit with all the circuit elements the same as the phasic neuron in **a**, except for the missing C$_{in}$ capacitor. **f–g**, Frequency-domain response of the tonic neuron to ZAP sweeps with an amplitude of 0.6 V, 0.7 V, 0.9 V, respectively. The integrator nature is reflected by the low-pass filter characteristics of the neuron response.

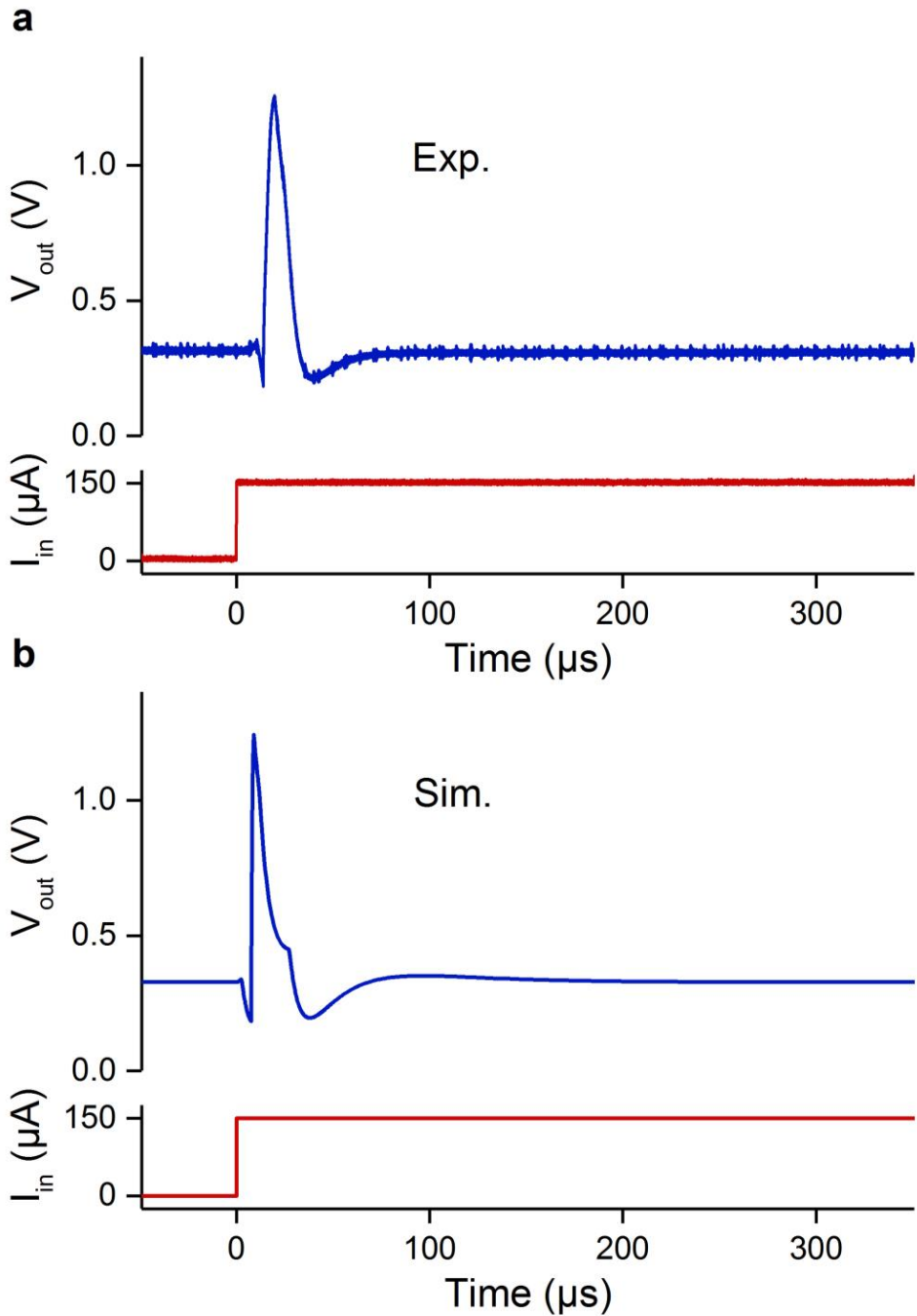

**Supplementary Figure 26. Phasic spiking (Class 3 excitable) behavior in a phasic VO$_2$ neuron circuit. a**, Experimentally measured phasic spiking behavior, showing that the neuron fires only a single spike at the onset of a d.c. input current, and then it remains quiescent even in the presence of the input current. The plotted input current (I$_{in}$) waveform is calculated from the monitored AWG voltage waveform, because in the phasic neuron circuit the load resistor R$_{L1}$ is replaced with a capacitor C$_{in}$. **b**, Simulated phasic spiking behavior of the same phasic VO$_2$ neuron circuit.

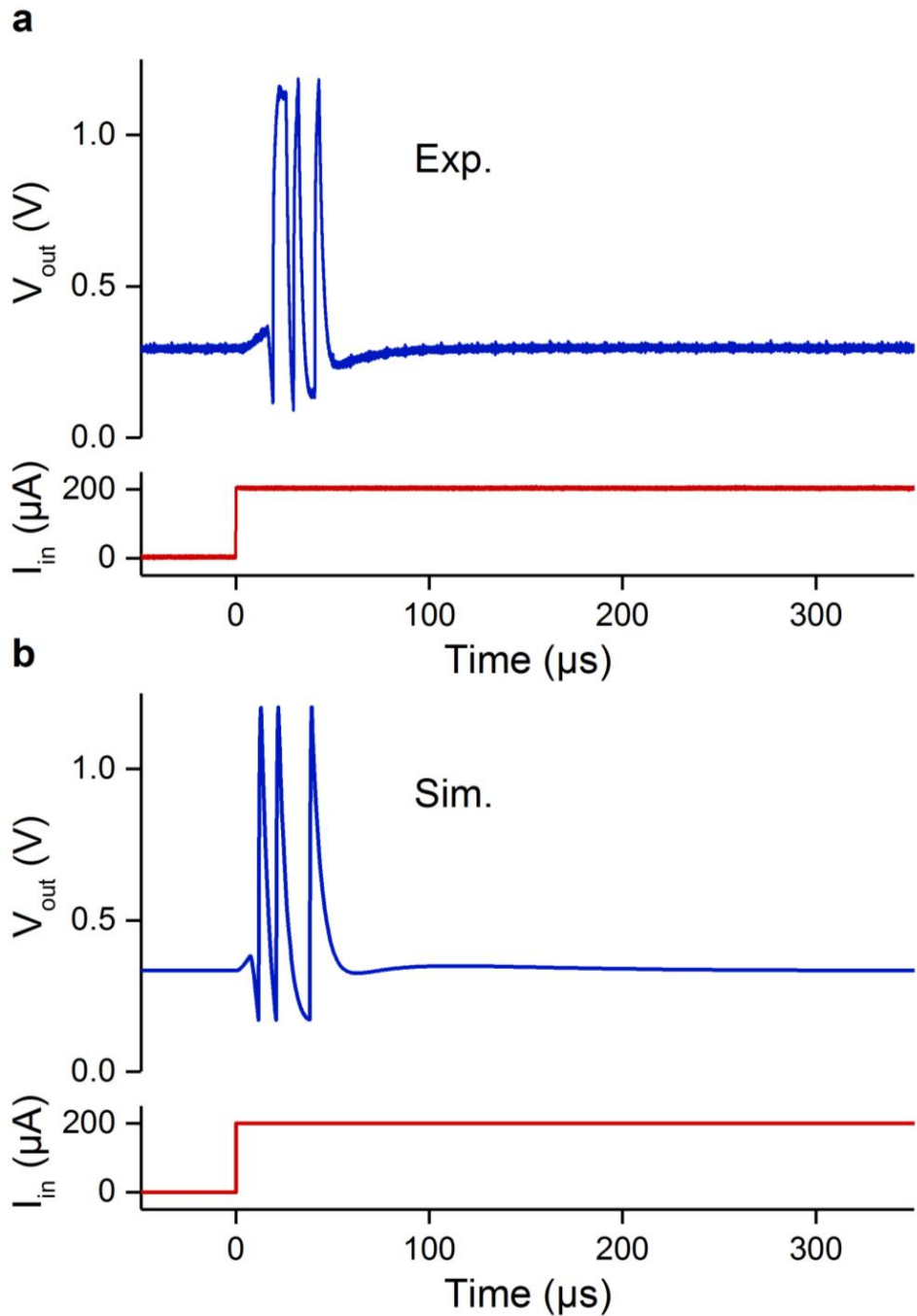

**Supplementary Figure 27. Phasic bursting behavior in a phasic VO$_2$ neuron circuit. a**, Experimentally measured phasic bursting behavior, showing that the neuron fires only a single period of burst spikes at the onset of a d.c. input current, and then it remains quiescent even in the presence of the input current. The plotted input current ($I_{in}$) waveform is calculated from the monitored AWG voltage waveform, because the load resistor $R_{L1}$ is replaced with a capacitor $C_{in}$ in the phasic neuron circuit. **b**, Simulated phasic bursting behavior of the same phasic VO$_2$ neuron circuit.

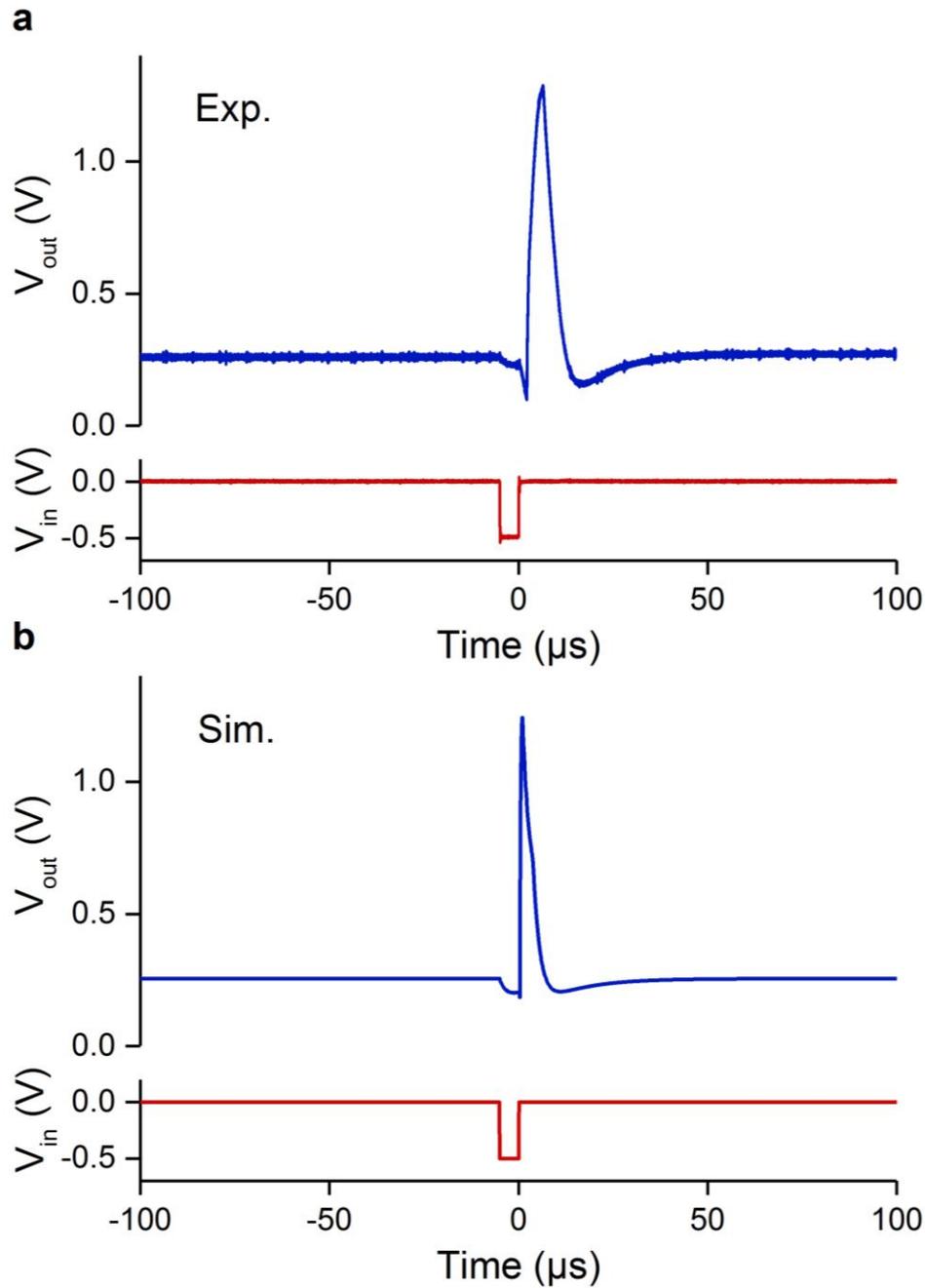

**Supplementary Figure 28. Rebound spike behavior in a phasic VO₂ neuron circuit. a**, Experimentally measured rebound spike behavior, showing that when the neuron receives and then is released from an inhibitory (negative) input, it fires a post-inhibitory (rebound) spike, in response to the release (the rise edge) of the inhibitory input waveform). **b**, Simulated rebound spike behavior of the same phasic VO₂ neuron circuit. In the case of excitatory input, a phasic spike is fired at the rise edge of the input current (see Supplementary Figure 26). In the case of inhibitory input, a rebound spike is fired also at the rise edge of the input current. Equipped with a capacitively-coupled input, the phasic neuron essentially acts as a rise-edge detector.

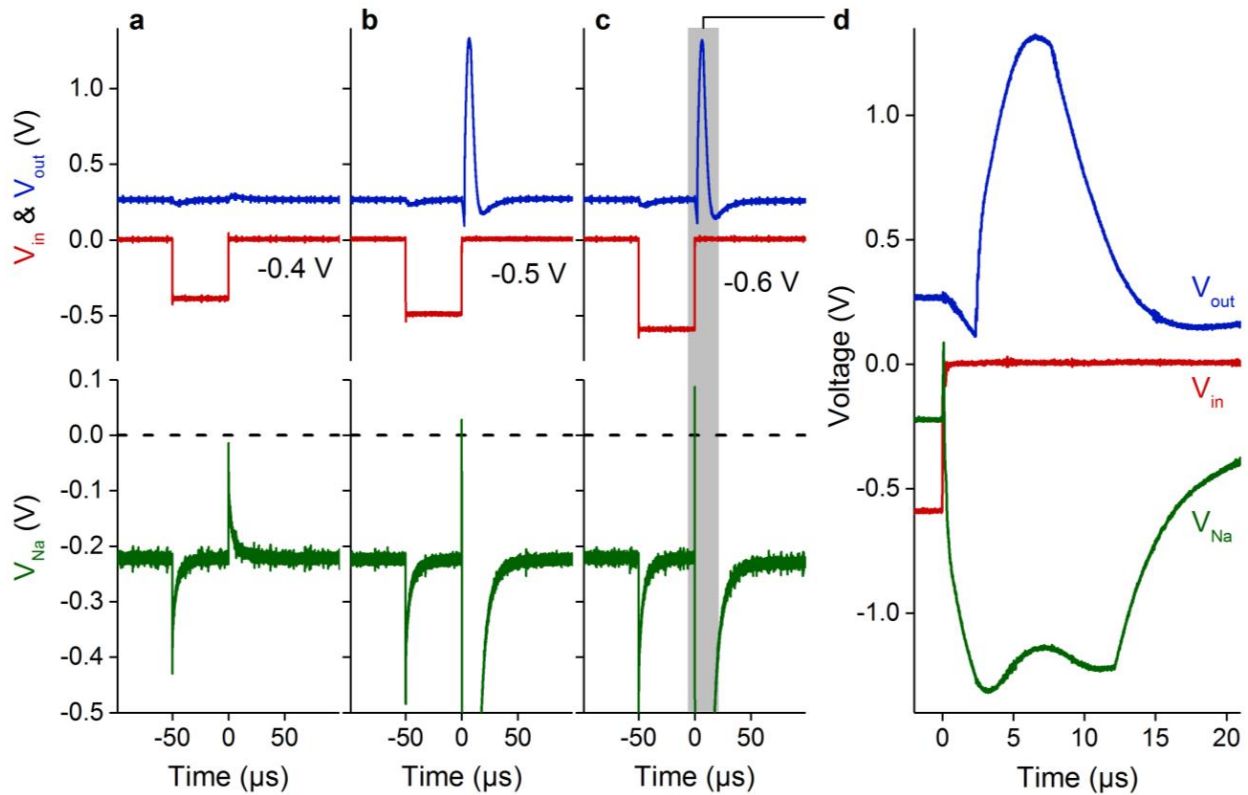

**Supplementary Figure 29. Experimental all-or-nothing characteristics in the rebound spike behavior of a phasic VO$_2$ neuron circuit.** Similar to the all-or-nothing response to an excitatory stimulus, for an inhibitory input, there also exists a threshold in its amplitude for a rebound spike to be fired when the input is released. **a**, The lack of response to a subthreshold inhibitory input pulse (-0.4 V), which is not strong enough for the neuron to fire a rebound spike at the rise edge of the input. It is noticed that the Na$^+$ channel membrane potential surges up at the rise edge of the input, but it still stays below zero and therefore does not trigger the Na$^+$ channel to open. **b–c**, A rebound spike is fired in response to a suprathreshold inhibitory input pulse of -0.5V and -0.6 V, respectively. In both cases, the Na$^+$ channel membrane potential surges above zero at the rise edge of the input, triggering the coordinated opening/closing of the Na$^+$ and K$^+$ channels and an action potential generation. **d**, a close-up view of the greyed-out area in **c**, showing more details in the time evolution of the input $V_{in}$, the Na$^+$ channel membrane potential $V_{Na}$, and the K$^+$ channel membrane potential (the neuron output) $V_{out}$.

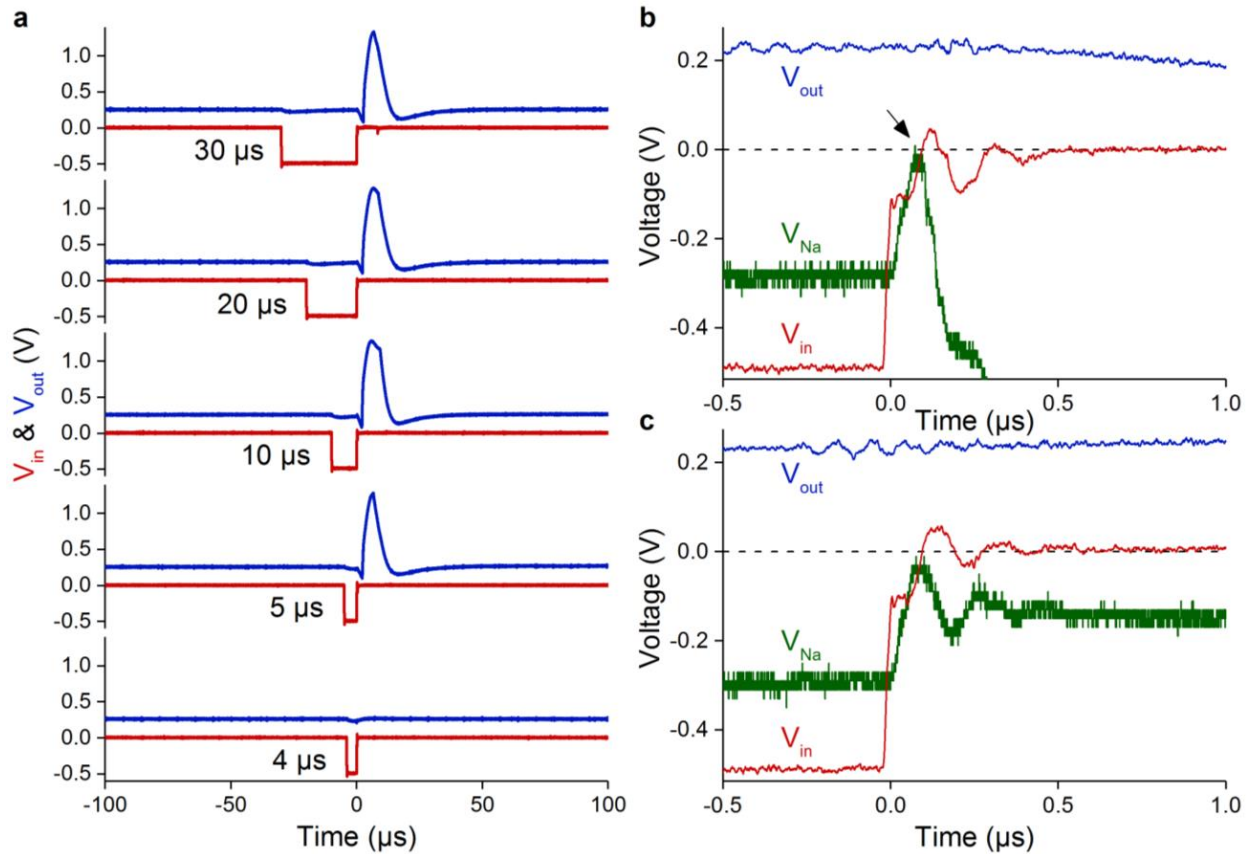

**Supplementary Figure 30. Experimental input-duration threshold characteristics in the rebound spike behavior of a phasic VO₂ neuron circuit. a**, From top to bottom, a rebound spike is fired in response to a suprathreshold inhibitory input pulse of -0.5V with a duration of 30 µs, 20 µs, 10 µs, and 5 µs, respectively, but the neuron does not fire if the duration is further shortened to 4 µs (the bottom panel). **b**, a close-up view of the case for 5 µs inhibitory input in **a**, showing that the Na$^+$ channel membrane potential $V_{Na}$ surges above zero at the rise edge of the input (arrow), triggering a rebound spike. **c**, a close-up view of the case for 4 µs inhibitory input in **a**, showing that the Na$^+$ channel membrane potential $V_{Na}$ surges but stays below zero at the rise edge of the input, and is thus incapable of triggering the rebound spike.

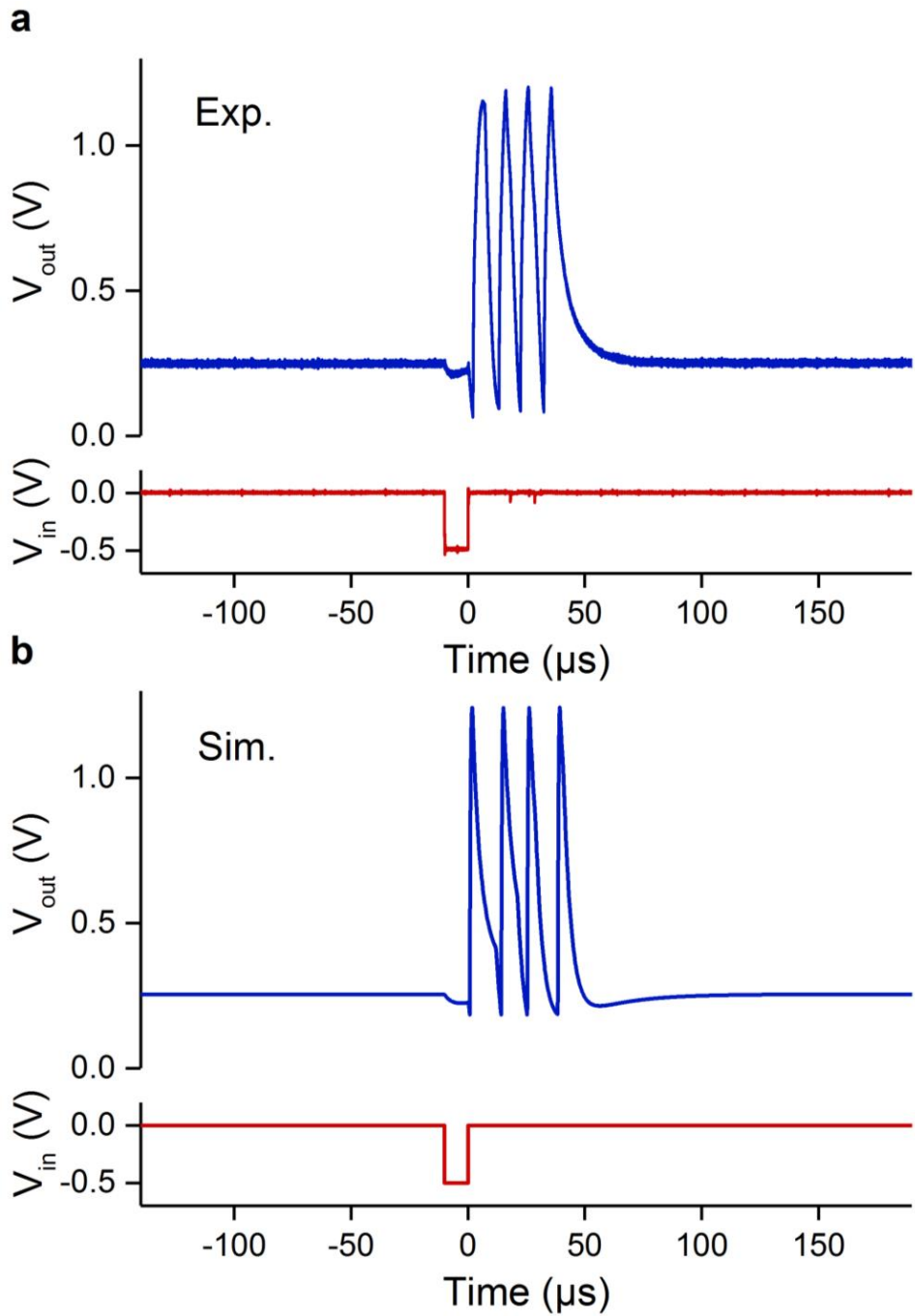

**Supplementary Figure 31. Rebound burst behavior in a phasic VO$_2$ neuron circuit. a**, Experimentally measured rebound burst behavior, showing that when the neuron receives and then is released from an inhibitory (negative) input, it fires a post-inhibitory (rebound) burst of spikes, in response to the release (the rise edge) of the inhibitory input waveform). **b**, Simulated rebound burst behavior of the same phasic VO$_2$ neuron circuit.

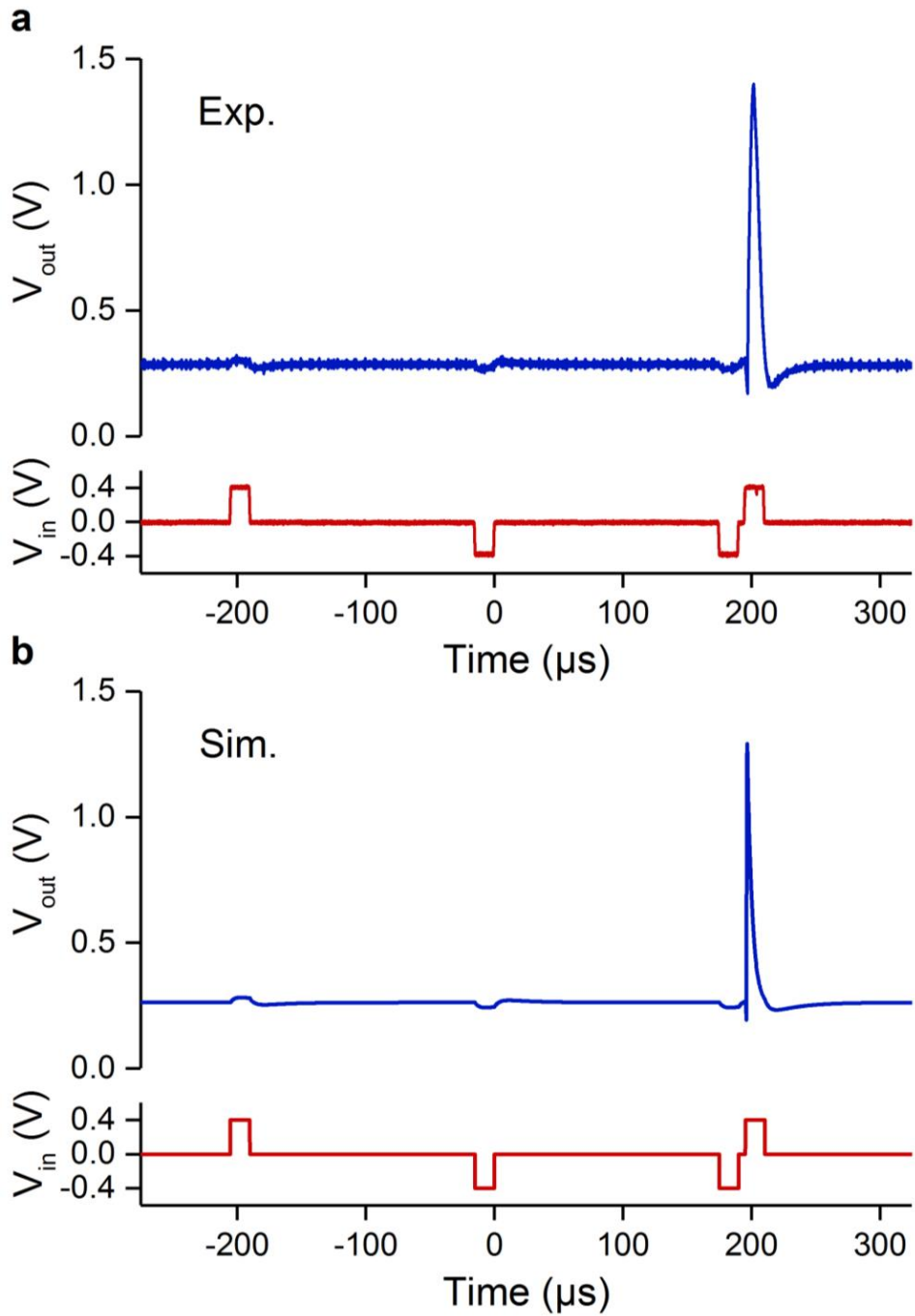

**Supplementary Figure 32. Threshold variability behavior in a phasic VO$_2$ neuron circuit. a**, Experimental threshold variability, showing that the neuron does not fire when it receives a brief subthreshold excitatory or inhibitory input pulse, but it fires a spike if the inhibitory input pulse is followed by an excitatory pulse (both are subthreshold) as long as the interval is short enough. The preceding inhibitory pulse lowers the threshold and makes the neuron more excitable. **b**, Simulated threshold variability of the same phasic VO$_2$ neuron circuit. Combining the rise edges of the inhibitory and excitatory pulses into one effective rise edge, threshold variability is caused by the same mechanism as the threshold seen in rebound spike (See Supplementary Fig. 29).

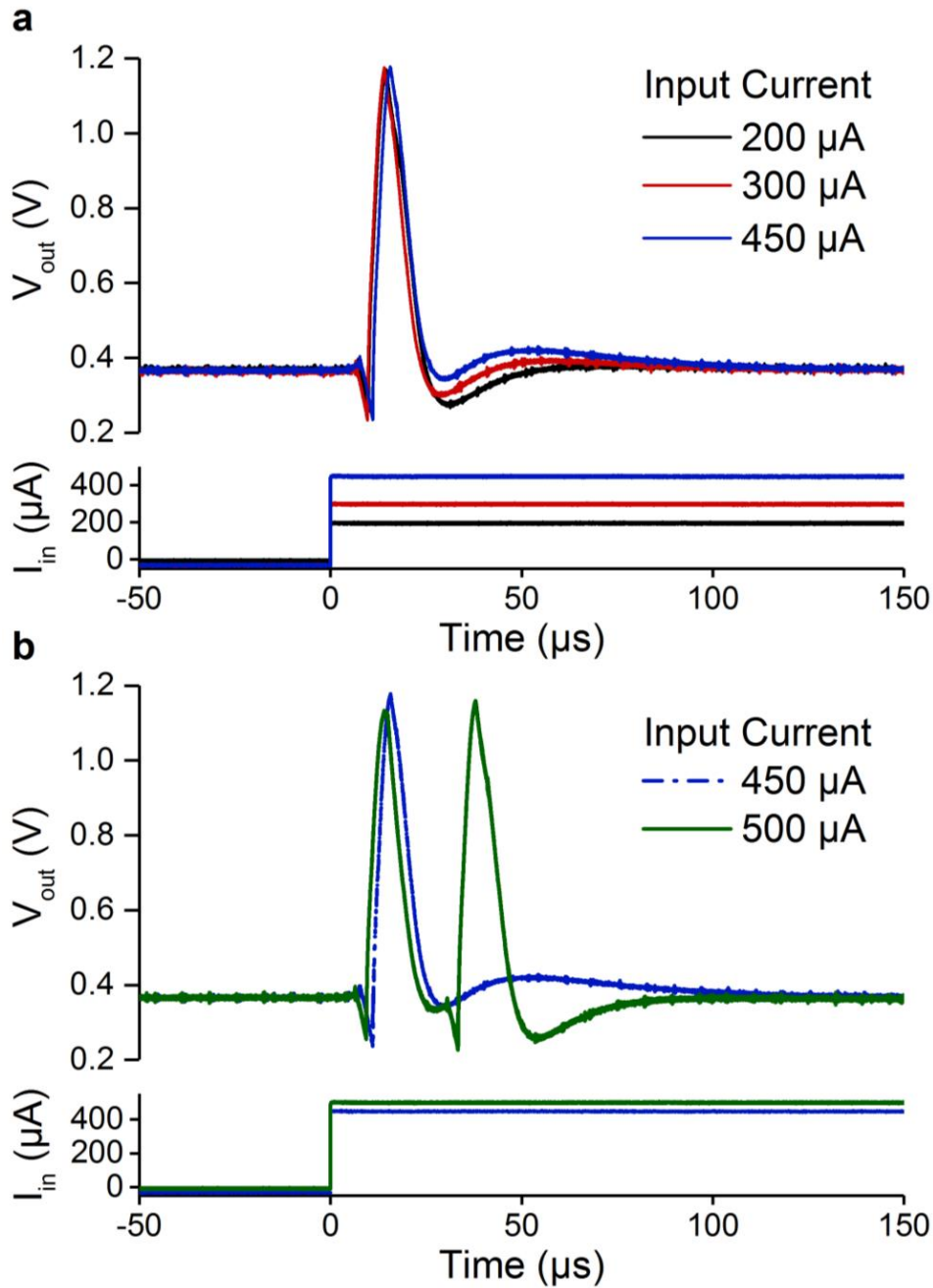

**Supplementary Figure 33. Experimental depolarizing after-potential behavior of a phasic VO₂ neuron circuit. a**, Single phasic spikes fired at the onset of d.c. input currents of 200 μA, 300 μA, and 450 μA levels. At relatively weaker input currents, the neuron membrane potential develops the commonly seen hyperpolarizing after-potential (HAP) that goes below the resting level. As the input strengthens, the HAP gradually weakens and morphs into a depolarizing after-potential (DAP) that goes above the resting level. **b**, Once a DAP is developed, the neuron has shortened refractory period and becomes superexcitable. A slightly stronger input, from 450 μA to 500 μA, causes the neuron to fire a second spike. The second spike is triggered by Na$^+$ channel reactivation when the relative refractory period is nullified by the formation of DAP.

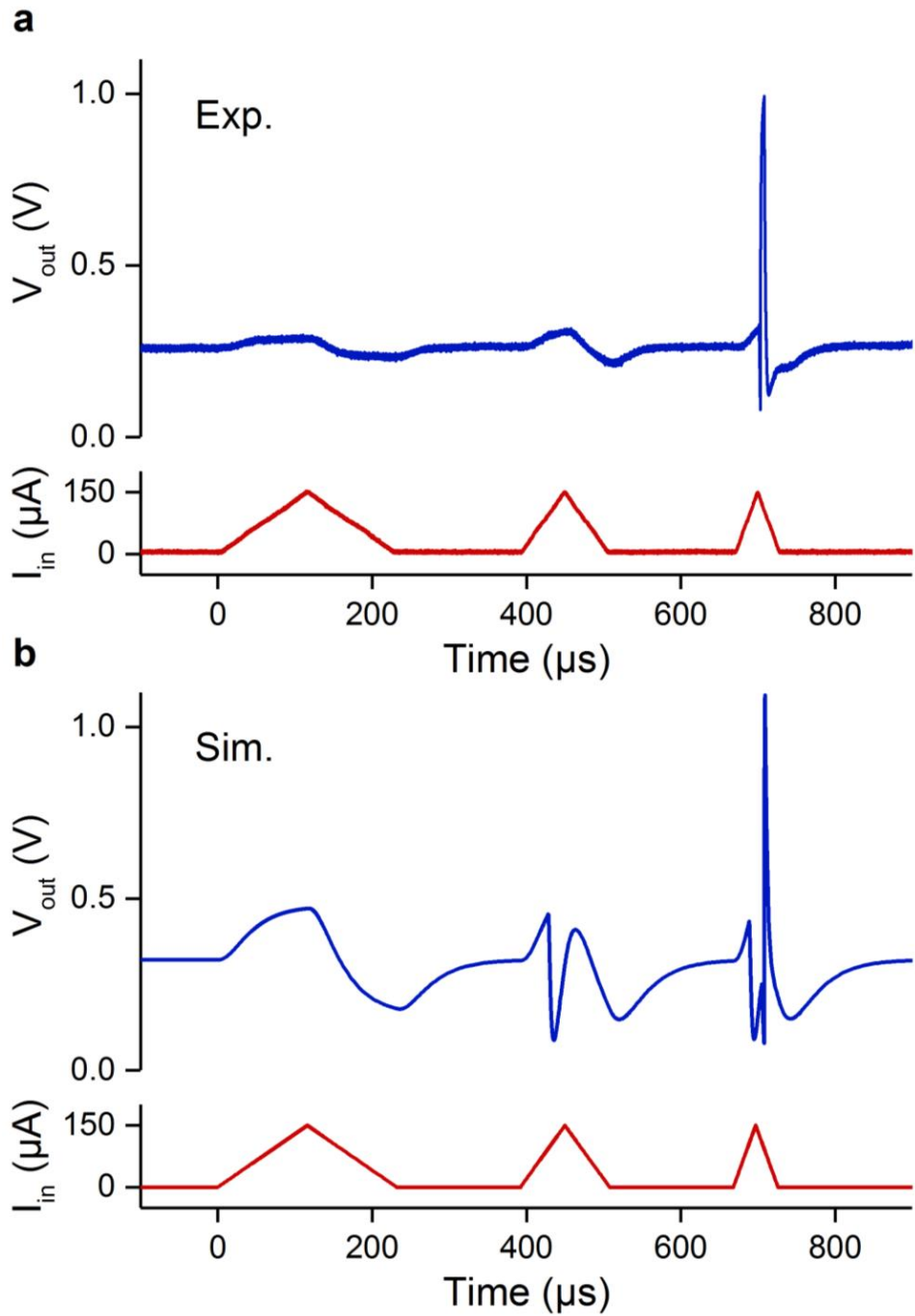

**Supplementary Figure 34. Accommodation behavior in a phasic VO₂ neuron circuit. a**, Experimental accommodation behavior, showing that a slowly ramped input current does not trigger the neuron to fire. In other words, the neuron accommodates the input change and becomes less excitable. A sharply ramped current, however, triggers a spike. All the current ramps have the same maximum amplitude of 150 µA. **b**, Simulated accommodation behavior of the same phasic VO₂ neuron circuit.

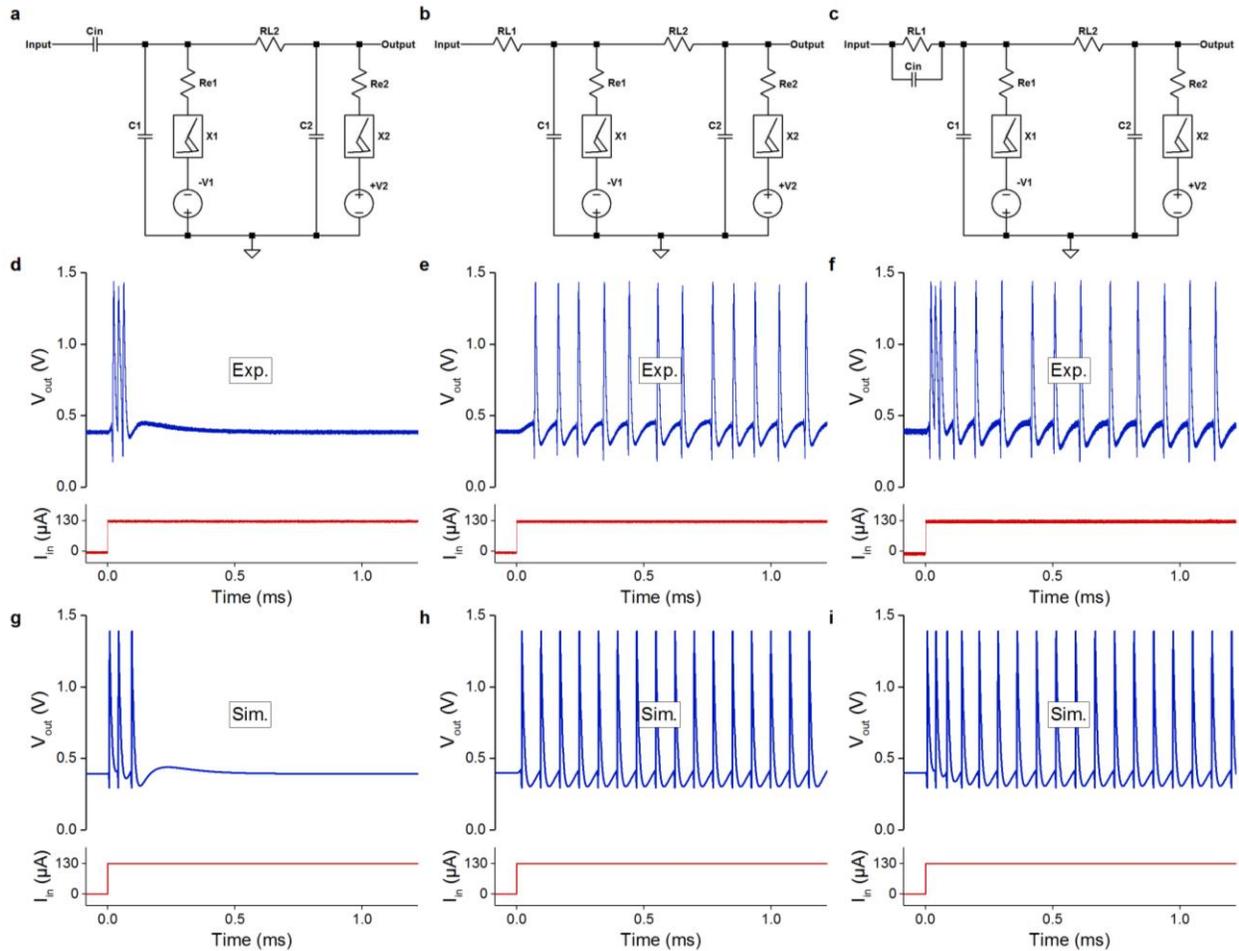

**Supplementary Figure 35. Mixed-mode spiking behavior of a mixed-mode VO$_2$ neuron circuit. a**, A phasic neuron with a capacitive coupling ($C_{in}$) to dendritic inputs. **b**, A tonic neuron with a resistive coupling ($R_{L1}$) to dendritic inputs. **c**, A mixed-mode neuron with both capacitive and resistive couplings ($C_{in}$ in parallel with $R_{L1}$) to dendritic inputs. Except for the difference in input impedance, the tested neuron circuits in **a–c** are identical, including the VO$_2$ devices and d.c. biases used. **d**, Phasic bursting measured from the phasic neuron circuit in **a**. **e**, Tonic spiking measured from the tonic neuron circuit in **b**. **f**, Mixed-mode spiking, i.e., phasic bursting followed by tonic spiking, measured from the mixed-mode neuron circuit in **c**. **g**, Simulated phasic bursting of the phasic neuron circuit in **a**. **h**, Simulated tonic spiking of the tonic neuron circuit in **b**. **i**, Simulated mixed-mode spiking of the mixed-mode neuron circuit in **c**.

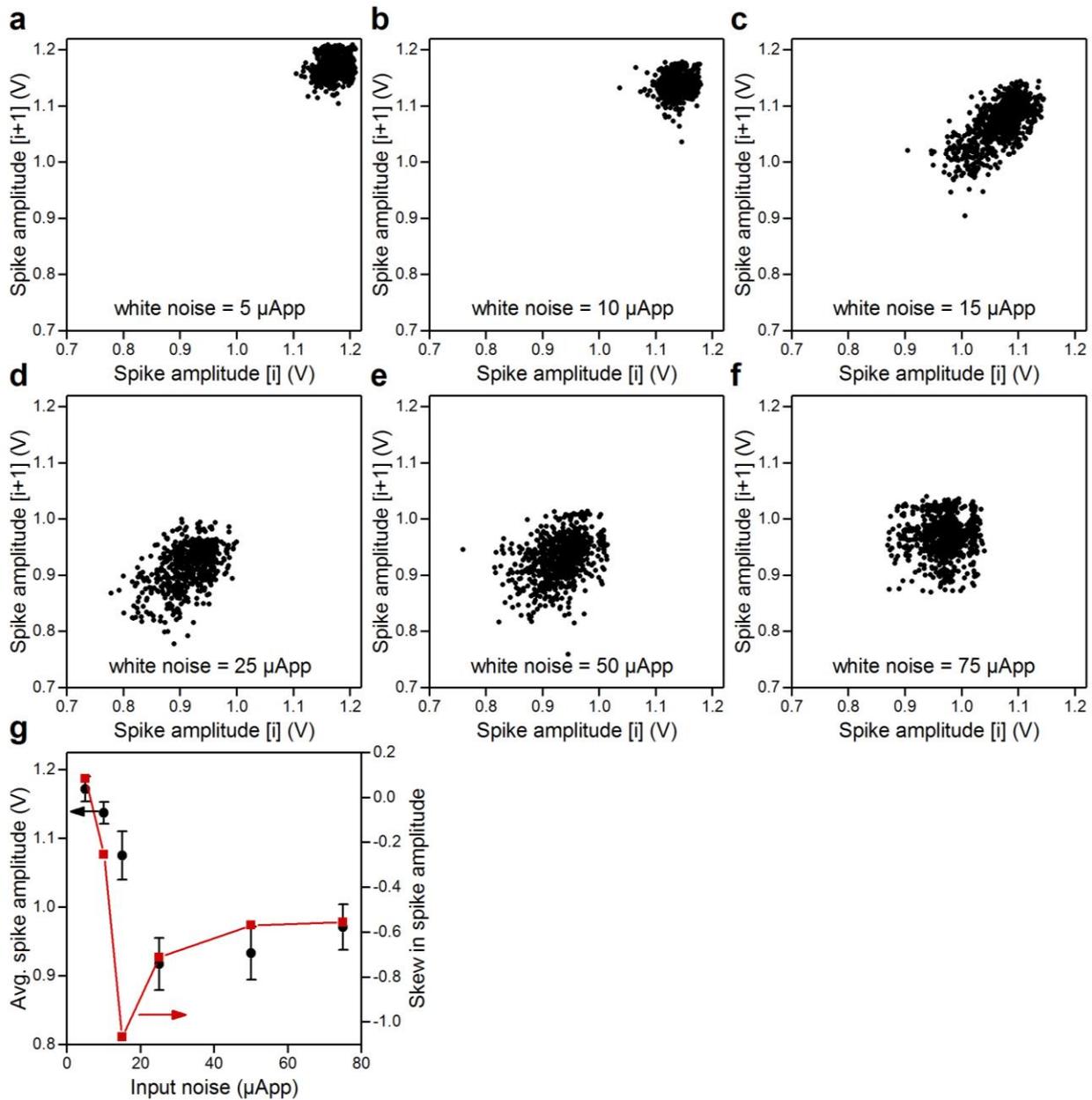

**Supplementary Figure 36. Experimental recurrence plots (Poincaré plots) of spike amplitudes in a tonic VO$_2$ neuron circuit (from the same data as Fig. 5). a–f**, Scatter recurrence plots of adjacent spike amplitudes at input white noise (peak-to-peak) levels of 5 µApp, 10 µApp, 15 µApp, 25 µApp, 50 µApp, and 75 µApp, respectively, showing that irregularities develop in both the spike timing (See Fig. 5) and the spike amplitude as the input stimulus becomes noisier. **g**, Dependences of the mean spike amplitude and the skewness in its distribution on the input noise level, showing similar trends of initially a fast decrease with the input noise, then a partial recovery if the input noise level is higher than ~20 µApp.

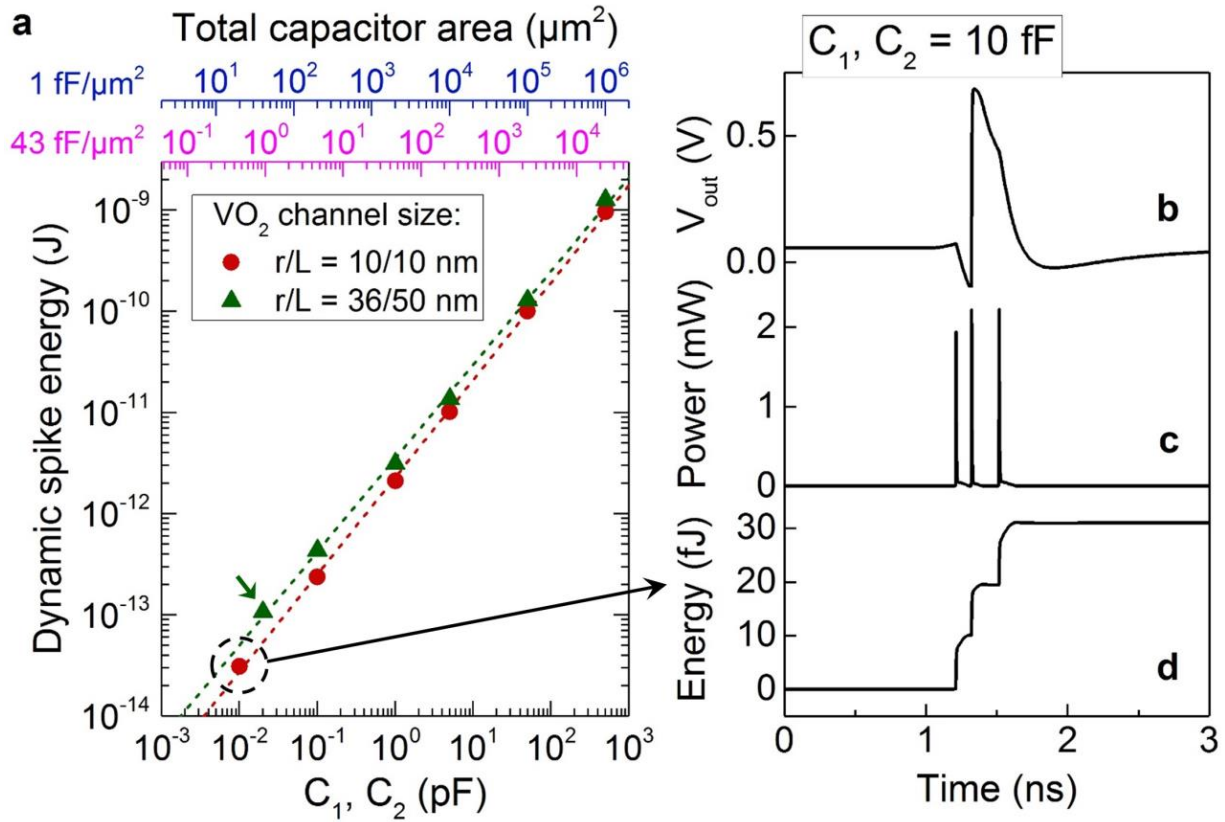

**Supplementary Figure 37. Simulated dynamic power scaling of a VO₂ neuron. a**, Simulated dynamic spike energy of a tonic VO₂ model neuron circuit vs. membrane capacitances, showing a nearly linear scaling with the capacitor values that also determine the neuron area. This is because the energy dissipated in a spike comes from the coordinated discharging of the two membrane capacitors. The dynamic spike energy is calculated by first summing the total power supplied by the $V_1$ and $V_2$ voltage sources, then integrating over time (see example in **b–d**). For simplicity, $C_1 = C_2$ was assumed in the simulations. Two sets of VO₂ channel dimensions of $r/L$ = 10/10nm and $r/L$ = 36/50 nm are compared. Here $r$ is the VO₂ channel radius, $L$ is the channel length (film thickness). The results show that the impact of the VO₂ channel dimensions on the neuron dynamic spike energy is relatively small. By aggressively shrinking the VO₂ volume by a factor of 18×, from $r/L$ = 36/50 nm to $r/L$ = 10/10 nm, the spike energy is only reduced by 24%. The top x axes show the calculated total capacitor area at capacitance density of 1 fF/μm² (blue) and 43 fF/μm² (magenta), respectively. A dynamic spike energy use <0.1 pJ/spike (green arrow) can be achieved at a total capacitor area of ~1 μm² by using 20 fF membrane capacitors, which can be realized by today's integrated high-$\kappa$ metal-insulator-metal (MIM) capacitors with a record-high capacitance density[21] of 43 fF/μm² (typical MIM capacitance density of high-$\kappa$ dielectrics is in the range of 15–20 fF/μm²). Note that at a given capacitor area, lower capacitance value (by using lower capacitance density) will result in a lower spiking energy and hence a better EE. This is clearly shown in the EE-area scaling trend lines of VO₂ neurons (See Supplementary Fig. 2). **b**, Time dependent neuron spike waveform at $C_1, C_2$ = 10 fF and $r/L$ = 10 nm/10 nm (the circled red dot in **a**). **c**, Time dependent total dynamic power supplied by the $V_1$ and $V_2$ voltage sources. **d**, Time dependent dynamic energy consumption, calculated by integrating the total dynamic power simulated in **c** over time.

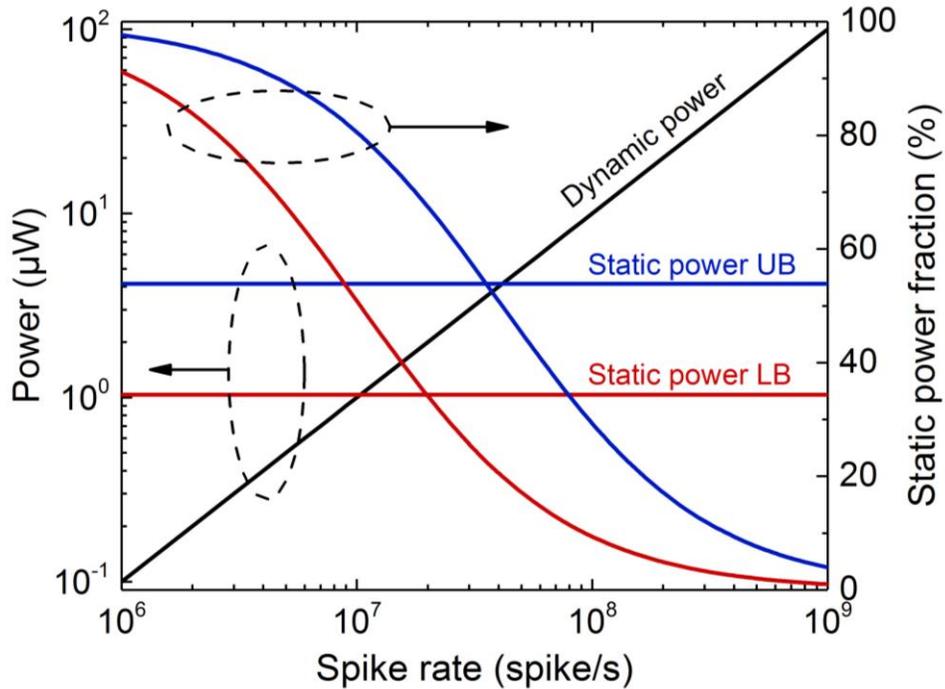

**Supplementary Figure 38. Simulated dynamic and static power scaling of a $VO_2$ neuron over its spike rate.** Left axis shows simulated dynamic and static power consumptions of a tonic $VO_2$ model neuron, and the right axis shows the percentage of static power in the total power consumption. Static power is dissipated by the neuron membrane leakage current in the resting state, i.e., leakage current drawn by d.c. biased $VO_2$ devices due to the finite resistivity of the insulating phase (~1 Ω·cm). With $V_{th}$ as the switching threshold voltage, the upper bound (UB) and lower bound (LB) of static power are calculated at d.c. bias $V_{dc} = V_{th}$ and $V_{th}/2$, respectively. Since the neuron only spikes if $(V_{in} + V_{dc}) \geq V_{th}$, the signal gain, capped by $V_{dc}/V_{in}$, is always smaller than $V_{dc}/(V_{th}-V_{dc})$. The gain will becomes less than 1 if $V_{dc}$ is less than $V_{th}/2$. While static power dissipation is independent of the spike rate, dynamic power dissipation is proportional to the spike rate. Therefore at low spike rates, static power may dominate the total power consumption. At sufficiently high spike rates, the static power makes only an insignificant contribution to the total power consumption, and is not expected to be of major concern for system level energy efficiency. The LB and UB of static power is less than 10 % of the total power at a spike rate higher than 100 MHz and 400 MHz, respectively, and the overall energy use is better than 0.11 pJ/spike. At 100 MHz spike rate, the single neuron total power consumption is 11 μW (LB) to 14 μW (UB). The SPICE simulation assumes a $VO_2$ channel of r/L = 10/10 nm, and the $VO_2$ model neuron has an energy use of 0.1 pJ/spike at $C_1, C_2$ = 38.3 fF (see the red dashed line in Supplementary Figure 37a for details).

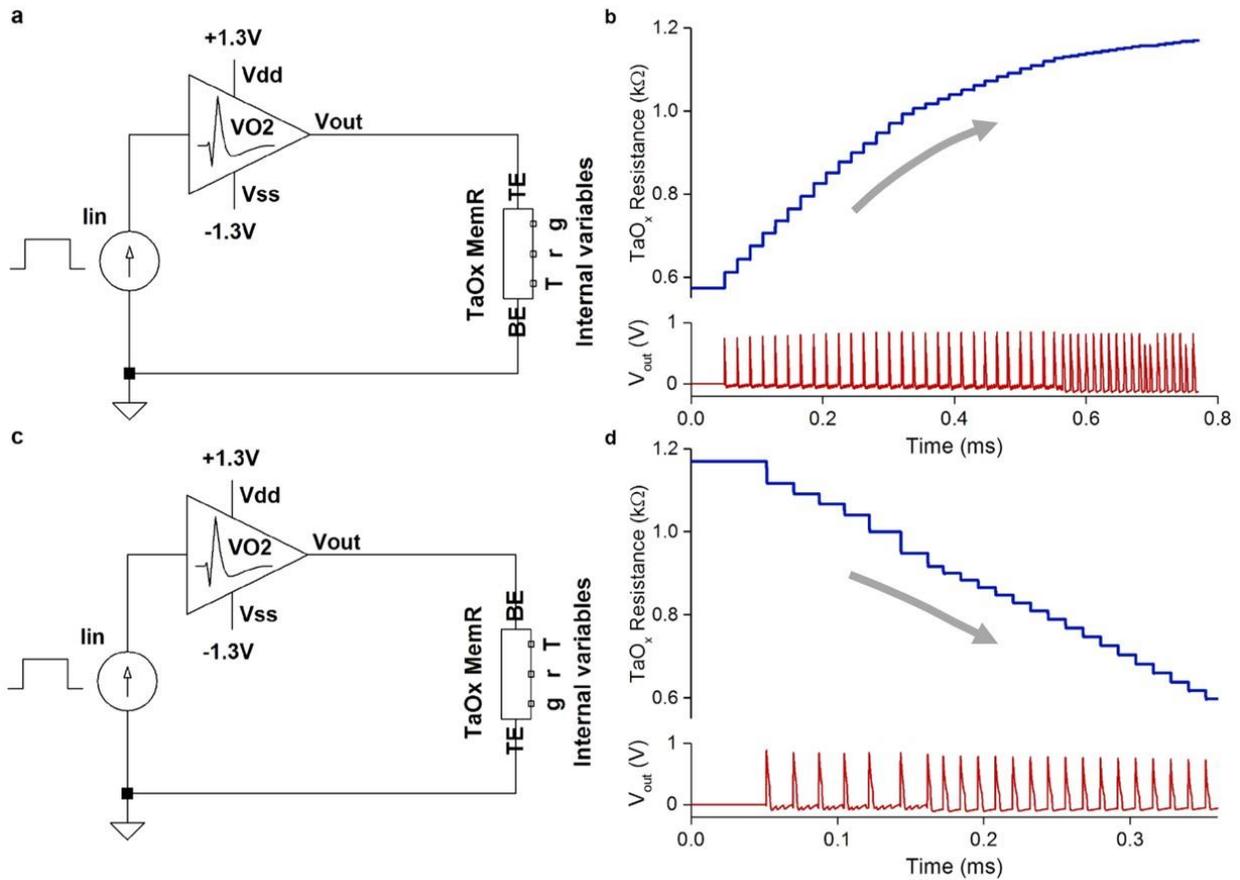

**Supplementary Figure 39. Simulated potentiation and depression of a TaO$_x$ memristor synapse by a VO$_2$ tonic neuron. a**, Circuit diagram of the simulated depression (reset) of a TaO$_x$ memristor with its top electrode connected to the output of a VO$_2$ tonic neuron circuit (the amplifier symbol). The VO$_2$ neuron fires a spike train in response to a square wave current input. **b**, Evolution of the TaO$_x$ device resistance over time in circuit **a**, showing that each step in resistance rise is caused by a presynaptic VO$_2$ neuron spike. **c**, Circuit diagram of the simulated potentiation (set) of a TaO$_x$ memristor with its bottom electrode connected to the VO$_2$ neuron output. **d**, Evolution of the TaO$_x$ device resistance over time in circuit **c**, showing that each step in resistance drop is caused by a presynaptic VO$_2$ neuron spike. The TaO$_x$ memristor SPICE model is from Ref. 22.

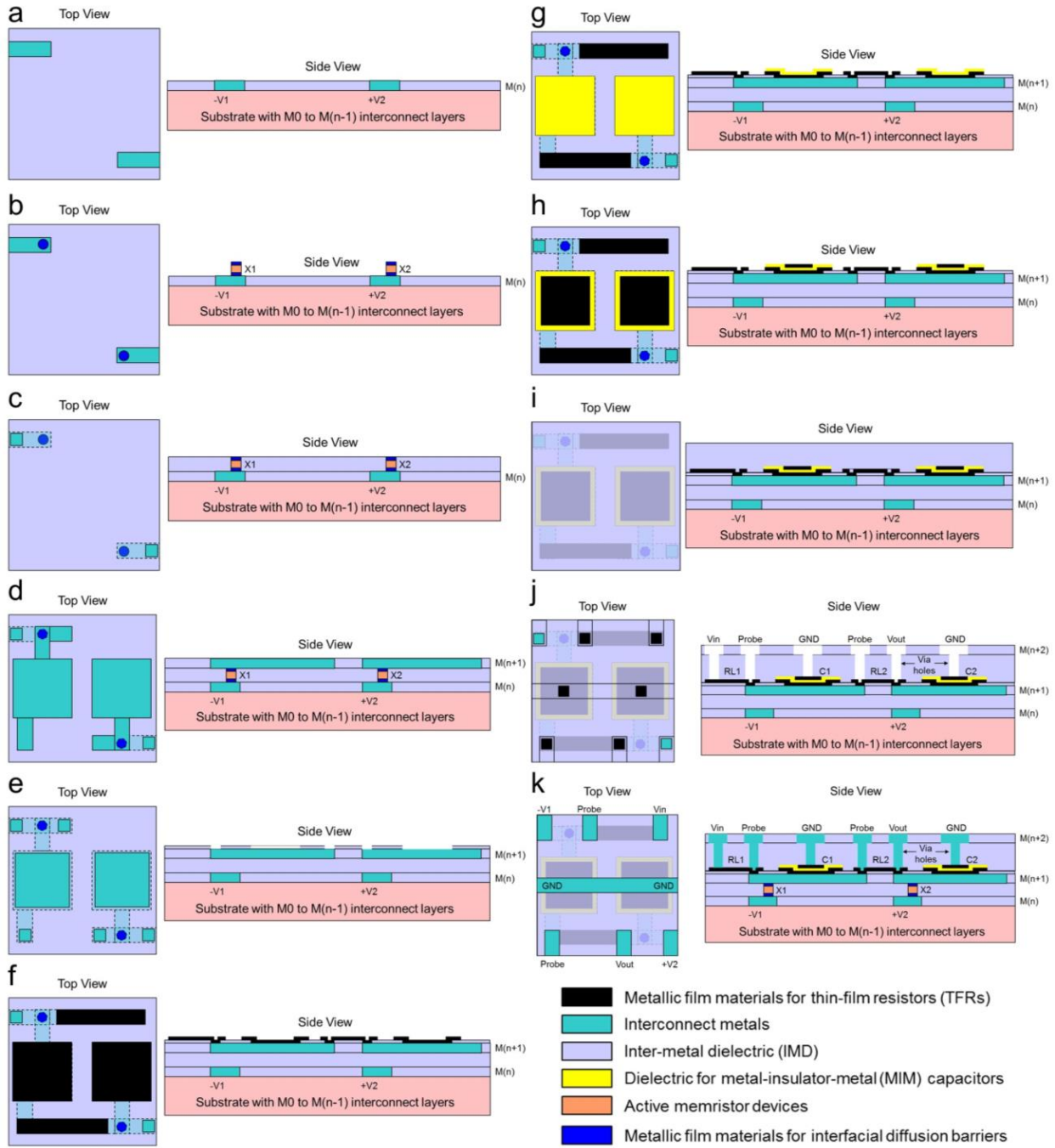

**Supplementary Figure 40. Schematic top view and side cross-section view (not to scale) of possible process steps to manufacture an integrated active memristor neuron circuit.** The complete integrated circuit only requires up to three layers of interconnect metals (M(n) to M(n+2)). The structures made in each process steps are: **a**, Bottom electrodes (BE) of memristors at M(n). **b**, Active memristor device stack that include metallic diffusion barriers. **c**, Inter-metal dielectric. d, Top electrodes (TE) of memristors (also as bottom electrodes of capacitors) at M(n+1). **e**, Dielectric passivation and contact openings of the TE. **f**, Thin-film resistors (also as the bottom contact layer for capacitors). **g**, Dielectric for capacitors. **h**, Top contact layer for capacitors. **i**, Inter-metal dielectric. **j-k**, Contact pads at M(n+2) by a dual-damascene process. Other methods, e.g. single damascene, can also be used.

**Supplementary Tables**

**Supplementary Table 1. Comparison of volumetric free energy cost for Mott phase transitions in NbO$_2$ and VO$_2$ materials.** Values for NbO$_2$ are from Table 1 in Ref. 16 (and references therein). At the same volume, the free energy cost for NbO$_2$ phase transition is 6.1 times that for VO$_2$. The total volumetric enthalpy change in a VO$_2$ nano-crossbar device with channel radius $r$ = 10 nm and length $L$ = 10 nm is merely 1.15 fJ.

| Material | NbO$_2$ | VO$_2$ |
|---|---|---|
| Mott transition T$_C$ | 1080 K | 340 K |
| Temperature rise | 800 K | 40 K |
| Volumetric heat capacity | 2.6×10$^6$ J·m$^{-3}$·K$^{-1}$ | 3.3×10$^6$ J·m$^{-3}$·K$^{-1}$ |
| Volumetric enthalpy of IMT | 1.6×10$^8$ J·m$^{-3}$ | 2.35×10$^8$ J·m$^{-3}$ |
| Total volumetric enthalpy change | 2.24×10$^9$ J·m$^{-3}$ | 3.67×10$^8$ J·m$^{-3}$ |
| Factor of improvement | - | 6.1× |

**Supplementary Table 2. VO₂ material and structural parameters used in SPICE model simulations.** In most cases, the Mott physics-based analytical compact SPICE model can faithfully reproduce experimental VO$_2$ device switching dynamics and neuron spiking behaviors (See all the simulated spiking patterns in Supplementary Figures). The model uses published VO$_2$ material properties. All the simulated neuron behaviors used the same VO$_2$ device model with a cylindrical-shaped VO$_2$ conduction channel of 56 nm in radius and 100 nm in length to match the actual VO$_2$ crystal volume in 100 × 100 nm$^2$ sized and 100 nm-thick nano-crossbar devices used in the experiments, and only varied the values of R, C elements. Series electrode resistance of 150-500 Ω, and parallel VO$_2$ channel leakage resistance of 13 to 17 kΩ were included in simulations to take into account their effects on the voltage drop across the memristors and the standby current in the insulating phase.

| Memristor Model Property | Symbol | Model Value | Unit | Reference |
|---|---|---|---|---|
| Heat capacity | $c_p$ | 3.30×10⁶ | J·m$^{-3}$·K$^{-1}$ | 23 |
| Enthalpy change (latent heat) of MIT | $\Delta h_{tr}$ | 2.35×10⁸ | J·m$^{-3}$ | 24 |
| Thermal conductivity of insulating phase | $\kappa$ | 3.5 | W·m$^{-1}$·K$^{-1}$ | 23 |
| Metallic phase electrical resistivity | $\rho_{met}$ | 3×10⁻⁶ | Ω·m | 23,24 |
| Insulating phase electrical resistivity | $\rho_{ins}$ | 1×10⁻² | Ω·m | 24 |
| Heating temperature | $\Delta T$ | 43 | K | 24 |
| Conduction channel radius | $r_{ch}$ | 56 | nm | Exp. |
| Conduction channel length | $L_{ch}$ | 100 | nm | Exp. |
| Series electrode resistance | $R_e$ | 150 to 500 | Ω | Exp. |
| Parallel leakage resistance | $R_{shunt}$ | 13k to 17k | Ω | Exp. |

**Supplementary Table 3. Experimental circuit parameters used in neuron spiking tests.** The VO$_2$ nano-crossbar devices X$_1$ and X$_2$ with the same nominal size of 100 × 100 nm$^2$ and film thickness of 100 nm are randomly selected from the same wafer. C$_1$ and C$_2$ values in the table are the values of discrete capacitors connected through coaxial cables to VO$_2$ devices. In simulations, stray capacitances in the setup, typically in the range of ~1 nF, are added to C$_1$ and C$_2$ values.

| Spiking behavior | Figure No. | $R_{L1}$ (kΩ) | $R_{L2}$ (kΩ) | $C_1$ (nF) | $C_2$ (nF) | $C_{in}$ (nF) | $V_1$ (V) | $V_2$ (V) | $X_1$ (ID) | $X_2$ (ID) |
|---|---|---|---|---|---|---|---|---|---|---|
| All-or-nothing | S11 | 6 | 6 | 2 | 2 | – | -1.35 | 1.35 | 5251-13 | 5251-9 |
| Refractory period | S12 | 5 | 5 | 5 | 5 | – | -1.6 | 1.6 | 5050-15 | 5050-7 |
| Absolute & relative refractory periods | S13 | 6 | 6 | 4 | 1 | – | -1.45 | 1.45 | 5352-1 | 5252-13 |
| Tonic spike | S14 | 5 | 5 | 5 | 2 | – | -1.5 | 1.5 | 5151-7 | 5151-3 |
| Tonic burst | S15 | 10 | 10 | 5–30 | 0 | – | -1.85 | 1.85 | 5051-9 | 5051-5 |
| Class 1 excitable | 4c | 5 | 5 | 5 | 5 | – | -1.5 | 1.5 | 5151-7 | 5151-3 |
| Class 2 excitable | 4b | 5 | 5 | 1 | 5 | – | -1.5 | 1.5 | 5151-7 | 5151-3 |
| Spike frequency adaptation (tonic) | S16 | 10 | 10 | 200 | 2 | – | -1.4 | 1.4 | 5251-13 | 5251-9 |
| Spike frequency adaptation (phasic) | S17 | – | 9 | 4 | 1.2 | 9 | -1.6 | 1.6 | 5351-11 | 5351-7 |
| Spike latency | S18 | 6 | 6 | 10 | 3 | – | -1.5 | 1.5 | 5352-1 | 5252-13 |
| Subthreshold oscillation | S19 | 5 | 5 | 2 | 3 | – | -1.4 | 1.4 | 5350-11 | 5350-7 |
| Integrator | S20 | 6 | 6 | 8.5 | 2 | – | -1.4 | 1.4 | 5251-13 | 5251-9 |
| Bistability | S21 | 0 | 7 | 1.5 | 2 | – | -1.58 | 1.58 | 5352-1 | 5252-13 |
| Inhibition-induced spike | S22 | 6 | 6 | 6 | 2 | – | -1.4 | 1.4 | 5251-13 | 5251-9 |
| Inhibition-induced burst | S23a | 6 | 6 | 35 | 0 | – | -1.4 | 1.4 | 5251-13 | 5251-9 |
| | S23b | 7 | 7 | 21 | 0 | – | -1.5 | 1.5 | 5049-3 | 4949-15 |
| Excitation block | S24 | 6 | 6 | 0 | 2 | – | -1.4 | 1.4 | 5251-13 | 5251-9 |
| Resonator | S25 | 5 | 7 | 5 | 0 | 5 | -1.5 | 1.5 | 5250-13 | 5250-9 |
| Phasic spike | S26 | – | 7 | 1 | 2 | 0.3 | -1.6 | 1.6 | 5352-1 | 5252-13 |
| Phasic burst | S27 | – | 7 | 4 | 0 | 0.3 | -1.6 | 1.6 | 5352-1 | 5252-13 |

(Continued on next page)

| Spiking behavior | Figure No. | $R_{L1}$ (kΩ) | $R_{L2}$ (kΩ) | $C_1$ (nF) | $C_2$ (nF) | $C_{in}$ (nF) | $V_1$ (V) | $V_2$ (V) | $X_1$ (ID) | $X_2$ (ID) |
| --- | --- | --- | --- | --- | --- | --- | --- | --- | --- | --- |
| Rebound spike | S28–S30 | – | 5.9 | 0 | 1 | 0.3 | -1.5 | 1.5 | 5352-1 | 5252-13 |
| Rebound burst | S31 | – | 5.9 | 0 | 0.5 | 0.3 | -1.5 | 1.5 | 5352-1 | 5252-13 |
| Threshold variability | S32 | – | 5.9 | 0 | 0.5 | 0.3 | -1.5 | 1.5 | 5352-1 | 5252-13 |
| Depolarizing after-potential | S33 | – | 6 | 0.9 | 2 | 0.3 | -1.3 | 1.3 | 5352-1 | 5252-13 |
| Accommodation | S34 | – | 7 | 1 | 0 | 0.3 | -1.68 | 1.68 | 5352-1 | 5252-13 |
| Mixed mode | S35 | 240 | 9 | 4 | 1.2 | 1 | -1.6 | 1.6 | 5351-11 | 5351-7 |
| Skipping | S36 | 7 | 7 | 1 | 1 | – | -1.5 | 1.5 | 5149-11 | 5149-7 |

**Supplementary Table 4. Biological fidelity and computational cost of neuron models in comparison with experimentally demonstrated biological fidelity of VO$_2$ neurons.** The neurocomputational properties of neuron models are adopted and augmented from Fig. 2 in Ref. 18. "# of FLOPS" is the approximate number of floating point operations needed to simulate the neuron model for a 1 ms duration using a digital computer. (+), (-) and empty square represents possessed, missing, and unconfirmed properties of the model. For VO$_2$ neurons, the only property that remains unconfirmed is chaos.

| Neuron Models | All-or-nothing firing | Refractory period | Excitation block | Biophysically meaningful | Tonic spiking | Phasic spiking | Tonic bursting | Phasic bursting | Mixed mode | Spike frequency adaptation | Class 1 excitable | Class 2 excitable | Spike latency | Subthreshold oscillations | Resonator | Integrator | Rebound spike | Rebound burst | Threshold variability | Bistability | DAP | Accommodation | Inhibition-induced spiking | Inhibition-induced bursting | Chaos | # of FLOPS |
|---|---|---|---|---|---|---|---|---|---|---|---|---|---|---|---|---|---|---|---|---|---|---|---|---|---|---|
| integrate-and-fire | + | + | - | - | + | - | - | - | - | - | + | - | - | - | - | + | - | - | - | - | - | - | - | - | - | 5 |
| integrate-and-fire with adapt. | + | + | - | - | + | - | - | - | - | + | + | - | - | - | - | + | - | - | - | - | + | - | - | - | - | 10 |
| integrate-and-fire-or-burst | + | + | - | - | + | + |  | + | - | + | + | - | - | - | - | + | + | + | - | + | + | - | - | - |  | 13 |
| resonate-and-fire | + | + | + | - | + | + | - | - | - | - | + | + | - | + | + | + | + | - | - | + | + | + | - | - | + | 10 |
| quadratic integrate-and-fire | + | + | - | - | + | - | - | - | - | - | + | - | + | - | - | + | - | - | + | + | - | - | - | - | - | 7 |
| Izhikevich (2003) | + | + | + | - | + | + | + | + | + | + | + | + | + | + | + | + | + | + | + | + | + | + | + | + | + | 13 |
| FitzHugh-Nagumo | * | + | + | - | + | + | - |  | - | - | + | - | + | + | + | - | + | - | + | + | - | + | + | - | - | 72 |
| Hindmarsh-Rose | + | + | + | - | + | + | + |  | - | + | + | + | + | + | + | + | + | + | + | + | + | + |  |  | + | 120 |
| Morris-Lecar | + | + | + | + | + | + | - |  | - | - | + | + | + | + | + | + |  | + | + | - | + | + | - | - |  | 600 |
| Wilson | + | + | + | - | + | + | + |  |  | + | + | + | + | + | + | + | + |  | + | + |  |  |  |  |  | 180 |
| Hodgkin-Huxley | * | + | + | + | + | + | + |  |  | + | + | + | + | + | + | + | + | + | + | + | + | + | + |  | + | 1200 |
| HRL VO$_2$ Neuron | + | + | + | + | + | + | + | + | + | + | + | + | + | + | + | + | + | + | + | + | + | + | + | + |  |  |

*Experimentally demonstrated*

**Supplementary Notes**

**Supplementary Note 1: VO$_2$ active memristor relaxation oscillator**

VO$_2$ is well-known for its first-order thermodynamically-driven Mott insulator-to-metal (IMT) phase transition with a critical temperature $T_C$ near 67 °C[25]. Joule heating produced by electrical current through a metal/VO$_2$/metal device generates Mott IMT-induced volatile hysteretic resistive switching and an NDR regime, which forms the basis to construct oscillators, amplifiers, and impulse circuits (neurons). Mott memristors, a type of active memristors based on Mott IMT, were previously realized by producing crystalline NbO$_2$ in an electroforming process from amorphous Nb$_2$O$_5$ films[26]. Electroformed devices suffer from large device variability that is undesirable for integration. In our case, electroform-free VO$_2$ active memristor nano-crossbar devices with typical device yield of 98–100 %, low-voltage (down to ~0.5 V), high-endurance, and low device variability (<13 % coefficient of variation in switching threshold voltage) are fabricated on CMOS-compatible 3-inch SiN$_x$-coated silicon substrates (See Methods). As discussed in the next section, Pearson-Anson (PA) relaxation oscillator is the prototype electronic circuit analogue for voltage-gated Na$^+$ or K$^+$ nerve membrane ion channels. Other ion channels, e.g. Cl$^-$ or Ca$^{2+}$, can also be emulated in a similar manner. If two such relaxation oscillators are coupled with proper impedance, the overall circuit can generate an action potential[26-28]. Supplementary Fig. 3 shows the circuit diagram and astable oscillator characteristics measured in a VO$_2$ relaxation oscillator. A one-to-one correspondence can be identified between the quasi d.c. V–I trace (force V, measure I) of the VO$_2$ device (without the capacitor) and the V–t and I–t waveforms of the astable oscillations under an external d.c. bias $V_{dc}$[17]. For astable oscillation to occur, the load line, defined by $V_{dc}$ and the load resistor $R_L$, must intersect the V–I curve in its NDR regime. A complete cycle of astable oscillation, from point (1) to (5), has four stages as explained by figure caption. The actual switching time scale of VO$_2$ is much faster than the rise time of the oscilloscope and cannot be measured. Values ranging from 100 fs to 5 ps have been measured by pump-probe methods[29,30].

Although many transition metal oxides exhibit Mott IMT, $T_C$ in many of these materials are well below 300 K (room temperature). Mott insulators with $T_C$ > 300 K, e.g. VO$_2$, Ti$_2$O$_3$, Ti$_3$O$_5$, NbO$_2$, SmNiO$_3$, LaCoO$_3$, are more suitable for electronic applications[31]. NbO$_2$ is a demonstrated material for spiking neurons[26]. However, its $T_C$ of 1080 K requires a large local temperature rise of 800 K to operate, which negatively impacts both power consumption and device longevity[31] (See Supplementary Table 1 for volumetric free energy cost of Mott IMT in NbO$_2$ and VO$_2$). We applied SPICE simulations of a VO$_2$-based relaxation oscillator to estimate the switching energy and switching time (speed) of Mott IMT[26]. The switching time of the phase transition is estimated from the rising edges of device current in each oscillation period. The VO$_2$ channel radius is varied while all the other model parameters, including the channel length (50 nm), are fixed. As shown in Supplementary Fig. 9, at the same channel dimensions, simulated Mott IMT switching in VO$_2$ is 100 times faster than in NbO$_2$, and only consumes about one-sixth (16 %) of the energy. We noted that the switching speed is a material-dependent parameter, and is not affected by the values of R, C passive elements. <1 fJ switching energy and <1 ps switching speed can be achieved at VO$_2$ channel radius of 7–15 nm, dimensions feasible for advanced-node lithography.

**Supplementary Note 2: Action potential generation in a VO$_2$ active memristor neuron**

The basic operational steps in experimental and simulated action potential (spike) generation are shown in Supplementary Fig. 10, with the analogous biological processes illustrated for pedagogical purpose. To be consistent with neuroscience convention, hyperpolarization means the membrane potential is driven toward negative direction, and depolarization is the opposite case. In the resting (quiescent) state, both the Na$^+$ and K$^+$ channels are closed ($X_1$ and $X_2$ are insulating). A resting potential (0.2–0.3 V) is produced by a small membrane leakage current flowing through the two oppositely-energized VO$_2$ devices in insulating state. Hyperpolarization will be triggered by the activation of the Na$^+$ channel if a suprathreshold input voltage or current stimulus (not shown) drives $X_1$ into a metallic state. The Na$^+$ channel membrane potential, $V_{Na}$, is pulled down close to the negative d.c. bias $-E_{Na}$, and its membrane capacitor $C_1$ gets discharged. The neuron output, i.e. the K$^+$ channel membrane potential $V_K$, also gets pulled down through the coupling of $R_{L2}$, but it remains above zero. The opening of Na$^+$ channel is almost instantaneous due to the ultrafast Mott IMT process (seen in the simulated Na$^+$ channel current), but the hyperpolarization of $V_{Na}$ is much slower as it is determined by the $C_1$ discharge time constant. Depolarization is then triggered by the activation of the K$^+$ channel. It occurs when the $V_{Na}$ hyperpolarization pulls $V_K$ down low enough to make the voltage across $X_2$ larger than its switching threshold. After $X_2$ switches to metallic state, $V_K$ gets pulled up close to the positive d.c. bias $+E_K$, and its membrane capacitor $C_2$ gets discharged. The time scale of hyperpolarization is determined by the $C_2$ discharge time constant. The spiking is finalized by a refractory (undershoot) period, during which the neuron is recovering and does not respond to the next input stimulus. The time scale for the action potential to fall, undershoot, then recover to resting is the longest, as $C_1$ and $C_2$ are slowly charged back to their resting states. In biological neurons, recovery is achieved by rebalancing the Na$^+$ and K$^+$ concentrations across the cell membrane by Na$^+$-K$^+$ pumps (conceptually shown by dashed lines) instead of voltage-gated Na$^+$ and K$^+$ protein channels (both are closed).

**Supplementary Note 3: Device modeling of VO₂ active memristors**

We used the same analytical mathematical equations developed by the authors of Ref. 16 to model the dynamics of VO₂ active memristors. The VO₂ model parameters are summarized in Supplementary Table 2. The main equations are relisted below:

$$v = R_{ch}(u) \cdot i \quad (S1)$$

$$\frac{du}{dt} = \left(\frac{d\Delta H}{du}\right)^{-1} \cdot (i^2 R_{ch}(u) - \Gamma_{th}(u)\Delta T) \quad (S2)$$

$$R_{ch}(u) = \frac{\rho_{ins} L_{ch}}{\pi r_{ch}^2}\left[1 + \left(\frac{\rho_{ins}}{\rho_{met}} - 1\right)u^2\right]^{-1} \quad (S3)$$

$$\Gamma_{th}(u) = 2\pi L_{ch} \kappa \left(\ln\frac{1}{u}\right)^{-1} \quad (S4)$$

$$\frac{d\Delta H}{du}(u) = \pi L_{ch} r_{ch}^2 \left[c_p \Delta T \frac{1 - u^2 + 2u^2 \ln u}{2u(\ln u)^2} + 2\Delta h_{tr} u\right] \quad (S5)$$

Eq. (S1) is the instantaneous Ohm's law relationship between current and voltage, wherein the VO₂ channel resistance $R_{ch}(u)$ is determined by a single state variable $u \triangleq r_{met}/r_{ch}$, i.e. the normalized radius of the metallic cylindrical conducting channel heated above the $T_C$ of Mott transition. Eq. (S2) is the first-order differential equation that drives the state dynamics. Eqs. (S3)–(S5) are equations for three auxiliary functions, including Eq. (S3) for the state-dependent resistance $R_{ch}(u)$, Eq. (S4) for the state-dependent thermal conductance $\Gamma_{th}(u)$, and Eq. (S5) for the differential change of enthalpy $d\Delta H/du$ with respect to the state $u$.

The SPICE compact model of VO₂ devices is constructed in a similar manner as outlined in the supplementary materials of Ref. 16. All the SPICE simulations were performed on a personal computer using the LTspice IV software.

**Supplementary Note 4: Dynamics equations of an active memristor neuron circuit**

Starting with the active memristor device model equations from Ref. 16, after applying Kirchhoff's voltage law (KVL) and Kirchhoff's current law (KCL), we derived the four coupled first-order ordinary differential equations (ODEs) that drive the dynamics of a model tonic active memristor neuron circuit (See Supplementary Fig. 41). Similar procedure can be applied to derive the model equations for phasic neuron circuits, which are not included here. The reference convention is that the potential inside the nerve cell is fixed at the ground level. A positive current flows toward the ground, and a positive voltage will produce a current flowing toward the ground. Following the convention in biological neuron models, the d.c. biases $-V_1$ and $+V_2$ applied on the two active memristors $X_1$ and $X_2$ are replaced by electromotive forces $E_1$ and $E_2$ (amplitudes only). Their polarities are taken care of when incorporated into equations. For simplicity, we temporarily assume that there is no external load that draws a current at the cell output, and only consider the situation of current clamp, i.e. an external input current $I$ is fed into the cell. The case for voltage clamp can be derived in a similar manner.

To simplify the expressions, let's define $\mathcal{H}(u) \triangleq (d\Delta H/du)^{-1}$ for the differential change of enthalpy, $Q(u) \triangleq \Gamma_{th}(u)\Delta T$ for the heat flux, and remove the subscript "*ch*" of channel resistance in $R_{ch}(u)$. The model equations for the two active memristors are rewritten as:

$$v_1 = R(u_1) \cdot i_1 \tag{S6}$$

$$\frac{du_1}{dt} = \mathcal{H}(u_1) \cdot (i_1^2 R(u_1) - Q(u_1)) \tag{S7}$$

$$v_2 = R(u_2) \cdot i_2 \tag{S8}$$

$$\frac{du_2}{dt} = \mathcal{H}(u_2) \cdot (i_2^2 R(u_2) - Q(u_2)) \tag{S9}$$

Currents flowing through the two membrane capacitors $C_1$ and $C_2$ are:

$$I_{C1} = C_1 \frac{d(v_1 - E_1)}{dt} = C_1 \frac{dv_1}{dt} \tag{S10}$$

$$I_{C2} = C_2 \frac{d(v_2 + E_2)}{dt} = C_2 \frac{dv_2}{dt} \tag{S11}$$

Applying KCL at the joint connecting $R_{L1}$, $C_1$, $X_1$ and $R_{L2}$, the external input current is

$$I = I_{C1} + i_1 + I_{C2} + i_2 \tag{S12}$$

Substituting $I_{C1}$, $i_1$, $I_{C2}$ and $i_2$ with Eqs. (S10), (S6), (S11) and (S8), we have

$$I = C_1 \frac{dv_1}{dt} + \frac{v_1}{R(u_1)} + C_2 \frac{dv_2}{dt} + \frac{v_2}{R(u_2)} \tag{S13}$$

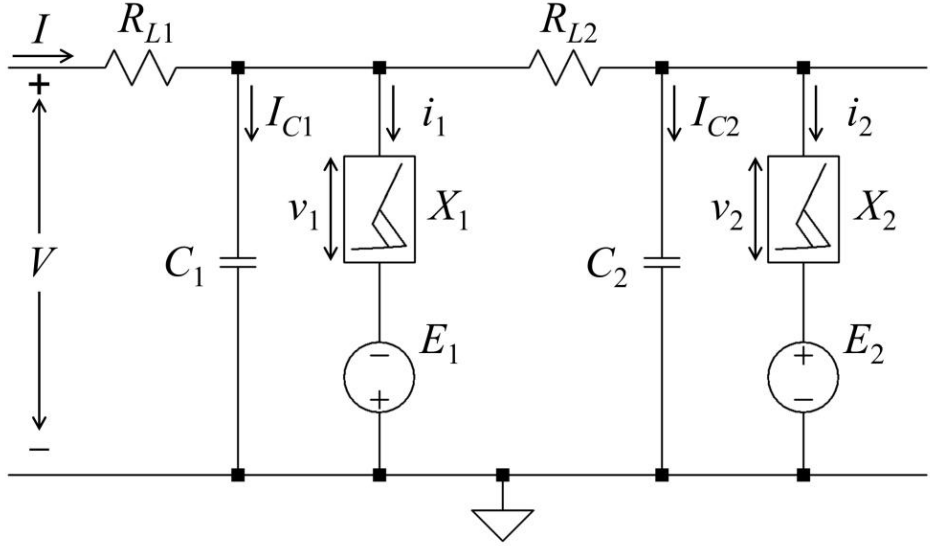

**Supplementary Figure 41. Circuit diagram of a tonic memristor neuron circuit with legends of voltages and currents to assist modeling.**

Applying KVL at the joint connecting $R_{L1}$, $C_1$, $X_1$ and $R_{L2}$,

$$v_1 - E_1 = \left(C_2 \frac{dv_2}{dt} + \frac{v_2}{R(u_2)}\right) R_{L2} + v_2 + E_2 \tag{S14}$$

Eq. (S14) is then rewritten in the form of a first-order ODE:

$$\frac{dv_2}{dt} = \frac{1}{R_{L2}C_2}\left[v_1 - \left(1 + \frac{R_{L2}}{R(u_2)}\right)v_2 - E_1 - E_2\right] \tag{S15}$$

Substituting $dv_2/dt$ in Eq. (13) with the formula of Eq. (S15), Eq. (13) can be rewritten in the form of a first-order ODE:

$$\begin{aligned}
\frac{dv_1}{dt} &= \frac{1}{C_1}\left\{I - \frac{v_1}{R(u_1)} - C_2\frac{dv_2}{dt} - \frac{v_2}{R(u_2)}\right\} \\
&= \frac{1}{C_1}\left\{I - \frac{v_1}{R(u_1)} - \frac{1}{R_{L2}}\left[v_1 - \left(1 + \frac{R_{L2}}{R(u_2)}\right)v_2 - E_1 - E_2\right] - \frac{v_2}{R(u_2)}\right\} \\
&= \frac{1}{C_1}\left\{I - \left(\frac{1}{R(u_1)} + \frac{1}{R_{L2}}\right)v_1 + \left(\frac{1}{R_{L2}} + \frac{1}{R(u_2)}\right)v_2 + \frac{E_1 + E_2}{R_{L2}} - \frac{v_2}{R(u_2)}\right\} \\
&= \frac{1}{C_1}\left\{I - \left(\frac{1}{R(u_1)} + \frac{1}{R_{L2}}\right)v_1 + \left(\frac{1}{R_{L2}}\right)v_2 + \frac{E_1 + E_2}{R_{L2}}\right\}
\end{aligned}$$

Multiplying both sides by $R_{L2}$, it becomes:

$$\frac{dv_1}{dt} = \frac{1}{R_{L2}C_1}\left[I \cdot R_{L2} - \left(1 + \frac{R_{L2}}{R(u_1)}\right)v_1 + v_2 + E_1 + E_2\right] \quad (S16)$$

Eqs. (S7), (S9), (S15) and (S16) are the four coupled first-order ODEs that solve the four state variables $(u_1, v_1, u_2, v_2)$ which drive the dynamics of the neuron circuit. They are grouped together as below

$$\begin{cases} \frac{dv_1}{dt} = \frac{1}{R_{L2}C_1}\left[I \cdot R_{L2} - \left(1 + \frac{R_{L2}}{R(u_1)}\right)v_1 + v_2 + E_1 + E_2\right] \\ \frac{dv_2}{dt} = \frac{1}{R_{L2}C_2}\left[v_1 - \left(1 + \frac{R_{L2}}{R(u_2)}\right)v_2 - E_1 - E_2\right] \\ \frac{du_1}{dt} = \mathcal{H}(u_1) \cdot \left(i_1^2 R(u_1) - Q(u_1)\right) \\ \frac{du_2}{dt} = \mathcal{H}(u_2) \cdot \left(i_2^2 R(u_2) - Q(u_2)\right) \end{cases}$$

Experimentally it's more convenient to probe the $Na^+$ and $K^+$ channel membrane potentials $V_{Na} = v_1 - E_1$ and $V_K = v_2 + E_2$ (note that $V_1$ and $V_2$ have been previously used to represent the d.c. biases). Substituting $v_1$ with $V_{Na}$ and $v_2$ with $V_K$ in the above equations, the dynamics equations can be recast as

$$\begin{cases} V'_{Na} = \frac{1}{C_1} \cdot \left[I - \left(\frac{1}{R_{L2}} + \frac{1}{R(u_1)}\right)V_{Na} + \frac{1}{R_{L2}}V_K - \frac{1}{R(u_1)}E_1\right] & (S17) \\ V'_K = \frac{1}{C_2} \cdot \left[\frac{1}{R_{L2}}V_{Na} - \left(\frac{1}{R_{L2}} + \frac{1}{R(u_2)}\right)V_K + \frac{1}{R(u_2)}E_2\right] & (S18) \\ u'_1 = \mathcal{H}(u_1) \cdot \left[\frac{(V_{Na} + E_1)^2}{R(u_1)} - Q(u_1)\right] & (S19) \\ u'_2 = \mathcal{H}(u_2) \cdot \left[\frac{(V_K - E_2)^2}{R(u_2)} - Q(u_2)\right] & (S20) \end{cases}$$

We noted that the dynamic equations (S17) and (S18) have been presented in Ref. 32. However, in that reference, the dynamic equations of state variables $u_1$ and $u_2$ are not used for reduced-dimension $V_{Na}$–$V_K$ nullcline analysis, instead a hard switching between two preset resistance values $R_{on}$ and $R_{off}$ were assumed. Such an overly-simplified approach unavoidably will miss some important aspects of the nonlinear dynamics of Eqs. (S7) and (S9). Applying Ohm's law, we rewrite Eqs. (S7) and (S9) as Eqs. (S19) and (S20), wherein the channel currents $i_1$ and $i_2$ are replaced by the corresponding membrane potentials $V_{Na}$ and $V_K$.

**Supplementary Note 5: Dynamic and static power scaling of VO$_2$ neurons**

We used SPICE simulations to analyze the dynamic and static power scaling of tonic VO$_2$ neurons (See Supplementary Figs. 37 and 38). The dynamic spike energy is calculated by first summing the total power supplied by the d.c. voltage sources, then integrating over time through the course of a spike.

In Supplementary Fig. 37, the dynamic spiking energy scales almost linearly with the capacitance of membrane capacitors, with a power-law fitted slope of 0.96 at $r/L$ = 10/10 nm and 0.924 at $r/L$ = 36/50 nm, respectively. The neuron area also scales linearly with the membrane capacitance as capacitor elements dominate the circuit area. It is therefore desirable to make smaller neurons to achieve higher (dynamic) spiking energy efficiency (EE) (See Supplementary Fig. 2). Note that area scaling of membrane capacitors is not a limiting factor for the neuron area scaling, because memristor neurons do not require a minimum value of membrane capacitors to operate, and therefore there is no size constraint posted by the requirement on certain membrane capacitance value. Another observation is that since VO$_2$ switching energy is extremely low (See Supplementary Table 1), only 1.15 fJ/device at $r/L$ = 10/10 nm, aggressive VO$_2$ device scaling is not needed: an 18-fold volume reduction, from $r/L$ = 36/50 nm to 10/10 nm, only reduces the neuron spike energy by ~24 %.

If only considering dynamic power consumption for the case of $r/L$ = 36/50 nm, <0.1 pJ/spike energy use can be achieved at a total capacitor area of ~1 µm$^2$ by using 20 fF membrane capacitors (see green arrow in Supplementary Fig. 37a), which can be realized by today's integrated high-$\kappa$ metal-insulator-metal (MIM) capacitors with a record-high capacitance density[21] of 43 fF/ µm$^2$ (typical MIM capacitance density of high-$\kappa$ dielectrics is in the range of 15–20 fF/µm$^2$). However, using high-$\kappa$ dielectric to boost the capacitance density is not a good strategy to achieve higher EE. At a given capacitor area, lower capacitance value (by using lower capacitance density) will result in a lower spiking energy and hence a better EE. This is clearly shown in the EE-area scaling trend lines of VO$_2$ neurons (See Supplementary Fig. 2) stimulated at capacitance density of 1, 10, and 43 fF/µm$^2$. At the same neuron (and capacitor) area, lower capacitance density translates into lower dynamic spiking energy and hence higher EE. At 1 fF/µm$^2$ capacitance density, VO$_2$ neurons show superior EE-area scaling than the best-case HH cells at neuron sizes smaller than 70 µm$^2$, and can surpass the estimated human brain EE of $1.8 \times 10^{14}$ spike/J (or 5.6 fJ/spike energy use) at neuron sizes smaller than 3 µm$^2$.

The static power consumption is dissipated by standby current through d.c. biased VO$_2$ devices due to the finite resistivity of the insulating phase (~1 Ω·cm)[24]. At low firing rates, static power may dominate the total power consumption. Since the dynamic power is proportional to the firing rate, while the static power remains a constant, the percentage of static power in total power decreases with firing rate. The lower and higher bounds of static power, estimated at d.c. bias of $V_{th}$ and $V_{th}/2$ ($V_{th}$ is the switching threshold) respectively, is <10 % of the total power at a firing rate higher than 100 MHz and 400 MHz, respectively, and the overall energy use is lower than 0.11 pJ/spike. We have not considered the possibility that the insulating-phase resistivity of VO$_2$ can be improved. Note that the neuron EE is not the only factor that determines the network power consumption. The firing rate, the synapse resistance, and the synapse/neuron ratio also need to be considered.

# Supplementary References

1. Esser, S. K. et al. Convolutional networks for fast, energy-efficient neuromorphic computing. *Proc. Nat. Acad. Sci.* **113**, 11441–11446 (2016).
2. Anonymous. GPU-based deep learning inference: a performance and power analysis. *Nvidia whitepaper* https://www.nvidia.com/content/tegra/embedded-systems/pdf/jetson_tx1_whitepaper.pdf (2015).
3. Ovtcharov, K. et al. Accelerating deep convolutional neural networks using specialized hardware. *Microsoft whitepaper* http://research.microsoft.com/apps/pubs/?id=240715 (2015).
4. Hauβmann, H. Comparing Google's TPUv2 against Nvidia's V100 on ResNet-50. *RiseML Blog* https://blog.riseml.com/comparing-google-tpuv2-against-nvidia-v100-on-resnet-50-c2bbb6a51e5e (2018).
5. Wijekoon, J. H. B. & Dudek, P. Compact silicon neuron circuit with spiking and bursting behavior. *Neural Netw.* **21**, 524–534 (2008).
6. Livi, P. & Indiveri, G. A current-mode conductance-based silicon neuron for Address-Event neuromorphic systems. *In IEEE Int. Symp. Circ. Sys.* https://doi.org/10.1109/ISCAS.2009.5118408 (2009).
7. Schemmel, J. et al. A wafer-scale neuromorphic hardware system for large-scale neural modeling. *2010 IEEE Int. Symp. Circ. Sys.* https://doi.org/10.1109/ISCAS.2010.5536970.
8. Yu, T., Park, J., Joshi, S., Maier, C. & Cauwenberghs, G. 65k-neuron integrate-and-fire array transceiver with address-event reconfigurable synaptic routing. *2012 IEEE Biomed. Circ. Sys. Conf.* https://doi.org/10.1109/BioCAS.2012.6418479 (2012).
9. Cruz-Albrecht, J. M., Yung, M. W. & Srinivasa, N. Energy-efficient neuron, synapse and STDP integrated circuits. *IEEE Trans. Biomed. Circuits Syst.* **6**, 246-256 (2012).
10. Joubert, A., Belhadj, B., Temam, O. & Heliot, R. Hardware spiking neurons design: analog or digital? *2012 IEEE Int. J. Conf. Neural Netw.* https://doi.org/10.1109/IJCNN.2012.6252600 (2012).
11. Park, J., Ha, S., Yu, T., Neftci, E. & Cauwenberghs, G. A 65k-neuron 73-Mevents/s 22-pJ/event asynchronous micro-pipelined integrate-and-fire array transceiver. *2014 IEEE Biomed. Circ. Sys. Conf.* https://doi.org/10.1109/BioCAS.2014.6981816 (2014).
12. Benjamin, B. V. et al. Neurogrid: a mixed-analog-digital multichip system for large-scale neural simulations. *Proc. IEEE* **102**, 699-716 (2014).
13. Merolla, P. A. et al. A million spiking-neuron integrated circuit with a scalable communication network and interface. *Science* **345**, 668-673 (2014).
14. Sengupta, B., Faisal, A. A., Laughlin, S. B. & Niven, J. E. The effect of cell size and channel density on neuronal information encoding and energy efficiency. *J. Cerebral Blood Flow & Metabolism* **33**, 1465–1473 (2013).
15. Hasler, J. M., B. Finding a roadmap to achieve large neuromorphic hardware systems. *Front. Neurosci.* **7**, 118 (2013).
16. Pickett, M. D. & Williams, R. S. Sub-100 fJ and sub-nanosecond thermally driven threshold switching in niobium oxide crosspoint nanodevices. *Nanotechnol.* **23**, 215202 (2012).
17. Gruver, G. W. *A study of one-port negative resistance oscillators utilizing four-layer diodes* M.S. thesis, Oklahoma State University, (1962).
18. Izhikevich, E. M. Which model to use for cortical spiking neurons? *IEEE Trans. Neural Netw.* **15**, 1063-1070 (2004).
19. Izhikevich, E. M. *Dynamical systems in neuroscience: The geometry of excitability and bursting*. (The MIT Press, 2007).

| | |
|---|---|
| 20 | Izhikevich, E. M., Desai, N. S., Walcott, E. C. & Hoppensteadt, F. C. Bursts as a unit of neural information: selective communication via resonance. *Trends Neurosci.* **26**, 161-167 (2003). |
| 21 | Ando, T. et al. CMOS compatible MIM decoupling capacitor with reliable sub-nm EOT high-k stacks for the 7 nm node and beyond. *2016 IEEE Int. Electron Dev. Meeting* https://doi.org/10.1109/IEDM.2016.7838382 (2016). |
| 22 | Kim, S., Du, C., Sheridan, P., Ma, W., Choi, S. & Lu, W. D. Experimental demonstration of a second-order memristor and its ability to biorealistically implement synaptic plasticity. *Nano Lett.* **15**, 2203–2211 (2015). |
| 23 | Oh, D.-W., Ko, C., Ramanathan, S. & Cahill, D. G. Thermal conductivity and dynamic heat capacity across the metal-insulator transition in thin film $VO_2$. *Appl. Phys. Lett.* **96**, 151906 (2010). |
| 24 | Berglund, C. N. G., H. J. Electronic properties of $VO_2$ near the semiconductor-metal transition. *Phys. Rev.* **185**, 1022-1033 (1969). |
| 25 | Eyert, V. The metal-insulator transitions of $VO_2$: A band theoretical approach. *Ann. Phys.* **11**, 650-702 (2002). |
| 26 | Pickett, M. D., Medeiros-Ribeiro, G. & Williams, R. S. A scalable neuristor built with Mott memristors. *Nat. Mater.* **12**, 114-117 (2013). |
| 27 | Crane, H. D. The neuristor. *IRE Trans. Elect. Comput.* **9**, 370-371 (1960). |
| 28 | Borghetti, J. et al. Oscillator circuitry having negative differential resistance. US Patent 8,324,976 B2. (2012). |
| 29 | Becker, M. F. et al. Femtosecond laser excitation of the semiconductor-metal phase transition in VO2. *Appl. Phys. Lett.* **65**, 1507-1509 (1994). |
| 30 | Cavalleri, A. et al. Femtosecond structural dynamics in $VO_2$ during an ultrafast solid-solid phase transition. *Phys. Rev. Lett.* **87**, 237401 (2001). |
| 31 | Yang, Z., Ko, C. & Ramanathan, S. Oxide electronics utilizing ultrafast metal-insulator transitions. *Annu. Rev. Mater. Res.* **41**, 337-367 (2011). |
| 32 | Lim, H. et al. Reliability of neuronal information conveyed by unreliable neuristor-based leaky integrate-and-fire neurons: a model study. *Sci. Rep.* **5**, 09776 (2015). |